\documentclass[aps,prd,twocolumn,amssymb,eqsecnum,showpacs,showkeyes,secnumarabic,graphics,floatfix,nofootinbib,tightenlines,longbibliography]{revtex4-1}
 
\usepackage{graphicx}
\usepackage{epsfig} 
\usepackage{bm}
\usepackage{cancel}  
\usepackage{dcolumn}  
\usepackage{amsmath} 
\usepackage{hhline}
\usepackage[utf8]{inputenc}
\usepackage{cancel}
\usepackage[dvipsnames]{xcolor}  
\usepackage{float}
\usepackage{longtable}
\usepackage[breaklinks=true,colorlinks=true,
linkcolor=blue,urlcolor=blue,citecolor=MidnightBlue,
bookmarks=true,bookmarksopenlevel=2]{hyperref}
\usepackage{mathrsfs}

\def \beq  {\begin{equation}}
\def \eeq  {\end{equation}}
\def \ber  {\begin{eqnarray}}
\def \eer  {\end{eqnarray}}

\begin{document}
\newcommand{\newc}{\newcommand}

\newc{\be}{\begin{equation}}
\newc{\ee}{\end{equation}}
\newc{\ba}{\begin{eqnarray}} 
\newc{\ea}{\end{eqnarray}}
\newc{\bea}{\begin{eqnarray*}}
\newc{\eea}{\end{eqnarray*}}
\newc{\D}{\partial}
\newc{\ie}{{\it i.e.} }
\newc{\eg}{{\it e.g.} }
\newc{\etc}{{\it etc.} }
\newc{\etal}{{\it et al.}}
\newc{\lcdm}{$\Lambda$CDM }
\newc{\lcdmnospace}{$\Lambda$CDM}
\newcommand{\nn}{\nonumber}
\newc{\ra}{\Rightarrow}
\newc{\omm}{$\Omega_{m}$ }
\newc{\ommnospace}{$\Omega_{m}$}
\newc{\fs}{$f\sigma_8$ }
\newc{\fsz}{$f\sigma_8(z)$ }
\newc{\fsnospace}{$f\sigma_8(z)$}
\newc{\plcdm}{Planck/$\Lambda$CDM }
\newc{\plcdmnospace}{Planck15/$\Lambda$CDM}
\newc{\wlcdm}{WMAP7/$\Lambda$CDM }
\newc{\wlcdmnospace}{WMAP7/$\Lambda$CDM}
\newcommand{\fss}{{\rm{\it f\sigma}}_8}

\title{Tension of the $E_G$ statistic and RSD data with Planck/$\Lambda$CDM  and implications for weakening gravity.} 

\author{F. Skara} 
\email{fskara@cc.uoi.gr} 

\author{L. Perivolaropoulos}
\email{leandros@uoi.gr} 

\affiliation{Department of Physics, University of Ioannina, 45110 Ioannina, Greece}

\date {\today}
 
\begin{abstract}
 
The  $E_G$ statistic is a powerful probe for detecting deviations from GR by combining weak lensing (WL), real-space clustering and redshift space distortion (RSD) measurements thus probing both the lensing and the growth effective Newton constants ($G_L$ and $G_{eff}$). We construct an up to date  compilation of $E_G$ statistic data including both redshift and scale dependence ($E_G(R,z)$). We combine this $E_G$ data compilation with an up to date compilation of  $f\sigma_8$ data from RSD observations to identify the current level of tension between the Planck/$\Lambda$CDM standard model based on general relativity and a general  model independent redshift evolution parametrization of $G_L$ and $G_{eff}$. Each \fs datapoint considered has been published separately in the context of independent analyses of distinct galaxy samples. However, there are correlations among the datapoints considered due to overlap of the analyzed galaxy samples. Due to these correlations the derived levels of tension of the best fit parameters with \plcdm are somewhat overestimated but this is the price to pay for maximizing the information encoded in the compilation considered.   
We find that the level of tension increases from about $3.5\sigma$ for the \fs data compilation alone to about $6\sigma$ when the $E_G$ data are also included in the analysis. The direction of the tension is the same as implied by the  $f\sigma_8$ RSD growth data alone (lower $\Omega_m$ and/or weaker effective Newton constant at low redshifts for both the lensing and the growth effective Newton constants ($G_L$ and $G_{eff}$)). These results further amplify the hints for weakening modified gravity discussed in other recent analyses \cite{Nesseris:2017vor,Kazantzidis:2018rnb,Perivolaropoulos:2019vkb,Kazantzidis:2019dvk}. 

\end{abstract} 
\maketitle
 
\section{INTRODUCTION}
\label{sec:Introduction}

The theory of general relativity (GR) and the standard $\Lambda$  cold dark matter ($\Lambda$CDM) \cite{Carroll:2000fy} cosmological model have been remarkably successful in explaining a wide array of observations \cite{Will:2014kxa} including the observed accelerating expansion of the universe \cite{Riess:1998cb,Perlmutter:1998np}. Despite of its  successes and  simplicity, the validity of the cosmological standard model $\Lambda$CDM  is currently under intense investigation. This is motivated by a range of profound theoretical and observational difficulties of the model. The most important theoretical difficulties of the $\Lambda$CDM model are the fine tuning \cite{Weinberg:1988cp,Martin:2012bt,Burgess:2013ara} and  coincidence problems \cite{1997cpp..conf..123S,Velten:2014nra}. The first of these problems corresponds to the large discrepancy between observations and quantum field theoretical predictions on the value of the cosmological constant $\Lambda$ while the second is associated with the coincidence between the observed vacuum energy density $\Omega_{\Lambda}$ and the matter density $\Omega_m$ which in the present epoch are of the same order of magnitude despite of their very different evolution during the cosmic history. 

A well known observational difficulty corresponds to the tension between the cosmic microwave background (CMB) measured value of the Hubble parameter $H_0$ \cite{Ade:2015xua,planck18} in the context of the $\Lambda$CDM  model and the local measurements from supernovae \cite{Riess:2016jrr,Riess:2018byc} and lensing time delay  indicators \cite{Birrer:2018vtm}, with local measurements suggesting a higher value. Another observational puzzle for \lcdm involves persisting indications from observational probes measuring the growth of matter perturbations that the observed growth is weaker than the growth predicted by the standard \plcdm  parameter values \cite{planck18}. Modified gravity (MG) models constitute a prime theoretical candidate to explain this tension.

The combination of cosmological observational probes is a powerful tool for the identification of signatures of MG \cite{Zhao:2008bn,Cai:2011wj,Joudaki:2011nw,Tereno:2010dt,Simpson:2012ra,Zhao:2015vta,Joudaki:2017zdt}. Such observational probes may be divided in two classes: geometric and dynamical (or  structure formation) probes  \cite{Bertschinger:2006aw,Nesseris:2006er,Basilakos:2013nfa,Ruiz:2014hma}.  Geometric observations measure cosmological distances using standard candles (e.g. Type Ia supernovae) and standard rulers (e.g. the horizon at the time of recombination probed through Baryon Acoustic Oscillations) and thus probe directly the cosmic metric, independent of the underlying theory of gravity. Dynamical observations probe the growth rate of cosmological perturbations and thus the gravitational laws and the consistency of GR with data provided the background geometry is known. 

Dynamical probes include cluster counts (CC) \cite{Rozo:2009jj,Rapetti:2008rm,Bocquet:2014lmj,Ruiz:2014hma}, weak lensing (WL) \cite{Schmidt:2008hc,kids1,cfhtlens,Joudaki:2017zdt,Troxel:2017xyo,kids2,des3,Abbott:2018xao} and redshift-space distortions (RSD) \cite{Samushia:2012iq,Macaulay:2013swa,Johnson:2015aaa,Nesseris:2017vor,Kazantzidis:2018rnb}. These probes are consistent with each other pointing either to a lower value of the matter density parameter $\Omega_{0m}$ in the context of GR or to weaker gravitational growth power  than the growth indicated by GR in  the  context  of a Planck$18/\Lambda$CDM background geometry at about $2-3\sigma$ level \cite{Macaulay:2013swa,Nesseris:2017vor,Kazantzidis:2018rnb}. Such weak growth may be  quantified by the parameter $\sigma_8$ which is the matter density rms  fluctuations within spheres of radius $8 h^{-1} Mpc$ and is determined by the amplitude of the primordial fluctuations power spectrum and by the growth rate of cosmological fluctuation. 

Various possible mechanisms have been proposed to slow down growth at low redshifts and thus reduce the above tension (see e.g. \cite{Kazantzidis:2019dvk}). Such mechanisms may be divided in two categories:  non-gravitational and gravitational. The former includes the effects of interacting dark energy models \cite{Kumar:2016zpg,Pourtsidou:2016ico,Barros:2018efl,Camera:2019vbp}, dynamical dark energy models \cite{Yang:2018qmz,Lambiase:2018ows}, running vacuum models \cite{Gomez-Valent:2017idt,Gomez-Valent:2018nib} and the effects of massive neutrinos \cite{DiazRivero:2019ukx}. The latter includes the effects of MG theories with a reduced (compared to GR) evolving effective Newton’s constant $G_{eff}$ at low redshifts \cite{Nesseris:2017vor,Kazantzidis:2018rnb}.  

The effects of MG \cite{Tsujikawa:2007gd,Hu:2007nk,Gannouji:2008wt,Capozziello:2011et,Clifton:2011jh,Nojiri:2010wj,Boubekeur:2014uaa,Cai:2018rzd,Perez-Romero:2017njc,Nojiri:2017ncd} models are  indistinguishable from GR at the geometric cosmological background level \cite{Bertschinger:2006aw,Song:2006ej,Brax:2008hh}. Signatures of MG can only be obtained by investigating the dynamics of cosmological perturbations \cite{Koyama:2015vza,Ishak:2018his} using specific statistics obtained through dynamical probe observables such as the two-point correlation of and power spectrum of the galaxy distribution, the RSD and WL. A useful  bias free statistic is the $f\sigma_8$ product of the rate of growth of matter density perturbations $f$ times  $\sigma_8$ discussed in more detail in what follows. An alternative observable statistic is the $E_G$ which was constructed to be independent of both the clustering bias factor $b$ and the parameter $\sigma_8$ on linear scales. This statistic   was proposed in 2007 \cite{Zhang:2007nk} and thereafter has been used several times to test MG theories \cite{Leonard:2015cba,delaTorre:2016rxm}. The expectation value of $E_G$ is equal to the ratio of the Laplacian of the sum of the Bardeen potentials \cite{Bardeen:1980kt}  $\Psi$ (the Newtonian potential) and $\Phi$ (the spatial curvature potential) $\nabla ^2(\Psi+\Phi)$ over the peculiar velocity divergence $\theta\equiv \nabla \cdot \frac{\vec{\upsilon}}{H(z)}$ (where  $\vec{\upsilon}$ is the peculiar velocity and $H(z)$ is the Hubble parameter in terms of the redshift $z$).   

The  $E_G$ statistic has been proposed as a model independent test of any MG theory \cite{Reyes:2010tr} and is constructed from three different probes of large scale structure (LSS): the galaxy-galaxy lensing (GGL), the galaxy clustering and the galaxy velocity field which leads to galaxy redshift distortions. Alternatively, $E_G$ may be constructed from galaxy-CMB lensing \cite{Ade:2015zua} instead of galaxy-galaxy lensing  as a more robust tracer of the lensing field at higher redshifts \cite{Pullen:2014fva,Pullen:2015vtb}. 

The first probe, the GGL (a special type of WL), is the slight distortion of shapes of source galaxies in the background of a lens galaxy, which arises from the gravitational deflection of light due to the gravitational potential of the lens galaxy along the line of sight (see for example \cite{Bartelmann:1999yn,Hoekstra:2003pn,Mandelbaum:2005wv,Kilbinger:2014cea}). This WL probe is sensitive to $\nabla^2(\Psi+\Phi)$, since relativistic particles collect equal contributions from the two Bardeen potentials which appear in the scalar perturbed  Friedmann-Lema\^{i}tre-Robertson-Walker (FLRW) metric in the Newtonian gauge \cite{Mukhanov:1990me,Ma:1995ey,EspositoFarese:2000ij} 
\be  
ds^2=-(1+2\Psi)dt^2+a^2(1-2\Phi)d \vec{x}^2
\label{metric} 
\ee  
where $a$  is the scale factor that is related to the redshift $z$ through $a=\frac{1}{1 +z}$.

The second probe, the galaxy clustering arises from the gravitational attraction of matter and is sensitive only to the potential $\Psi$. Similarly, the third probe, the galaxy velocity field, is quantified by measuring redshift space distortions (RSD) \cite{Kaiser:1987qv,Hamilton:1997zq,boss,Beutler:2012px} (an illusory anisotropy that distorts the distribution of galaxies in redshift space generated by their peculiar motions falling towards overdense regions). This  important probe of LSS is sensitive to the rate of growth of matter density perturbations $f$ which depends on the theory of gravity and provides measurements of $f\sigma_8$ that depends on the potential $\Psi$. 

\begin{table*}    
\centering  
\caption{\small  Planck$18$/$\Lambda$CDM parameters values \cite{planck18}  based on TT,TE,EE+lowE+lensing likelihoods.}
\label{planck18} 
\begin{tabular}{ccc c } 
\hhline{====}
   & \\
  Parameter&& &Planck$18$/$\Lambda$CDM   \\
    & \\
 \hline 
    & \\
  $\Omega_bh^2$&&&$0.02237\pm 0.00015$   \\
   $\Omega_ch^2$&&&$0.1200\pm 0.0012$   \\ 
   $n_S$&&&$0.9649\pm 0.0042$   \\
    $H_0$ $[kms^{-1}Mpc^{-1}]$&&&$67.36\pm 0.54$   \\
  $\Omega_{0m}$&&&$0.3153\pm 0.0073$   \\
   $w$&&&$-1$\\
   $\sigma_8$&&&$0.8111\pm 0.0060$ \\
 \hhline{====}   
\end{tabular} 
\end{table*} 

In most MG theories the potentials $\Phi$ and $\Psi$ obey generalized Poisson equations like the GR Newtonian potential where the MG effects are encoded in generalized space-time dependent effective Newton constants. These generalized Newton constants for the potential $\Psi$ and for the lensing combination $\Psi+\Phi$ are usually described by two parameters: the effective Newton's constant parameter  $\mu$ and the light deflection parameter $\Sigma$. In the  modified Poisson equations \cite{Bertschinger:2011kk} the $\mu$ and $\Sigma$ are connected with the potentials $\Psi$ and $\Psi+\Phi$ respectively. In GR the value of $\mu$ and  $\Sigma$ coincides with unity while in a MG model $\mu$ and  $\Sigma$ can be in general functions of both time and scale \cite{Bertschinger:2008zb,Zhao:2008bn}.  Using $f\sigma_8$ and $E_G$ datasets  constraints can be imposed on the parameters $\mu$ and  $\Sigma$ \cite{Zhao:2010dz,Daniel:2010ky,Song:2010fg,Daniel:2010yt,Simpson:2012ra,Hu:2013aqa,Ferte:2017bpf}). Such analyses have revealed various levels of tension of the best fit forms of  $\mu$ and  $\Sigma$ with the GR prediction of unity showing hints that these parameters may be less than unity implying weaker growth of perturbations than that predicted in GR. The goal of the present analysis is to extend these studies and use an updated data compilation for both the $f\sigma_8$ and $E_G$ statistics to identify the current level of tension with GR implied by these data compilations.
  
In particular, we address the following questions:
\begin{itemize}
\item
What are efficient phenomenological redshift dependent parametrizations of the generalized normalized Newton constants $\mu(z)$ and $\Sigma(z)$ that are consistent with solar system and nucleosynthesis constraints that indicate that GR is restored at high $z$ and at the present time in the solar system?
\item  
What are the constraints imposed by the $E_G$ and $f\sigma_8$ updated data compilations on the parameters of the above parametrizations and do these constraints amplify the hints for weakening gravity at low $z$ implied by the \fs data alone as indicated by previous studies?
\end{itemize}
The plan of this paper is the following: In the next Section \ref{sec:Theory} we present a brief review of the theoretical  expression for $E_G$. We also present phenomenologically motivated parametrizations for $\mu$ and  $\Sigma$ and describe how we use them in order to probe possible deviations from GR on cosmological scales. In Section \ref{sec:constraints} we use compilations  of  \fs and $E_G$ data along with the theoretical  expressions for $f\sigma_8$ and $E_G$ which involve $\mu$ and  $\Sigma$  to derive constraints on these parameters and to identify  the  tension  level  between the  \plcdm parameter  values  favoured  by  Planck  2018 \cite{planck18} shown in Table \ref{planck18} and the corresponding parameter values favored by the two datasets. Finally in Section \ref{sec:Discussion} we conclude, summarize and discuss the implications and possible extensions of our analysis. \\

\section{THEORETICAL BACKGROUND }
\label{sec:Theory}   
\subsection{$E_G$ statistic}
\label{Egform}

The $E_G$ statistic \cite{Zhang:2007nk,Reyes:2010tr} is designed as a probe of the ratio of the Bardeen potentials of the perturbed FRW metric (\ref{metric}) in such a way as to be independent  of the effects of galaxy bias at linear order. It is defined as the ratio of the cross correlation power spectrum $P_{g\nabla^2(\Phi+\Psi)}$ between lensing maps (cosmic shear or CMB) and galaxy positions, over the the cross-correlation power spectrum $P_{g\theta}$ between galaxies and velocity divergence field $\theta$
\be
 E_G \equiv \frac{P_{g\nabla^2 (\Phi+\Psi)}}{P_{g\theta}}
\label{EG0}
\ee
In Fourier space the $E_G$ statistic may also be expressed as \cite{Zhang:2007nk}
\be 
E_G(l,\Delta l)=\frac{C_{\kappa g} (l,\Delta l)}{3 H_0^2 a^{-1} \sum_{\alpha} q_{\alpha}(l,\Delta l) P_{vg}^{\alpha}}
\ee
where $H_0$ is the Hubble parameter today, $l$ is the magnitude of two-dimensional wavenumber of the on-sky Fourier space, $C_{\kappa g} (l,\Delta l)$ is the galaxy-galaxy lensing cross correlation power spectrum in bins of $\Delta l$, $P_{vg}^{\alpha}$ is the galaxy-velocity cross correlations power spectrum between $k_{\alpha}$ and $k_{\alpha +1}$ (where $k$ three-dimensional wavenumber of the on-sky Fourier transform with $k_1 < k_2 <... < k_{\alpha} < ...$) and $q_{\alpha}(l,\Delta l)$ is the weighting function defined accordingly. 

The corresponding expectation value of $E_G$, averaged over $l$ is the the ratio of the Laplacian of the gravitational scalar
potentials  $\Psi$ and $\Phi$ which appear in the scalar perturbed  Friedmann-Lema\^{i}tre-Robertson-Walker (FLRW) metric Eq. (\ref{metric}) over the peculiar velocity divergence \cite{Leonard:2015cba} 
\be 
\langle E_G \rangle= \left[\frac{\nabla ^2(\Psi+\Phi)}{3 H_0^2 a^{-1} \theta}\right]_{k=l/\bar{\chi},\bar{z}} 
\label{Egexp}
\ee
where  $\bar{\chi}$ is the comoving mean distance corresponding to the mean redshift $\bar{z}$. 

In $\Lambda$CDM cosmology and assuming that the velocity field is generated under linear perturbation theory, the peculiar velocity divergence is connected to the growth rate $f$ as $\theta=f \delta$ \cite{Bernardeau:2001qr}
where $\delta\equiv \frac{\delta\rho}{\rho}$ is the matter overdensity field, $\rho$ is the matter density of the background, $f(a)\equiv \frac{d\ln D(a)}{d\ln a}$ is the linear growth rate of structure and $D(a)\equiv\frac{\delta(a)}{\delta(a=1)}$ the growth factor. 

In the case of GR and in the absence of any anisotropic stress the Bardeen potentials are equal ($\Psi=\Phi$) and the gravitational field equations reduce to Poisson equations of the form 
\be
\nabla ^2\Phi= \nabla ^2\Psi=4\pi G a^2\rho\delta=\frac{3}{2}H_0^2\Omega_{0m}a^{-1}\delta
\label{poisson}
\ee
where $\Omega_{0m}=\Omega_m(z=0)$ is the matter density parameter today and  the second equality is straightforwardly derived assuming non-relativistic matter species and using the equations $H_0^2=\frac{8\pi G \rho_{c,0}}{3}$, $\rho=\rho_0 a^{-3}$ and $\Omega_{0m}=\frac{\rho_0}{\rho_{c,0}}$ (with $\rho_0$ the matter density today and $\rho_{c,0}$ the critical density today).

Therefore within GR Eq.(\ref{poisson}),  the Eq.(\ref{Egexp}) reduce to
\be 
E_G=\frac{\Omega_{0m}}{f(z)} 
\label{EgGR}   
\ee  
where $f$ is well approximated as $f(z)\simeq\Omega_m^{\gamma}(z)$ with the growth index $\gamma$ in a narrow range near 0.55, for a wide variety of dark-energy models in GR \cite{1980lssu.book.....P,Lahav:1991wc,Wang:1998gt,Linder:2005in,Polarski:2007rr,Linder:2007hg,Gannouji:2008jr,Nesseris:2015fqa,Polarski:2016ieb}. Note that $E_G$ in GR  is scale independent (Eq.(\ref{EgGR})). This is not necessarily the case in the context of MG theories  where the growth rate $f$ may be strongly scale dependent even on subhorizon scales.

\subsection{The effective Newton's constant parameter $\mu$ and the light deflection parameter $\Sigma$} 
\label{param}

The gravitational slip parameter $\eta$ describes the possible inequality \cite{Pogosian:2007sw,Jain:2010ka} of the two Bardeen potentials that may occur in MG theories. It is defined as 
\be 
\eta(a,k)=\frac{\Phi(a,k)}{\Psi(a,k)} 
\label{slip}
\ee  
Clearly an observation of $\eta\neq 1$ would indicate physics beyond GR.  In this case the gravitational field equations at linear level take the form of  Poisson equations that generalize Eqs. (\ref{poisson}).   At linear level, in MG models, using the perturbed metric (\ref{metric}) and the gravitational field equations the following  phenomenological equations emerge \cite{Amendola:2007rr,Bertschinger:2008zb,Pogosian:2010tj,Daniel:2012kn,Huterer:2013xky,Johnson:2015aaa,Mueller:2016kpu,Perenon:2019dpc} for the scalar perturbation potentials
\be 
k^2(\Psi+\Phi)=-8\pi G_N\Sigma(a,k)a^2\rho\Delta 
\label{poissonsigma} 
\ee 
\be  
k^2\Psi=-4\pi G_N\mu(a,k)a^2\rho\Delta 
\label{poissonmu} 
\ee
where $\rho$ is the matter density of the background, $\Delta$ the comoving matter density contrast defined as $\Delta\equiv\delta+3Ha(1+w)\upsilon/k$ which is gauge-invariant \cite{Pogosian:2010tj}, $w =p/\rho$  is the equation-of-state parameter  and $\upsilon^i=-\nabla^i u$  is the irrotational component of the velocity field. Also $\mu$ and $\Sigma$ are the generalized growth and lensing effective Newton constants. They are in general functions of time and scale encoding the possible modifications of General Relativity defined as\footnote{Note that, in the literature $\mu$ and $\Sigma$ are also  referred to as $G_M$  and $G_L$ (e.g. in Refs. \cite{Mueller:2016kpu,Nesseris:2017vor})  or as $G_{matter}$  and $G_{light}$ (e.g. in Refs. \cite{Daniel:2012kn,Johnson:2015aaa}).}
\bea 
\mu(a,k) &\equiv& \frac{G_{eff}(a,k)}{G_N}  \\ 
\Sigma(a,k)&\equiv& \frac{G_L(a,k)}{G_N}
\eea
with $G_N$ is the Newton's constant as measured by local experiments, $G_{eff}$ is the effective Newton's constant which is related to the growth of matter perturbation and $G_L$ is related to the lensing of light (the propagation of relativistic particles, such as photons when  they traverse equal regions of space and time along null geodesics experiencing gravitational lensing collecting equal contributions from two gravitational potentials $\Psi$ and $\Phi$). Using the gravitational slip Eq.(\ref{slip}) and the ratios of the Poisson equations  (\ref{poissonsigma}), (\ref{poissonmu}) defined above the two LSS functions  $\mu$ and $\Sigma$ are related via  
\be 
\Sigma(a,k)=\frac{1}{2}\mu(a,k)\left[1+\eta(a,k)\right]
\ee
In GR which  predicts  a  constant  homogeneous $G_{eff}=G_N$, we obtain $\mu=1$, $\eta=1$ and $\Sigma=1$.

Notice that Eqs. (\ref{poissonsigma}) and (\ref{poissonmu}) indicate that a possible observation of reduced gravitational growth of the Bardeen potentials may be interpreted either as reduced strength of gravitational interaction (reduced $\mu$ and/or $\Sigma$) or due to reduced matter density $\rho$ (or $\Omega_{0m}$). In the context of a fixed value of matter density determined by geometric probes of the cosmological background, the reduced gravitational growth could be either interpreted as a tension within the \lcdm parameter value for the matter density or as a hint for weakening gravity. Indeed, such hints of weaker than expected gravitational growth of the Bardeen potentials has been observed at low redshifts by a wide range of dynamical probes including RSD observations \cite{Macaulay:2013swa,Johnson:2015aaa,Nesseris:2017vor,Kazantzidis:2018rnb}, WL \cite{kids1,Joudaki:2017zdt,Troxel:2017xyo,kids2,des3,Abbott:2018xao} and CC data \cite{Rozo:2009jj,Rapetti:2008rm,Bocquet:2014lmj,Ruiz:2014hma}. In most cases this weak growth has been interpreted as a tension for the parameters $\sigma_8$ and $\Omega_{0m}$ which are found by dynamical probes to be smaller than the values indicated by geometric probes in the context of \lcdm .

The observables $f\sigma_8(a,k)$ and $E_G(a,k)$ can probe directly the gravitational strength functions $\mu(a,k)$ and $\Sigma(a,k)$. In particular \fs is easily expressed in terms of the amplitude $\sigma_8$ and the matter overdensity $\delta$ using the matter overdensity evolution equation (see e.g. \cite{EspositoFarese:2000ij})
\be  {\ddot \delta}+2H\dot{\delta}-4\pi G_N \mu(a,k) \rho\delta\simeq 0
\label{eq:odedeltat}
\ee
where the dot denotes differentiation with respect to cosmic time $t$. In terms of redshift  Eq. (\ref{eq:odedeltat}) takes the form \cite{EspositoFarese:2000ij,Nesseris:2017vor}
\begin{widetext}
\begin{equation}
\delta''(z) + \left(\frac{(H(z)^2)'}{2~H(z)^2} -
{1\over 1+z}\right)\delta'(z)
-{3\over 2} \frac{(1+z)~\Omega_{0m}~ \mu(z,k)
}{H(z)^2/H_0^2}~\delta(z)=0
\label{eq:odedeltaz}
\end{equation}
where primes denote differentiation with respect to the redshift. 
While in terms of the scale factor we have \cite{Tsujikawa:2007gd,Nesseris:2015fqa,Arjona:2018jhh}  
\be
\delta''(a)+\left(\frac{3}{a}+\frac{H'(a)}{H(a)}\right)\delta'(a)
-\frac{3}{2}
\frac{\Omega_{0m} \mu(a,k)}
{a^5 H(a)^2/H_0^2} 
\delta(a)=0 
\label{eq:odedeltaa}
\ee
\end{widetext}
here primes denote differentiation with respect to the scale  factor.
In Eqs. (\ref{eq:odedeltaz}), (\ref{eq:odedeltaa}) possible  deviations  from  GR  are expressed by allowing for a scale and redshift-dependent $\mu=\mu(z,k)$. In the present section and in section \ref{scalindep} we ignore scale dependence due to the lack of good quality scale dependent \fs and $E_G$ data. However, in section \ref{scaledep} we discuss the scale dependence of $E_G$ data. 

For a given parametrization of $\mu(a)$ and initial conditions deep in the matter era where GR is assumed to be valid leading to $\delta \sim a$ equations (\ref{eq:odedeltaz}), (\ref{eq:odedeltaa}) may be easily solved numerically leading to a predicted form of $\delta(a)$ for a given $\Omega_{0m}$ and background expansion $H(z)$. In the context of the present analysis we assume a \lcdm backgroung $H(z)$
\be 
H^2(z) =H_0^2\left[\Omega_{0m}(1+z)^3 +(1-\Omega_{0m})\right]
\label{eq:hzwcdm} 
\ee 
Once the evolution of $\delta$ is known, the observable product $f\sigma_8(a)\equiv f(a)\cdot \sigma(a)$  can be obtained using the definitions 
\be
f(a)\equiv\frac{ d\ln \delta(a)}{ d\ln a} 
\label{fa}
\ee 
\be 
\sigma(a)\equiv\sigma_8 \frac{\delta(a)}{\delta(a=1)} 
\ee
where $\sigma(a)$ is the redshift dependent rms  fluctuations of the linear density field within spheres of radius $R=8 h^{-1} Mpc$ and $\sigma_8$ is its value today.
Thus, we have 
\be
\fss(a,\sigma_8,\Omega_{0m},\mu)=\frac{\sigma_8}{\delta(a=1)}~a~\delta'(a,\Omega_{0m},\mu) 
\label{eq:fs8}
\ee
This theoretical prediction may now be used to compare with the observed \fs data and obtain fits for the parameters $\Omega_{0m}$, $\sigma_8$ and $\mu(z)$ (assuming a specific parametrization of $\mu(z)$). 

The lensing gravity parameter $\Sigma(z)$ can be fit in the context of specific parametrizations using its connection with the $E_G(a)$ observable as \cite{Amendola:2012ky,Motta:2013cwa,Pinho:2018unz}

\be 
E_G(a,\Omega_{0m},\mu,\Sigma)=\frac{\Omega_{0m}\Sigma(a)}{f(a,\Omega_{0m},\mu)}
\label{Egth}
\ee
This equation assumes that the redshift of the lens galaxies can be approximated by a single value while $E_G$ corresponds to average value along the line of sight \cite{Pinho:2018unz}. In the context of Eq. (\ref{Egth}) and assuming a specific parametrization for $\mu$ and $\Sigma$, the theoretical prediction for $E_G$ may be used to compare with the observed $E_G$ datapoints and lead to constraints on $\Omega_{0m},\mu,\Sigma$. These constraints may be considered either separately from those of the \fs data or jointly by combining the $E_G$ and \fs datasets. The allowed range of these parameters may then be compared with the standard \plcdm parameter values $\mu=1$, $\Sigma=1$, $\Omega_{0m}=0.315\pm 0.0073$, $\sigma_8=0.811\pm0.006$ to identify the likelihood of \plcdm in the context of the dynamical probe data $E_G$ and \fs. This plan is implemented in what follows in the context of specific parametrizations describing the possible evolution of $\mu$ and $\Sigma$.

On  scales  much  smaller  than  the Hubble scale for most  modified  gravity  models  the scale  dependence  of $\mu$ and $\Sigma$ is  weak. For example in scalar-tensor (ST) model (for $k\gg aH$) $\mu$  is independent of the scale \cite{Sanchez:2010ng}. Thus, we start by considering scale independent parametrizations  for $\mu$ and $\Sigma$ which reduce to the GR value at early times and at the present time as indicated by solar system (ignoring possible screeing effects) and  Big Bang Nucleosynthesis constraints ($\mu=1$ and $\mu'=0$ for $a=1$ and $\mu=1$ for $a\ll1$) \cite{Muller:2005sr,Pitjeva:2013chs,Gannouji:2006jm}. Such parametrizations are  of the form \cite{Nesseris:2006hp,Nesseris:2017vor,Kazantzidis:2018rnb}
\begin{widetext}
\be 
\mu =1+g_a(1-a)^n-g_a(1-a)^{2n}=1+g_a(\frac{z}{1+z})^n-g_a(\frac{z}{1+z})^{2n}
\label{muaz} 
\ee
\be 
\Sigma =1+g_b(1-a)^m-g_b(1-a)^{2m}=1+g_b(\frac{z}{1+z})^m-g_b(\frac{z}{1+z})^{2m}
\label{sigmaaz}  
\ee
 \end{widetext}
where  $g_a$ and $g_b$ are  parameters  to   be   fit and $n$ and $m$ are integer parameters with  $n\geq2$ and $m\geq2$ which we set equal to $2$ in the present analysis.

\begin{figure}  
\begin{centering}
\includegraphics[width=0.45\textwidth]{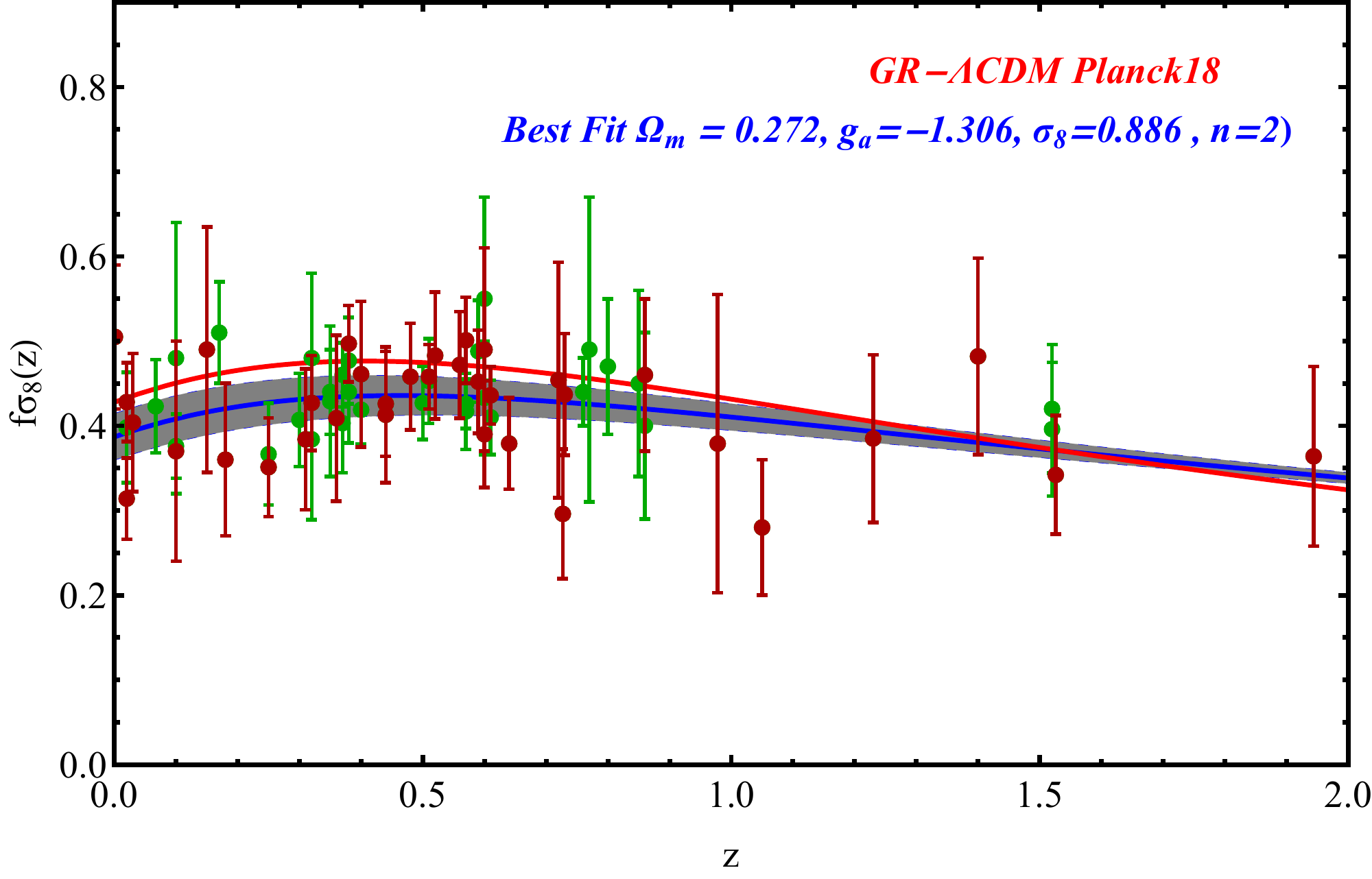}
\par\end{centering}
\caption{The $f\sigma_8(z)$ data compilation from Table \ref{tab:data-rsd}  used in the present analysis. The  subset of the data with less correlation is indicated with dark red. The red curve shows the Planck18/$\Lambda$CDM   prediction (parameter values $\Omega_{0m}=0.315$, $g_a=0$, $\sigma_8=0.811$), the blue curve shows the best fit of the $f\sigma_8(z)$ in the context of parametrizations Eq.(\ref{muaz}) with a \lcdm background (parameter values $\Omega_{0m}=0.272$, $g_a=-1.306$, $\sigma_8=0.886$) and the shaded regions correspond to $1\sigma$ confidence level around the best fit (see also Table \ref{bestfitpar}).}
\label{figfs8z}  
\end{figure} 

\begin{figure} 
\begin{centering}
\includegraphics[width=0.45\textwidth]{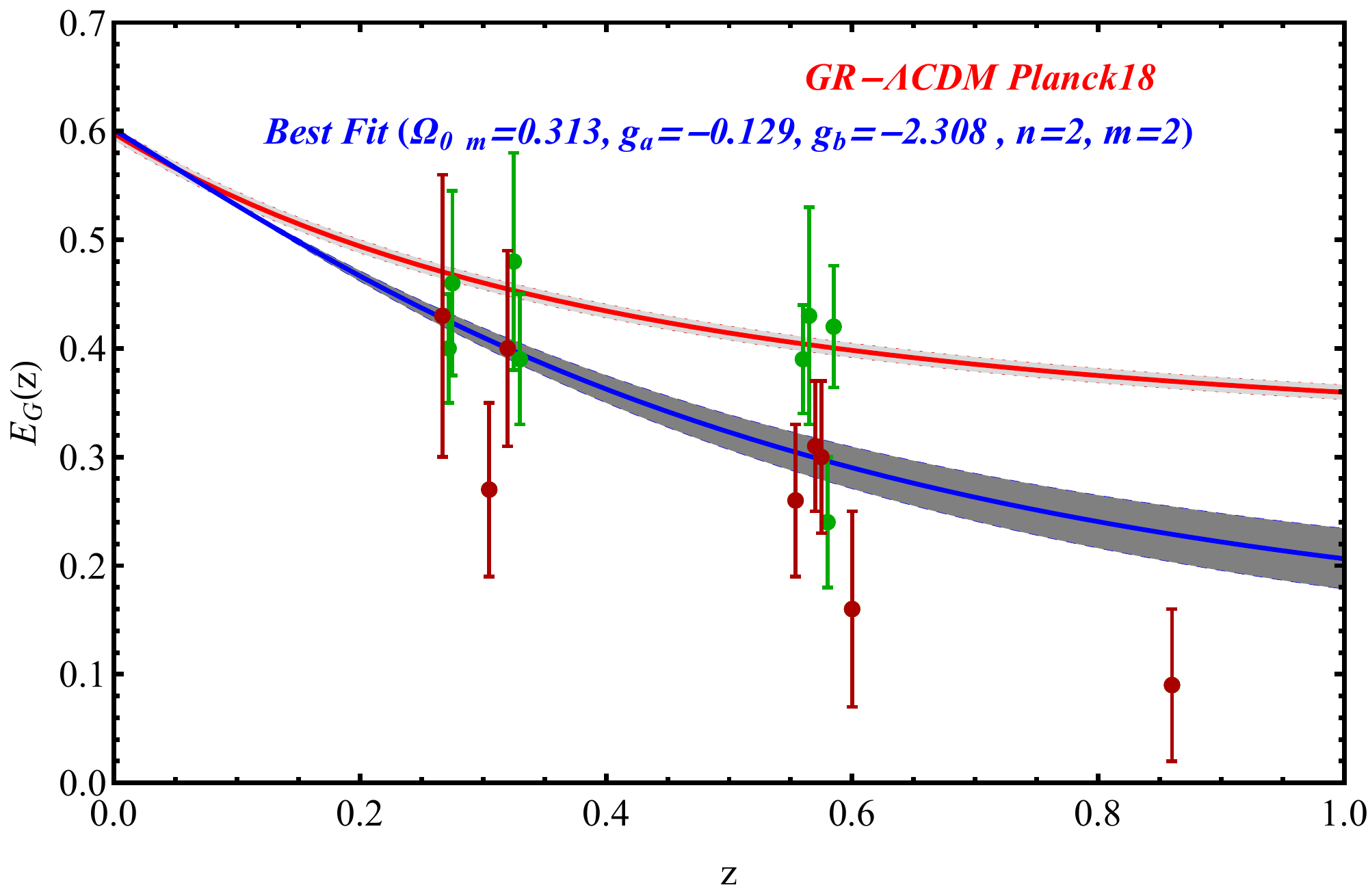}
\par\end{centering}
\caption{The $E_G(z)$ data compilation from Table \ref{tab:data-EG}  (scales $3<R<150 h^{-1} Mpc$) used in the present analysis. The  subset of the data with less correlation is indicated with dark red. The red curve shows the theoretical prediction based on the  Planck18/$\Lambda$CDM   parameter values ($\Omega_{0m}=0.315$, $\sigma_8=0.811$, $\mu=1$, $\Sigma=1$), the blue curve shows the best fit theoretical prediction based on the parametrizations (\ref{muaz}) and (\ref{sigmaaz}) with parameter values ($\Omega_{0m}=0.313$, $g_a=-0.129$, $g_b=-2.308$). Notice that the best fit is significantly below the \plcdm theoretical prediction and implies weaker gravity ($\mu<1$ and $\Sigma<1$) at the $4.6\sigma$ level (see also Table \ref{bestfitpar}). } 
\label{figegz}  
\end{figure} 

\section{Observational Constraints }
\label{sec:constraints}   
\subsection{Scale Independent Analysis}
\label{scalindep}

The $f\sigma_8(z)$ and $E_G(z)$ updated data compilations used in our analysis are shown in Tables \ref{tab:data-rsd}  and \ref{tab:data-EG} of the Appendix \ref{sec:Appendix_B} along with the references where each datapoint was originally published. The datapoints are also shown in Figs. \ref{figfs8z} and \ref{figegz}  along with curves corresponding to the \plcdm prediction and the best fit parameter values.
As it can be seen the datapoints from the various surveys are consistent with each other at any given redshift and at $1\sigma$ level. Clearly, in both cases the data appear to favor lower values of \fs and $E_G$ than the values corresponding to the \plcdm parameters.
This trend combined with the indications for a Planck/$\Lambda$CDM  background from geometric probes may be interpreted as a need for a new degree of freedom which in our approach is coming from MG. In addition, we see that there is no tension between different $f\sigma_8$ datapoints. Instead, there is a combined trend of the datapoints to be in tension with the Planck/$\Lambda$CDM  prediction. This tension disappears when we keep the same $\Lambda$CDM background but allow for a MG evolution of the effective Newton's constant. In fact, this trend may be shown to be translated into a trend for lower values for the gravitational parameters $\mu$ and $\Sigma$ and is quantified through a detailed maximum likelihood analysis.
 
Each $f\sigma_8(z)$ and $E_G(z)$ datapoint of the compilations of Tables \ref{tab:data-rsd}  and \ref{tab:data-EG} has been published separately in the context of independent analyses of distinct galaxy samples and lensing data. However, the correlations among the datapoints considered due to overlap of the analyzed galaxy samples may lead to an amplification of the existing trends indicated by the data and an amplification of the existing tension of the best fit parameters with \plcdm. Despite of this fact we have chosen to keep the relatively large number of distinct published datapoints in order to maximize the information encoded in the compilations considered keeping in mind that this may lead to an artificial amplification of the trends that already exist in the data.

An additional motivation for keeping the full set of published datapoints is that it is not always clear which one of the correlated points is more suitable to keep. Ignoring one of the correlated points arbitrarily or simply based on time of publication criteria could lead to loss of useful information or selection bias.

Keeping the full set of points does not significantly change the results and the level of tension between the growth data best fit parameter values corresponding to MG and Planck/$\Lambda$CDM  best fit in the context of GR. In order to demonstrate the validity of the above reasons we have repeated our analysis for a subset of the  $f\sigma_8$ and $E_G$ data  where we have removed most earlier data that were subject to correlations with more recent data as indicated with bold font in the index of the Tables  \ref{tab:data-rsd} and \ref{tab:data-EG} and as shown in Figs. \ref{figfs8z} and \ref{figegz} with dark red. The result was a data compilation of about half the $f\sigma8$  and $E_G$  datapoints with significantly less correlation. The results of the statistical analysis of this dataset are presented in Appendix \ref{sec:Appendix_A} and indicate a minor reduction of the overall tension.\\

\begin{figure*}
\begin{centering}
\includegraphics[width=0.97\textwidth]{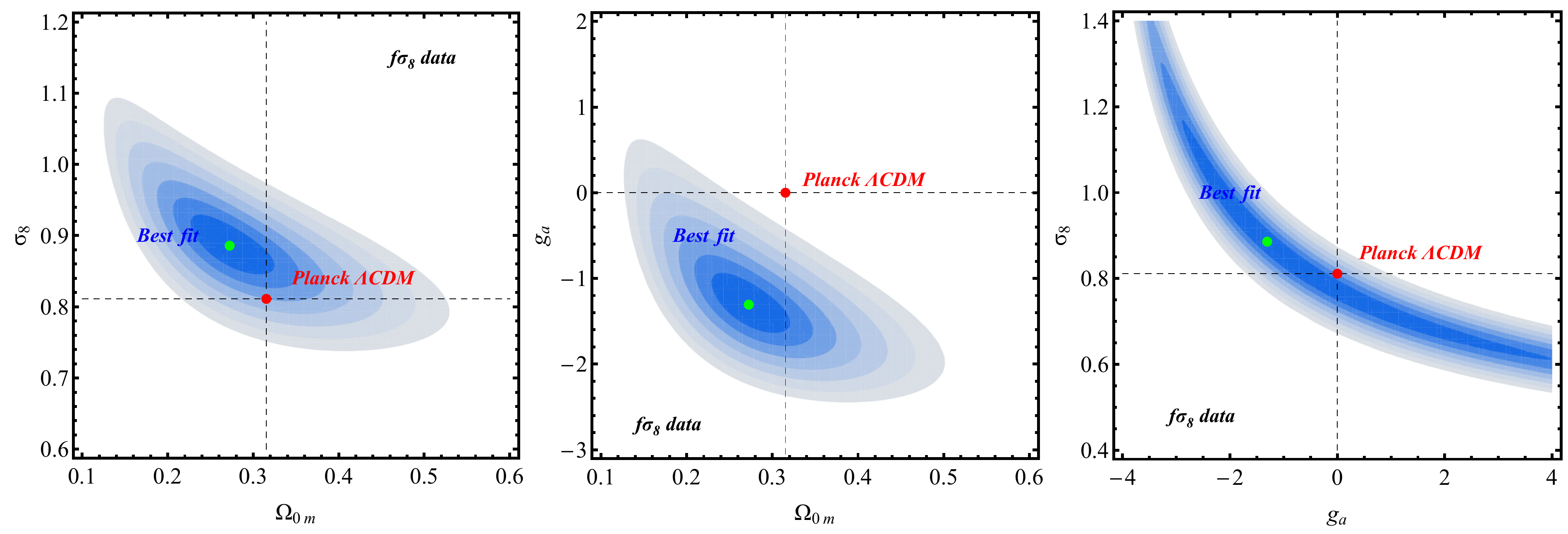}
\par\end{centering}
\caption{The three $1\sigma$ - $7\sigma$ confidence contours  in 2D projected parameter spaces of the parameter space ($\Omega_{0m},\sigma_8,g_a$) in the context of parametrization Eq.(\ref{muaz}) with $n=2$ including the fiducial correction factor Eq. (\ref{corr}). The RSD data $f\sigma_8(z)$ from Table \ref{tab:data-rsd} of  the Appendix \ref{sec:Appendix_B} was used. The third  parameter in each contour  was fixed to the best fit value. The red and green dots describe the Planck18/$\Lambda$CDM best fit and the best-fit values from data.}
\label{contfs82fid}  
\end{figure*}
\begin{figure*}
\begin{centering}
\includegraphics[width=0.97\textwidth]{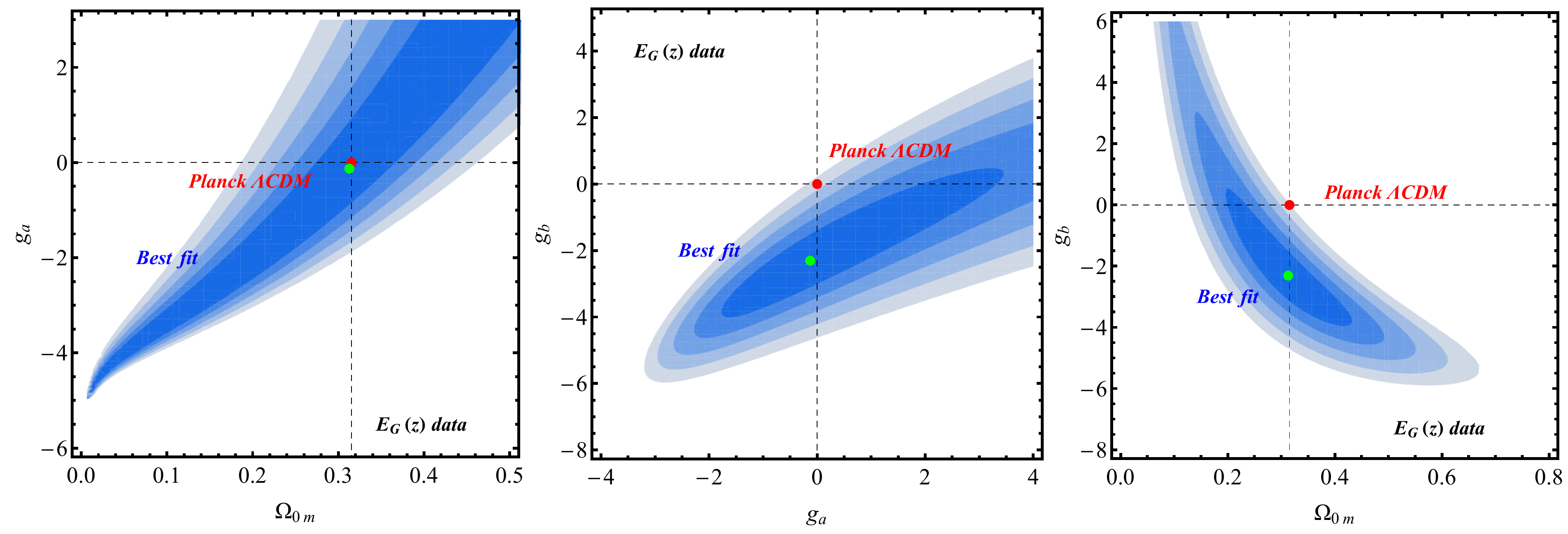}
\par\end{centering}
\caption{The three $1\sigma$ - $5\sigma$ confidence contours  in 2D projected parameter spaces  of the parameter space ($\Omega_{0m},g_a,g_b$) in the context of parametrizations Eqs.(\ref{muaz}) and (\ref{sigmaaz}) with $n=2, m=2$. The data $E_G(z)$ from Table \ref{tab:data-EG} of  the Appendix \ref{sec:Appendix_B} were used. The third  parameter in each contour  was fixed to the best fit value. The red and green dots describe the Planck18/$\Lambda$CDM best fit and the best-fit values from data. } 
\label{contEgz} 
\end{figure*}

\begin{figure*} 
\begin{centering}
\includegraphics[width=0.87\textwidth]{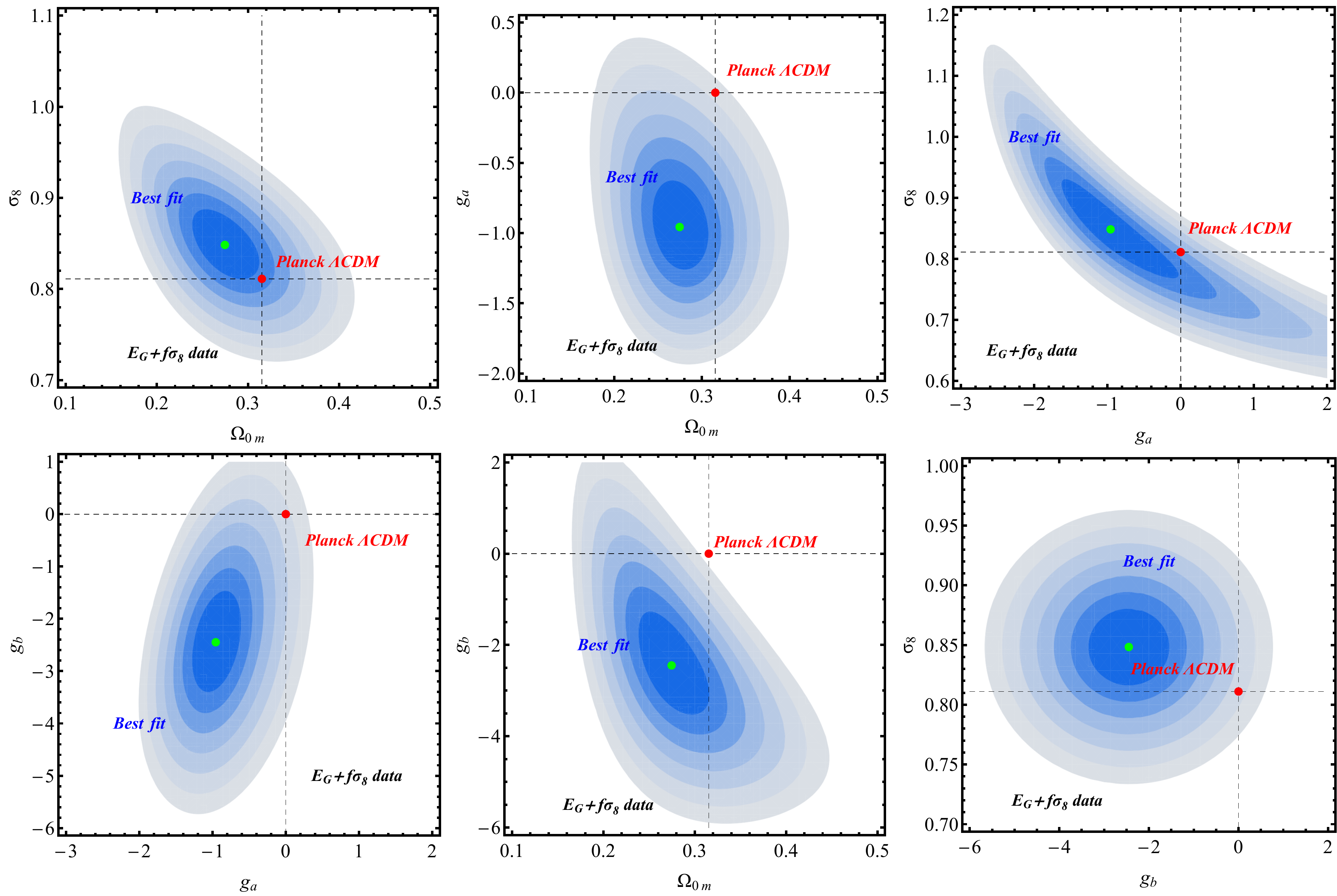}
\par\end{centering}
\caption{The six $1\sigma$ - $7\sigma$ confidence contours  in  2D projected parameter spaces  of the parameter space ($\Omega_{0m},\sigma_8,g_a,g_b$) in the context of parametrizations Eqs.(\ref{muaz})and (\ref{sigmaaz}) with $n=2$ and $m=2$ including the fiducial correction factor Eq. (\ref{corr}). The data $E_G(z)$ and $f\sigma_8(z)$ from Tables \ref{tab:data-EG} and \ref{tab:data-rsd} of  the Appendix \ref{sec:Appendix_B} was used. The third and the forth parameter in each contour  were fixed to the best fit values. The red and green dots describe the Planck18/$\Lambda$CDM best fit and the best-fit values from data.}
\label{contEgfs82fid}   
\end{figure*} 

\begin{figure*}
\begin{centering}
\includegraphics[width=0.87\textwidth]{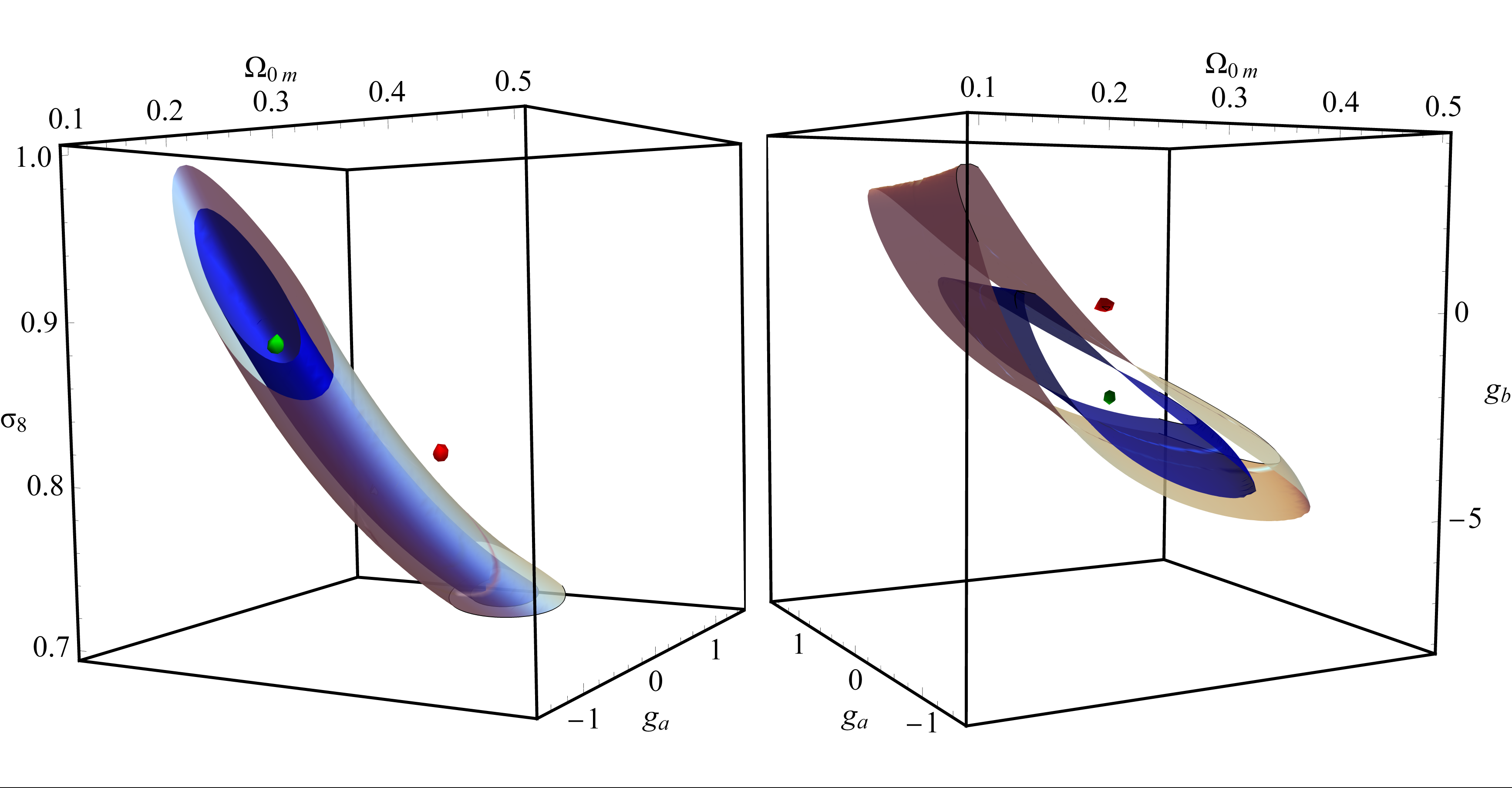}
\par\end{centering}
\caption{Left: The  $1\sigma$ - $2\sigma$ confidence contour  of the parameter space ($\Omega_{0m},\sigma_8,g_a$) in the context of parametrization Eq.(\ref{muaz}) with $n=2$ including the fiducial correction factor Eq. (\ref{corr}). The RSD data $f\sigma_8(z)$ from Table \ref{tab:data-rsd} of  the Appendix \ref{sec:Appendix_B} was used. The red and green dots describe the Planck18/$\Lambda$CDM best fit and the best-fit values from data.
Right: The  $1\sigma$ - $2\sigma$ confidence contour  of the parameter space ($\Omega_{0m},g_a,g_b$) in the context of parametrizations Eqs.(\ref{muaz}) and (\ref{sigmaaz}) with $n=2$. The data $E_G(z)$ from Table \ref{tab:data-EG} of  the Appendix \ref{sec:Appendix_B} was used. The red and green dots describe the Planck18/$\Lambda$CDM best fit and the best-fit values from data. The 3D contours include only the surfaces in 3D while the intermediate space is not filled. Thus, the white gaps that appear in the right figure between the surfaces, simply correspond to the white background seen from behind.}
\label{cont3d}  
\end{figure*}

For the construction of the likelihood contours of the model parameters in the context of the \fs and $E_G$ datasets we construct $\chi^2_{f\sigma_8}$ and $\chi^2_{E_G}$
For the construction of $\chi^2_{f\sigma_8}$ we use the vector \cite{Kazantzidis:2018rnb}
\be 
V_{f\sigma_8}^i(z_i,p)\equiv f\sigma_{8,i}^{obs}-\frac{f\sigma_8^{th}(z_i,p)}{q(z_i,\Omega_{0m},\Omega_{0m}^{fid})}
\ee 
where $f\sigma_{8,i}^{obs}$ is the the value of the $i$th datapoint, with $i= 1,...,N_{f\sigma_8}$ (where $N_{f\sigma_8}=66$ corresponds to the total number of datapoints of Table \ref{tab:data-rsd}) and $f\sigma_8^{th}(z_i,p)$ is the theoretical prediction, both at  redshift $z_i$. The parameter vector $p$ corresponds to the parameters $\sigma_8,\Omega_{0m},g_a$ of Eq. (\ref{eq:fs8}) with the parametrization (\ref{muaz}).
The fiducial Alcock-Paczynsk correction factor $q$ \cite{Macaulay:2013swa,Nesseris:2017vor,Kazantzidis:2018rnb}  is defined as 
\be  
q(z_i,\Omega_{0m},\Omega_{0m}^{fid})=\frac{H(z_i)d_A(z_i)}{H^{fid}(z_i)d_A^{fid}(z_i)}
\label{corr} 
\ee
where  $H(z), d_A(z)$ correspond to the Hubble parameter and the angular diameter distance of the true cosmology and the superscript $^{fid}$ indicates the fiducial cosmology used in each survey to convert angles and redshift to distances for evaluating the correlation function. As shown in Table \ref{bestfitpar}, the effects of this correction factor are less than about $10\%$ in the derived best fit parameter values. 
Thus we obtain $\chi_{f\sigma_8}^2$ as  
\be
\chi_{f\sigma_8}^2(\Omega_{0m},\sigma_8,g_a)=V_{f\sigma_8}^iF_{f\sigma_8,ij}V_{f\sigma_8}^j
\label{chif}
\ee
where $F_{f\sigma_8,ij}$ is  the  Fisher  matrix (the  inverse  of  the covariance matrix $C_{f\sigma_8,ij}$ of the data) which is assumed to be diagonal with the exception of the $3\times3$ WiggleZ subspace (see \cite{Kazantzidis:2018rnb} for more details on this compilation).

Similarly, for the construction of $\chi_{E_G}^2$, we consider the vector
\be  
V_{E_G}^i(z_i,p)\equiv E_{G,i}^{obs}-E_G^{th}(z_i,p)
\ee 
where $E_{G,i}^{obs}$ is the the value of the $i$th datapoint, with $i= 1,...,N_{E_G}$ (where $N_{E_G}=16$ corresponds to the total number of datapoints of Table \ref{tab:data-EG}), while $E_G^{th}(z_i,p)$ is the theoretical prediction (Eq. (\ref{Egth})), both at redshift $z_i$. The parameter vector $p$ corresponds to the parameters of Eq. (\ref{Egth}) with the parametrization (\ref{muaz}) namely $\Omega_{0m}$, $g_a$, $g_b$.

Thus we obtain $\chi_{E_G}^2$ as 
\be
\chi_{E_G}^2 (\Omega_{0m},g_a,g_b)=V_{E_G}^iF_{E_G,ij}V_{E_G}^j
\label{chieg}
\ee
where $F_{E_G,ij}$ is  the  Fisher  matrix also assumed to be diagonal.

By minimizing $\chi_{f\sigma_8}^2$,  $\chi_{E_G}^2$ separately and combined as $\chi_{tot}^2\equiv \chi_{f\sigma_8}^2+\chi_{E_G}^2$ we obtain the constraints on the parameters $\Omega_{0m}$, $\sigma_8$, $g_a$, $g_b$ shown in Figs. \ref{contfs82fid}, \ref{contEgz} and \ref{contEgfs82fid} respectively. Each one of these Figures corresponds to a 2D projection that goes through the best fit parameter point in parameter space of the full three or four dimensional contour plot in each case. The full number of parameters (three or four) was assumed when constructing the contour 2D projections. Previous studies \cite{Nesseris:2017vor,Kazantzidis:2018rnb}  have considered similar 2D projections that go through the \plcdm best fit parameter point in the higher dimensional parameter space. This later choice tends to change somewhat (in most projections it is increased) the apparent tension between the best fit MG parameter values and the best fit \plcdm parameters in the 2D projection parameter subspaces. This 2D tension may be in some cases misleading due to projection effects and thus in Table (\ref{sigmadif}) we stress the tension in the full 3D or 4D parameter space. \\

The tension level between the best fit MG parameter values and the \plcdm best fit parameter values is significant in both the 2D projection parameter spaces shown in Figs \ref{contfs82fid}, \ref{contEgz} and \ref{contEgfs82fid} and in the higher 3D parameter space likelihood surfaces shown in Fig. \ref{cont3d}.  The best fit parameter values obtained in the context of the datasets considered and the tension levels in both the 2D projections and in the full 3D-4D parameter spaces are shown in Tables \ref{bestfitpar} and \ref{sigmadif} respectively. In these Tables we also show the cases corresponding to fits without including the correction factor (\ref{corr}) in the \fs data demonstrating that there is a small change in the best fit parameter values.\\
\begin{widetext}
\begin{center}      
\begin{table*}     
\centering  
\begin{tabular}{c c  ccc ccccccc} 
\hhline{============}
   & \\
Param.&Planck$18$/$\Lambda$CDM &&Dataset &&Dataset  &&Dataset &&Datasets &&Datasets  \\
   & &&$f\sigma_8(z)$.&&$f\sigma_8(z)$ && $E_G(z)$ &&   $f\sigma_8(z)+E_G(z)$&&$f\sigma_8(z)+E_G(z)$ \\
  & && corr.&&no corr.&& &&   corr.&&no corr.  \\
 \hline 
    & \\
  $\Omega_{0m}$&$0.3153\pm 0.0073$&& $0.272\pm 0.019$  &&$0.263\pm 0.015$&&$0.313\pm 0.024$&& $0.275\pm 0.015$   && $0.264\pm 0.012$ \\
  $\sigma_8$&$0.8111\pm 0.0060$&& $0.886\pm 0.015$   &&$0.90\pm 0.016$&&     &&      $0.848\pm 0.015$       &&$0.879\pm 0.015$ \\
   $g_a$&0&& $-1.306\pm 0.140$  &&$-1.331\pm 0.138$&&$-0.129\pm 0.490$&&   $-0.957\pm 0.144$       &&$-1.115\pm 0.137$ \\
    $g_b$&0&& &&  &&$-2.308\pm 0.423$&&  $-2.448\pm 0.414$       &&$-2.422\pm 0.416$ \\
 \hhline{============}   
\end{tabular} 
\caption{\small Planck18/$\Lambda$CDM  based  on  TT,TE,EE+lowE+ lensing likelihoods best fit \cite{planck18} and the best-fit values from data.}
\label{bestfitpar} 
\end{table*}  
\end{center} 

\begin{center}    
\begin{table*}     
\centering   
\begin{tabular}{c| c c c| c c c c c c}     
\hhline{==========}

& \multicolumn{3}{c|}{Space}&\multicolumn{6}{c}{2D Projected  Space }   \\
Dataset&($\Omega_{0m},\sigma_8,g_a$)&($\Omega_{0m},g_a,g_b$)&($\Omega_{0m},\sigma_8,g_a,g_b$)&($\Omega_{0m},\sigma_8$)&($\Omega_{0m},g_a$)& ($\sigma_8,g_a$)& ($g_a,g_b$)&($\Omega_{0m},g_b$) &($\sigma_8,g_b$) \\
    & \multicolumn{3}{c|}{ }&\multicolumn{6}{c}{} \\
 \hline 
 & \multicolumn{3}{c|}{ }&\multicolumn{6}{c}{} \\
  $f\sigma_8(z)$ corr.& $3.70\sigma$ & &  & $3.00\sigma$ & $\sim8\sigma$ &$2.08\sigma$ &  & &  \\
 $f\sigma_8(z)$ no corr.& $4.15\sigma$ & &  & $2.75\sigma$ & $\sim8\sigma$ &$1.13\sigma$ &  & &  \\
  $E_G(z)$ &   &$4.57$& &   &$0.002\sigma$&    &$4.45\sigma$&$4.94\sigma$&   \\
  $E_G(z)$+$f\sigma_8(z)$ corr.&  &  &$6.03$ &$1.47\sigma$ &$6.39\sigma$&$2.59\sigma$ &$5.74\sigma$&$7.74\sigma$ &$5.58\sigma$\\
  $E_G(z)$+$f\sigma_8(z)$no corr.&  &  &$6.33$ &$2.17\sigma$ &$\sim8\sigma$&$2.16\sigma$ &$7.53\sigma$&$\sim8\sigma$ &$6.74\sigma$\\
 \hhline{==========}   
\end{tabular}  
\caption{\small Sigma differences of the best fit contours from Planck18/$\Lambda$CDM.} 
\label{sigmadif} 
\end{table*}   
\end{center}
\end{widetext}
The following comments can be made on the results shown in Figs. \ref{contfs82fid}, \ref{contEgz} and \ref{contEgfs82fid} and Tables \ref{bestfitpar} and \ref{sigmadif}:
\begin{itemize}
\item
The left part of Table \ref{sigmadif} shows the tension level in the full 3D or 4D parameter space.  
The tension level between \plcdm and best fit MG model parametrizations (\ref{muaz}) and (\ref{sigmaaz}) in the context of the \fs data is significant (about $3.5\sigma$) but is is less than the corresponding tension obtained using the $E_G$ statistic data (more than $4\sigma$). In fact for the combined \fs + $E_g$ dataset the tension level increases to close to $6\sigma$! This significant tension level comes independently from both the \fs and $E_G$ data and hints towards weaker gravity ($\mu$ and $\Sigma$ lower than 1) compared to the predictions of GR at low $z$. We stress however that this extreme level of tension is partly due to correlations among the considered datapoints which necessarily exist in our compilations.
\item
The weaker than expected gravitational growth indicated by the data is expressed as both a lower best fit $\Omega_{0m}$ than expected from \lcdm and as negative best fit values for the gravitational strength evolution parameters $g_a$ and $g_b$ (see e.g. Fig. \ref{contEgfs82fid}).
\item 
Ignoring the fiducial model correction factor of Eq. (\ref{corr}) in most cases tends to slightly increase the tension level (compare e.g. the last two lines of Table \ref{sigmadif}). Thus the consideration of this correction in our analysis is a conservative approach. 
\end{itemize}
 
\begin{widetext}
\begin{figure*} 
\begin{centering}
\includegraphics[width=0.97\textwidth]{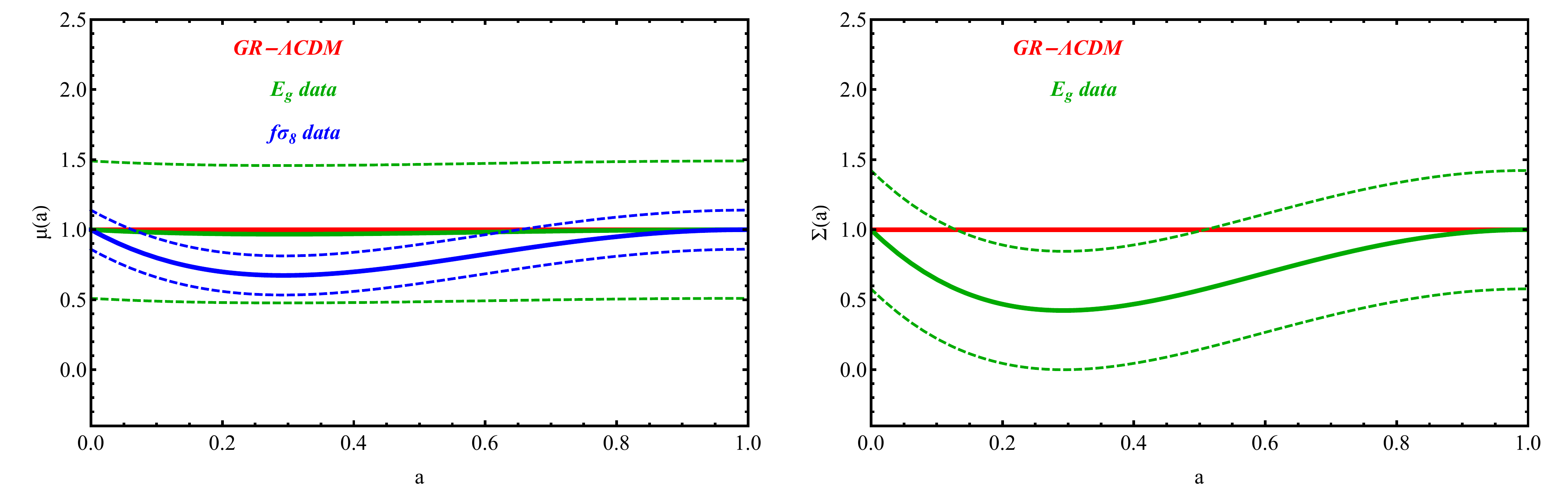}
\par\end{centering}
\caption{ Evolution of  $\mu$ and  $\Sigma$  as  functions of the scale factor $a$ considering the best fit values for $g_a$  and $g_b$ in the context of parametrizations Eqs.(\ref{muaz}) and  (\ref{sigmaaz}) with $n=2, m=2$. The data $E_G(z)$ and $f\sigma_8(z)$ from Tables \ref{tab:data-EG} and \ref{tab:data-rsd} of  the Appendix \ref{sec:Appendix_B} was used.  The dashed curves correspond to $1\sigma$ deviations of the parameters  $\mu$ and $\Sigma$. The red lines correspond to the GR-$\Lambda$CDM model.}  
\label{figms} 
\end{figure*}

\begin{figure*} 
\begin{centering} 
\includegraphics[width=0.97\textwidth]{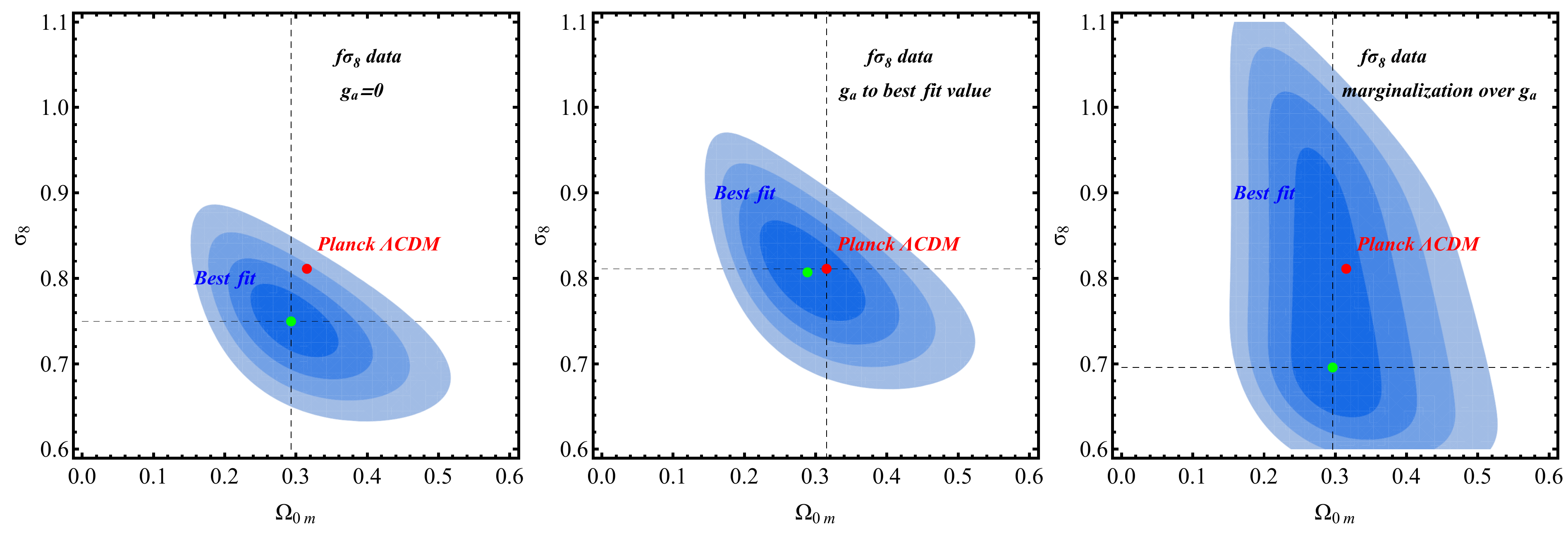}
\par\end{centering}
\caption{The confidence contours  of the parameter space ($\sigma_8$-$\Omega_m$) in the context of GR (left panel) and in the presence of the MG parameter $g_a$  (fixing $w=-1$) in the context of parametrization Eq.(\ref{muaz})  with $n=2$. We have considered both the case of a marginalized MG parameter value (right panel) and the case of setting $g_a$ to its best fit value (middle panel). The red and green dots describe the Planck18/$\Lambda$CDM best fit and the best-fit values from data.  The  $f\sigma_8(z)$ data compilations of datapoints with less correlation from Table \ref{tab:data-rsd} of  the Appendix \ref{sec:Appendix_B} was used. Notice the reduction of tension between the growth data best fit and the \plcdm parameter values when the MG degree of freedom is introduced.}  
\label{figga}  
\end{figure*} 
 
\begin{figure*} 
\begin{centering} 
\includegraphics[width=0.97\textwidth]{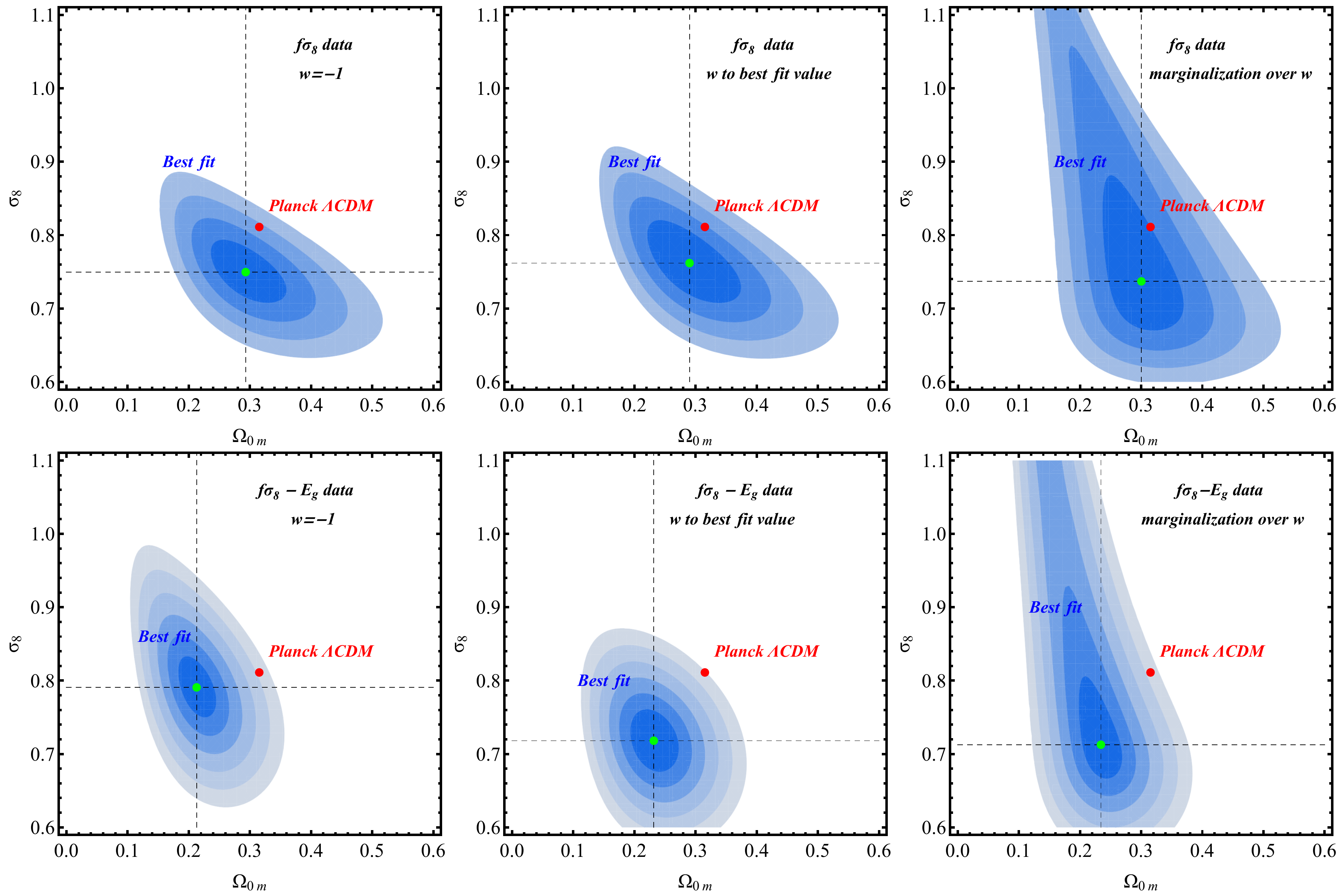}
\par\end{centering}
\caption{The confidence contours  of the parameter space ($\sigma_8$-$\Omega_m$)  in the context of GR (left panels) and in the presence of the $w$ parameter (fixing $g_a=0$ and $g_b=0$). We have considered both the case of a marginalized $w$ ($[-1.5,-0.5]$) parameter value (right panels) and the case of setting $w$ to its best fit value ($-0.94$ and $-1.29$  from $f\sigma_8(z)$ and $f\sigma_8(z)+E_G(z)$ data respectively) (middle panels). The red and green dots describe the Planck18/$\Lambda$CDM best fit and the best-fit values from data. The $E_G(z)$ and $f\sigma_8(z)$ data compilations of datapoints with less correlation from Tables \ref{tab:data-EG} and \ref{tab:data-rsd} of  the Appendix \ref{sec:Appendix_B} was used. Notice that the reduction of tension between the best fit parameter values and \plcdm is less efficient when the $w$ degree of freedom (modified background expansion rate) is introduced compared to the MG degree of freedom $g_a$ shown in Fig. \ref{figga}.}  
\label{figw}  
\end{figure*}

\end{widetext}

The trend for weaker gravity at low redshifts is also evident in Fig. \ref{figms} which shows the best fit form of $\mu(a)$ and $\Sigma(a)$ in the context of each dataset.

Also, the likelihood contours in the $\sigma_8$-$\Omega_m$ parameter space obtained using the growth data in the presence of the MG parameter $g_a$ and in the context of GR ($g_a=0$) are shown in Fig. \ref{figga}. We have considered both the case of a marginalized MG parameter value and the case of setting $g_a$ to its best fit value.  Clearly the tension level between the best fit parameter values and \plcdm decreases significantly in the presence of the MG parameter $g_a$.\\

The introduction of additional parameters of any type would in general widen the likelihood contours and thus reduce the tension between growth data and geometric/CMB data. In general a faster expansion rate ($w<-1$) would tend to reduce the growth rate of perturbations in agreement with dynamical observables. However, geometric observables (SnIa, BAO etc.) do not allow significant deviations of the expansion rate from $\Lambda$CDM. Thus the most efficient way to produce a weaker growth of perturbations is the introduction of evolution of the MG parameters $\mu$ and $\Sigma$. In Fig \ref{figw} we have demonstrated this effect by fixing $g_a=0$, $g_b=0$ and constructing the $\sigma_8$-$\Omega_m$ contours with $w=-1$ and $w$ free to vary in a range ($[-1.5,-0.5]$) consistent with geometric probes. The reduction of the tension in this case is significantly smaller compared to the introduction of MG degrees of freedom.\\
   
\subsection{Scale Dependent Data Compilations}
\label{scaledep}

Scale dependent parametrizations for $\mu$ and $\eta$ can describe a large class of MG models \cite{Bertschinger:2008zb,Pogosian:2010tj}. For example  a scale dependent class of parametrizations  predicted by  scalar-tensor  theories for $\mu$ and $\eta$  is of the form \cite{Ade:2015rim,DiValentino:2015bja}
\be  
\mu (a,k)=1+f_1(a)\frac{1+c_1(\lambda H /k)^2}{1+(\lambda H /k)^2}  
\label{muak}
\ee 
\be 
\eta (a,k)=1+f_2(a)\frac{1+c_2(\lambda H /k)^2}{1+(\lambda H /)^2} 
\label{sigmaak} 
\ee  
where $f_1$ and $f_2$ are properly chosen functions that depend on the scale factor.
Thus a physically motivated scale dependent generalization of the  parametrizations (\ref{muaz}) and (\ref{sigmaaz}) for  $\mu$ and $\Sigma$ may be written as
\be 
\mu (R,z)=1+\left[g_a(\frac{z}{1+z})^n-g_a(\frac{z}{1+z})^{2n}\right]\frac{1+s_a(\lambda H R)^2}{1+(\lambda H R)^2}
\label{muRz}
\ee
\be 
\Sigma (R,z)=1+\left[g_b(\frac{z}{1+z})^m-g_b(\frac{z}{1+z})^{2m}\right]\frac{1+s_b(\lambda H R)^2}{1+(\lambda H R)^2} 
\label{sigmaRz} 
\ee
\begin{widetext}
\begin{figure*} 
\begin{centering}
\includegraphics[width=0.97\textwidth]{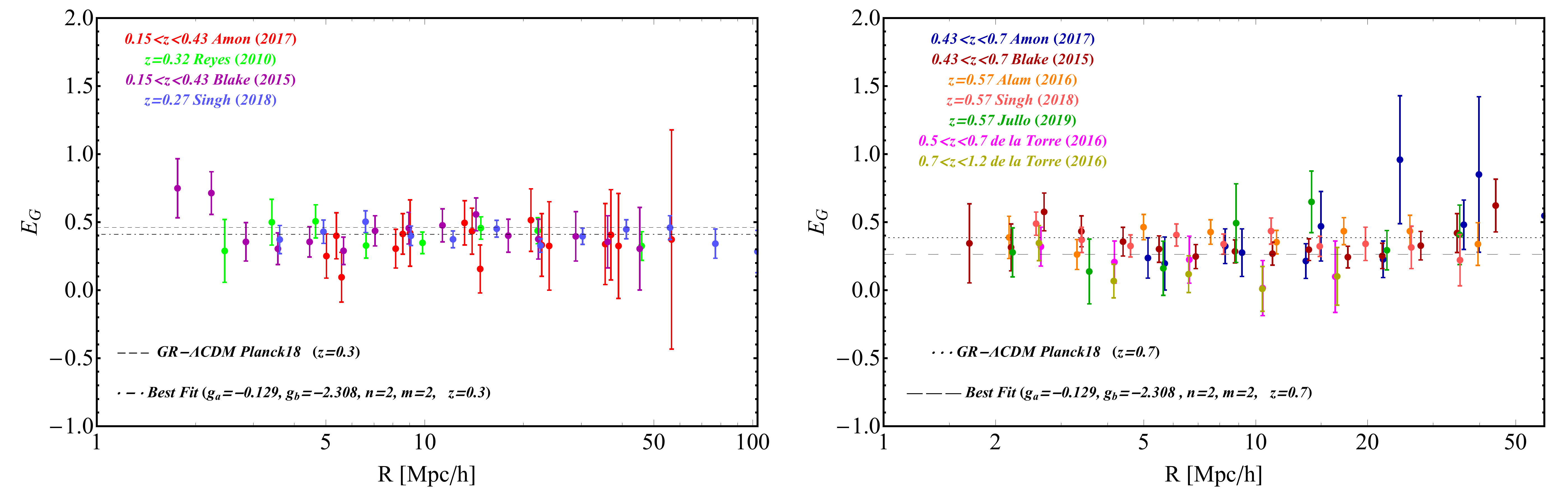}
\par\end{centering}
\caption{Measurements of $E_G$ as a function of scale $R$ in the range $0.15<z<0.43$ (left panel) and $0.43<z<1.2$ (right panel). The data $E_G(R)$  from Tables \ref{tab:data-EGRlow} and \ref{tab:data-EGRhigh} of  the Appendix \ref{sec:Appendix_B} was used. The dashed black line shows the Planck18/$\Lambda$CDM prediction at $z=0.3$, the dotted black line shows the Planck18/$\Lambda$CDM prediction at $z=0.7$, while  the dotdashed black line and the large dashed black line shows the best fit of the $E_G$ in the context of parametrizations Eqs.(\ref{muaz})  and (\ref{sigmaaz}) at $z=0.3$  and at $z=0.7$ respectively.}  
\label{figEgRerr2} 
\end{figure*}
\end{widetext}
where $s_a$, $s_b$ and $\lambda$ are parameters to be determined from a proper scale dependent dataset. Such a scale dependent data compilation for the statistic $E_G$ in two redshift ranges is shown in Fig. \ref{figEgRerr2}  and in Tables \ref{tab:data-EGRlow} and \ref{tab:data-EGRhigh} for low and high $z$ respectively in the Appendix \ref{sec:Appendix_B}.  The analysis of this compilation may be performed in the context of the scale dependent parametrizations (\ref{muRz}) and (\ref{sigmaRz}). Clearly as shown in Fig. \ref{figEgRerr2}, for both low and high $z$ the scale independent MG parametrizations of Eqs.(\ref{muaz})  and (\ref{sigmaaz}) at $z=0.3$  and at $z=0.7$, lead to a best fit value of $E_G$ that is lower compared to the \plcdm prediction. The full scale dependent analysis leads to similar levels of tension as those indicated in Table \ref{sigmadif} for the scale independent case and will be presented in detail elsewhere.

 \section{CONCLUSIONS-DISCUSSION} 
\label{sec:Discussion}   

We have used up to date compilations of $E_G$ and \fs data (Tables \ref{tab:data-rsd} and \ref{tab:data-EG}) based on WL and RSD observations to obtain updated estimates of the tension between the \plcdm best fit parameter values and the best fit parameter values obtained in the context of an effective MG gravity model allowing for properly parametrized evolution of the growth and lensing gravitational constants $\mu$ and $\Sigma$. The scale independent parametrizations (Eqs.(\ref{muaz}), (\ref{sigmaaz})) of $\mu$ and $\Sigma$ depend on the parameters $g_a$ and $g_b$ respectively and are by construction consistent with GR at early times and at present as indicated by nucleosynthesis and solar system constraints assuming no screening is present. We have assumed a flat \lcdm expansion background and we thus fit the parameters  ($\Omega_{0m},\sigma_8,g_a,g_b$).

We find that the $E_G$ data amplify the previously well known indications for low $\Omega_{0m}$ and/or weaker gravity ($\mu<1$) at low $z$  and favor weaker gravity for both the growth and the lensing gravitational constants ($\mu<1$ and $\Sigma <1$). The tension level between the \plcdm parameter values $(\Omega_{0m},\sigma_8,g_a,g_b)=(0.31,0.81,0,0)$ and the best fit parameter values obtained using the combined $E_G+f\sigma_8$ dataset $(\Omega_{0m},\sigma_8,g_a,g_b)=(0.28,0.85,-0.96,-2.45)$ is $6\sigma$ which is significantly larger compared to the tension obtained when only the \fs dataset is used ($3.7\sigma$ as shown in Table \ref{sigmadif}). 
 Even though the absolute magnitude of the derived tension is overestimated due to the correlations  among the datapoints  the amplified trend for weaker gravity at low $z$ is clearly indicated by both the \fs and $E_G$ data compilations and appears to be stronger for the case of the $E_G$ data.

If this trend has some physical origin and is not due only to data systematics or physical effects in the context of GR, there are significant implications for theoretical models. In particular $f(R)$ theories generically predict stronger gravity at low $z$ compared to its present time\cite{Gannouji:2018ncm} (thus the prediction is $\mu(z)>1$, $g_a>0$) and therefore if the identified tension has physical origin this can not be attributed to an $f(R)$ MG gravity theory for any expansion background. Similarly minimal scalar tensor theories \cite{Gannouji:2018ncm,Perivolaropoulos:2019vkb}, Horndeski theories \cite{Linder:2018jil,Arjona:2019rfn} and beyond Horndeski Gleyzes-Langlois-Piazza-Vernizzi (GLPV) theories \cite{Tsujikawa:2015mga} can only produce weaker gravity at low $z$ under very specific and in some cases unnatural conditions\cite{Tsujikawa:2007gd}. For example minimal scalar-tensor theories would require the existence of a phantom comsological background expansion (equation of state parameter $w<-1$)\cite{Gannouji:2018ncm,Perivolaropoulos:2019vkb}. 
In fact, a very large class of MG models, the scalar-tensor Horndeski models, are not consistent with the observational indications of weakening gravity. In fact as stated in Ref. \cite{Amendola:2019laa} (p. 12), $\mu$ for stable Horndeski models  is always larger than, or equal to, 1 so that matter perturbations in viable Horndeski models always grow faster than the corresponding GR models with the same backgrounds. Thus these MG models (which include $f(R)$ gravities) are unable to account for the weakening and would provide a worse fit than GR to the $f\sigma_8$/$E_G$ data. The search for MG models that can account for the observed indications for weakening gravity is thus an interesting extension of the present analysis.

A partial cause of the $E_G$ data tension with \plcdm is lensing magnification. As shown in \cite{Dizgah:2016bgm,Ghosh:2018ijm} the effects of lensing magnification modify the galaxy-galaxy lensing correlations as well as galaxy-galaxy correlations and as a consequence introduce systematic errors in the estimate of $E_G$ while making it bias dependent. The effect is small for redshifts smaller than 1 (about $5-10\%$) but it can become as large as $20-40\%$ for redshifts $z\simeq 1.5$.
Thus, this systematic contribution can be relevant already for Dark Energy Survey (DES) \cite{des2,Drlica-Wagner:2017tkk,Hoyle:2017mee,Elvin-Poole:2017xsf,des3} and certainly for higher redshift surveys. However, the magnitude of lensing contribution at the redshifts of the data compilation we are using ($z<1$) is not large enough to significantly reduce the identified tension which exists even at the level of the RSD data alone. The systematic effect discussed in  \cite{Dizgah:2016bgm,Ghosh:2018ijm} is important especially for upcoming surveys like Euclid \cite{euclid2} which probe higher redshifts even though even in that case it may not be large enough to be the only source the observed tension.
An interesting feature of our compilation is the  scale dependence the $E_G(R,z)$ data. This  may be used to probe the parameters of scale dependent MG $\mu$ and $\Sigma$ parametrizations which are well motivated physically. We plan to present such constraints elsewhere using upcoming and more extensive data  able to constrain the required larger parameter space that appears in scale dependent $\mu$ and $\Sigma$ parametrizations. A key question to address is whether the addition of scale dependence in the parametrizations can improve significantly the overall fit. No such indications are currently known \cite{Ade:2015rim} but this may well change using more extensive and accurate scale dependent $E_G$ and \fs data.\\
The introduction of the MG parameters $\mu$ and $\Sigma$ along with the variation of the parameters $\Omega_m$ and $\sigma_8$ leads to a model (MG-$\Lambda$CDM) that is a much better fit to the growth $f\sigma_8$ and $E_G$ data than the Planck/$\Lambda$CDM model in the context of GR. We have called this effect a 'tension' between the new best fit parameter values (MG-$\Lambda$CDM) and the GR-Planck/$\Lambda$CDM parameter  values (from Planck18 fit) which are $5-6 \sigma$ away from the new best fit parameter values. On the other hand, the MG parameters do not seem to change significantly the fit of the Planck data as indicated in Ref. \cite{Kazantzidis:2019dvk} and in Planck18 \cite{planck18} which indicate that pure CMB data appear to favor GR. Thus, the particular parametrization we have used does not seem to significantly reduce the tension between CMB and growth/weak lensing data since MG gravity appears to be favored by growth/weak lensing but not by the CMB. This is an issue we plan to investigate in more detail in the future by considering e.g. different MG parametrizations for the evolution of the $\mu$ and $\Sigma$ parameters that will not only improve the fit to the  $f\sigma_8$ /$E_G$ data but also improve the fit to the CMB data where some tensions are already evident (e.g. the lensing anomaly discussed in Planck18 \cite{planck18}).\\

\textbf{Supplemental Material:} The Mathematica file used for the numerical analysis and for construction of the figures can be found in \cite{suppl}.\\

\section*{ACKNOWLEDGEMENTS}
We thank Savvas Nesseris and Ruth Durrer for useful comments. This  article has benefited from COST Action CA15117 (CANTATA), supported by COST (European Cooperation in Science and Technology).\\
\begin{widetext}
\begin{center}       
\begin{table*}     
\centering  
\begin{tabular}{c c  ccc ccccccc} 
\hhline{============}
   & \\
Param.&Planck$18$/$\Lambda$CDM &&Dataset &&Dataset  &&Dataset &&Datasets &&Datasets  \\
   & &&$f\sigma_8(z)$.&&$f\sigma_8(z)$ && $E_G(z)$ &&   $f\sigma_8(z)+E_G(z)$&&$f\sigma_8(z)+E_G(z)$ \\
  & && corr.&&no corr.&& &&   corr.&&no corr.  \\
 \hline 
    & \\
  $\Omega_{0m}$&$0.3153\pm 0.0073$&& $0.289\pm 0.032$  &&$0.283\pm 0.028$&&$0.285\pm 0.044$&& $0.288\pm 0.026$   && $0.282\pm 0.023$ \\
  $\sigma_8$&$0.8111\pm 0.0060$&& $0.807\pm 0.024$   &&$0.819\pm 0.025$&&     &&      $0.795\pm 0.024$       &&$0.810\pm 0.024$ \\
   $g_a$&0&& $-0.767\pm 0.299$  &&$-0.826\pm 0.293$&&$-0.621\pm 0.914$&&   $-0.627\pm 0.291$       &&$-0.723\pm 0.281$ \\
    $g_b$&0&& &&  &&$-3.510\pm 0.605$&&  $-3.562\pm 0.601$       &&$-3.563\pm 0.601$ \\
 \hhline{============}   
\end{tabular} 
\caption{\small Planck18/$\Lambda$CDM  based  on  TT,TE,EE+lowE+ lensing likelihoods best fit \cite{planck18} and the best-fit values from data  compilation of datapoints with less correlation.}
\label{bestfitparsub} 
\end{table*}  
\end{center}  

\begin{center}    
\begin{table*}     
\centering   
\begin{tabular}{c| c c c| c c c c c c}     
\hhline{==========}

& \multicolumn{3}{c|}{Space}&\multicolumn{6}{c}{2D Projected  Space }   \\
Dataset&($\Omega_{0m},\sigma_8,g_a$)&($\Omega_{0m},g_a,g_b$)&($\Omega_{0m},\sigma_8,g_a,g_b$)&($\Omega_{0m},\sigma_8$)&($\Omega_{0m},g_a$)& ($\sigma_8,g_a$)& ($g_a,g_b$)&($\Omega_{0m},g_b$) &($\sigma_8,g_b$) \\
    & \multicolumn{3}{c|}{ }&\multicolumn{6}{c}{} \\
 \hline 
 & \multicolumn{3}{c|}{ }&\multicolumn{6}{c}{} \\
  $f\sigma_8(z)$ corr.& $2.39\sigma$ & &  & $0.22\sigma$ & $2.19\sigma$ &$1.79\sigma$ &  & &  \\
 $f\sigma_8(z)$ no corr.& $2.55\sigma$ & &  & $0.19\sigma$ & $2.19\sigma$ &$1.48\sigma$ &  & &  \\
  $E_G(z)$ &   &$5.06$& &   &$0.01\sigma$&    &$4.04\sigma$&$6.28\sigma$&   \\
  $E_G(z)$+$f\sigma_8(z)$ corr.&  &  &$5.69$ &$0.36\sigma$ &$1.35\sigma$&$1.59\sigma$ &$4.31\sigma$&$6.12\sigma$ &$5.12\sigma$\\
  $E_G(z)$+$f\sigma_8(z)$no corr.&  &  &$5.78$ &$0.31\sigma$ &$2.21\sigma$&$1.33\sigma$ &$4.54\sigma$&$6.38\sigma$ &$5.29\sigma$\\
 \hhline{==========}   
\end{tabular}  
\caption{\small Sigma differences of the best fit contours from Planck18/$\Lambda$CDM. The $E_G(z)$ and $f\sigma_8(z)$ data compilations of datapoints with less correlation from Tables \ref{tab:data-EG} and \ref{tab:data-rsd} of  the Appendix \ref{sec:Appendix_B} was used.} 
\label{sigmadifsub} 
\end{table*}   
\end{center}  
\end{widetext} 

\appendix
\section{ANALYSIS OF SUBSETS OF DATAPOINTS WITH LESS CORRELATION} 
\label{sec:Appendix_A}

In this Appendix we present the results of the statistical analysis of the $f\sigma_8(z)$ and $E_G(z)$ data compilations of datapoints with less correlation. These subsets of the data  are indicated with bold font in the index of the Tables \ref{tab:data-rsd}  and \ref{tab:data-EG} of the Appendix \ref{sec:Appendix_B}. Using these subsets of the data and  repeating our analysis we obtain the best fit parameter values and the tension levels in both the 2D projections and in the full 3D-4D parameter spaces as shown in Tables \ref{bestfitparsub} and \ref{sigmadifsub} respectively.\\

These results indicate that even though the tension level for the combined ($E_G$+$f\sigma_8$ ) reduces somewhat (from $6\sigma$ to about $5.5\sigma$) it remains high enough to cause concerns for the self consistency of the Planck/$\Lambda$CDM model and indications for the presence of weakening gravity. \\

\section{DATA USED IN THE ANALYSIS} 
\label{sec:Appendix_B}

In this appendix we present the data used in the analysis. \\

\begin{widetext}
\begin{longtable}{ | c | c | c | c | c | c | c | }
\caption{The \fs updated data compilation of Ref. \cite{Kazantzidis:2018rnb} used in the present analysis. The subset of the datapoints with less correlation  is indicated with bold font in the index.} 
\label{tab:data-rsd}\\
\hline
    Index & Dataset & $z$ & $f\sigma_8(z)$ & Refs. & Year & Fiducial Cosmology \\ 
\hline
\hline
1 & SDSS-LRG & $0.35$ & $0.440\pm 0.050$ & \cite{Song:2008qt} &  30 October 2006 &$(\Omega_{m},\Omega_K,\sigma_8$)$=(0.25,0,0.756)$\cite{Tegmark:2006az} \\

2 & VVDS & $0.77$ & $0.490\pm 0.18$ & \cite{Song:2008qt}  & 6 October 2009 & $(\Omega_{m},\Omega_K,\sigma_8)=(0.25,0,0.78)$ \\

3 & 2dFGRS & $0.17$ & $0.510\pm 0.060$ & \cite{Song:2008qt}  &  6 October 2009 & $(\Omega_{m},\Omega_K)=(0.3,0,0.9)$ \\

\textbf{4} &2MRS &0.02& $0.314 \pm 0.048$ &  \cite{Davis:2010sw}, \cite{Hudson:2012gt}& 13 November 2010 & $(\Omega_{m},\Omega_K,\sigma_8)=(0.266,0,0.65)$ \\

5 & SnIa+IRAS &0.02& $0.398 \pm 0.065$ & \cite{Turnbull:2011ty}, \cite{Hudson:2012gt}& 20 October 2011 & $(\Omega_{m},\Omega_K,\sigma_8)=(0.3,0,0.814)$\\

\textbf{6} & SDSS-LRG-200 & $0.25$ & $0.3512\pm 0.0583$ & \cite{Samushia:2011cs} & 9 December 2011 & $(\Omega_{m},\Omega_K,\sigma_8)=(0.276,0,0.8)$  \\

7 & SDSS-LRG-200 & $0.37$ & $0.4602\pm 0.0378$ & \cite{Samushia:2011cs} & 9 December 2011 & \\

8 & SDSS-LRG-60 & $0.25$ & $0.3665\pm0.0601$ & \cite{Samushia:2011cs} & 9 December 2011 & $(\Omega_{m},\Omega_K,\sigma_8)=(0.276,0,0.8)$ \\

9 & SDSS-LRG-60 & $0.37$ & $0.4031\pm0.0586$ & \cite{Samushia:2011cs} & 9 December 2011 &\\

\textbf{10} & WiggleZ & $0.44$ & $0.413\pm 0.080$ & \cite{Blake:2012pj} & 12 June 2012  & $(\Omega_{m},h,\sigma_8)=(0.27,0.71,0.8)$ \\

\textbf{11} & WiggleZ & $0.60$ & $0.390\pm 0.063$ & \cite{Blake:2012pj} & 12 June 2012 &  \\

\textbf{12} & WiggleZ & $0.73$ & $0.437\pm 0.072$ & \cite{Blake:2012pj} & 12 June 2012 &\\

13 & 6dFGS& $0.067$ & $0.423\pm 0.055$ & \cite{Beutler:2012px} & 4 July 2012 & $(\Omega_{m},\Omega_K,\sigma_8)=(0.27,0,0.76)$ \\

14 & SDSS-BOSS& $0.30$ & $0.407\pm 0.055$ & \cite{Tojeiro:2012rp} & 11 August 2012 & $(\Omega_{m},\Omega_K,\sigma_8)=(0.25,0,0.804)$ \\

15 & SDSS-BOSS& $0.40$ & $0.419\pm 0.041$ & \cite{Tojeiro:2012rp} & 11 August 2012 & \\

16 & SDSS-BOSS& $0.50$ & $0.427\pm 0.043$ & \cite{Tojeiro:2012rp} & 11 August 2012 & \\

17 & SDSS-BOSS& $0.60$ & $0.433\pm 0.067$ & \cite{Tojeiro:2012rp} & 11 August 2012 & \\

18 & VIPERS& $0.80$ & $0.470\pm 0.080$ & \cite{delaTorre:2013rpa} & 9 July 2013 & $(\Omega_{m},\Omega_K,\sigma_8)=(0.25,0,0.82)$  \\

19 & SDSS-DR7-LRG & $0.35$ & $0.429\pm 0.089$ & \cite{Chuang:2012qt}  & 8 August 2013 & $(\Omega_{m},\Omega_K,\sigma_8$)$=(0.25,0,0.809)$\cite{Komatsu:2010fb}\\

\textbf{20} & GAMA & $0.18$ & $0.360\pm 0.090$ & \cite{Blake:2013nif}  & 22 September 2013 & $(\Omega_{m},\Omega_K,\sigma_8)=(0.27,0,0.8)$ \\

21& GAMA & $0.38$ & $0.440\pm 0.060$ & \cite{Blake:2013nif}  & 22 September 2013 & \\

22 & BOSS-LOWZ& $0.32$ & $0.384\pm 0.095$ & \cite{Sanchez:2013tga}  & 17 December 2013  & $(\Omega_{m},\Omega_K,\sigma_8)=(0.274,0,0.8)$ \\

23 & SDSS DR10 and DR11 & $0.32$ & $0.48 \pm 0.10$ & \cite{Sanchez:2013tga} &   17 December 2013 & $(\Omega_{m},\Omega_K,\sigma_8$)$=(0.274,0,0.8)$\cite{Anderson:2013zyy}\\

24 & SDSS DR10 and DR11 & $0.57$ & $0.417 \pm 0.045$ & \cite{Sanchez:2013tga} &  17 December 2013 &  \\

\textbf{25} & SDSS-MGS & $0.15$ & $0.490\pm0.145$ & \cite{Howlett:2014opa} & 30 January 2015 & $(\Omega_{m},h,\sigma_8)=(0.31,0.67,0.83)$ \\

\textbf{26} & SDSS-veloc & $0.10$ & $0.370\pm 0.130$ & \cite{Feix:2015dla}  & 16 June 2015 & $(\Omega_{m},\Omega_K,\sigma_8$)$=(0.3,0,0.89)$\cite{Tegmark:2003uf} \\

\textbf{27} & FastSound& $1.40$ & $0.482\pm 0.116$ & \cite{Okumura:2015lvp}  & 25 November 2015 & $(\Omega_{m},\Omega_K,\sigma_8$)$=(0.27,0,0.82)$\cite{Hinshaw:2012aka} \\

28 & SDSS-CMASS & $0.59$ & $0.488\pm 0.060$ & \cite{Chuang:2013wga} & 8 July 2016 & $\ \ (\Omega_{m},h,\sigma_8)=(0.307115,0.6777,0.8288)$ \\

\textbf{29} & BOSS DR12 & $0.38$ & $0.497\pm 0.045$ & \cite{Alam:2016hwk} & 11 July 2016 & $(\Omega_{m},\Omega_K,\sigma_8)=(0.31,0,0.8)$ \\

\textbf{30} & BOSS DR12 & $0.51$ & $0.458\pm 0.038$ & \cite{Alam:2016hwk} & 11 July 2016 & \\

\textbf{31} & BOSS DR12 & $0.61$ & $0.436\pm 0.034$ & \cite{Alam:2016hwk} & 11 July 2016 & \\

32 & BOSS DR12 & $0.38$ & $0.477 \pm 0.051$ & \cite{Beutler:2016arn} & 11 July 2016 & $(\Omega_{m},h,\sigma_8)=(0.31,0.676,0.8)$ \\

33 & BOSS DR12 & $0.51$ & $0.453 \pm 0.050$ & \cite{Beutler:2016arn} & 11 July 2016 & \\

34 & BOSS DR12 & $0.61$ & $0.410 \pm 0.044$ & \cite{Beutler:2016arn} & 11 July 2016 &  \\

35 &VIPERS v7& $0.76$ & $0.440\pm 0.040$ & \cite{Wilson:2016ggz} & 26 October 2016  & $(\Omega_{m},\sigma_8)=(0.308,0.8149)$ \\

\textbf{36} &VIPERS v7 & $1.05$ & $0.280\pm 0.080$ & \cite{Wilson:2016ggz} & 26 October 2016 &\\

\textbf{37} &  BOSS LOWZ & $0.32$ & $0.427\pm 0.056$ & \cite{Gil-Marin:2016wya} & 26 October 2016 & $(\Omega_{m},\Omega_K,\sigma_8)=(0.31,0,0.8475)$\\

38 & BOSS CMASS & $0.57$ & $0.426\pm 0.029$ & \cite{Gil-Marin:2016wya} & 26 October 2016 & \\

\textbf{39} & VIPERS  & $0.727$ & $0.296 \pm 0.0765$ & \cite{Hawken:2016qcy} &  21 November 2016 & $(\Omega_{m},\Omega_K,\sigma_8)=(0.31,0,0.7)$\\

\textbf{40} & 6dFGS+SnIa & $0.02$ & $0.428\pm 0.0465$ & \cite{Huterer:2016uyq} & 29 November 2016 & $(\Omega_{m},h,\sigma_8)=(0.3,0.683,0.8)$ \\

41 &VIPERS PDR2& $0.60$ & $0.550\pm 0.120$ & \cite{Pezzotta:2016gbo} & 16 December 2016 & $(\Omega_{m},\Omega_b,\sigma_8)=(0.3,0.045,0.823)$ \\

42 & VIPERS PDR2& $0.86$ & $0.400\pm 0.110$ & \cite{Pezzotta:2016gbo} & 16 December 2016 &\\

43 & SDSS DR13  & $0.1$ & $0.48 \pm 0.16$ & \cite{Feix:2016qhh} & 22 December 2016 & $(\Omega_{m},\sigma_8$)$=(0.25,0.89)$\cite{Tegmark:2003uf} \\

\textbf{44} & 2MTF & 0.001 & $0.505 \pm 0.085$ &  \cite{Howlett:2017asq} & 16 June 2017 & $(\Omega_{m},\sigma_8)=(0.3121,0.815)$\\

45 & VIPERS PDR2 & $0.85$ & $0.45 \pm 0.11$ & \cite{Mohammad:2017lzz} & 31 July 2017  &  $(\Omega_b,\Omega_{m},h)=(0.045,0.30,0.8)$ \\

\textbf{46} & BOSS DR12 & $0.31$ & $0.384 \pm 0.083$ &  \cite{Wang:2017wia} & 15 September 2017 & $(\Omega_{m},h,\sigma_8)=(0.307,0.6777,0.8288)$\\

\textbf{47} & BOSS DR12 & $0.36$ & $0.409 \pm 0.098$ &  \cite{Wang:2017wia} & 15 September 2017 & \\

\textbf{48} & BOSS DR12 & $0.40$ & $0.461 \pm 0.086$ &  \cite{Wang:2017wia} & 15 September 2017 & \\

\textbf{49} & BOSS DR12 & $0.44$ & $0.426 \pm 0.062$ &  \cite{Wang:2017wia} & 15 September 2017 & \\

\textbf{50} & BOSS DR12 & $0.48$ & $0.458 \pm 0.063$ &  \cite{Wang:2017wia} & 15 September 2017 & \\

\textbf{51} & BOSS DR12 & $0.52$ & $0.483 \pm 0.075$ &  \cite{Wang:2017wia} & 15 September 2017 & \\

\textbf{52} & BOSS DR12 & $0.56$ & $0.472 \pm 0.063$ &  \cite{Wang:2017wia} & 15 September 2017 & \\

\textbf{53} & BOSS DR12 & $0.59$ & $0.452 \pm 0.061$ &  \cite{Wang:2017wia} & 15 September 2017 & \\

\textbf{54} & BOSS DR12 & $0.64$ & $0.379 \pm 0.054$ &  \cite{Wang:2017wia} & 15 September 2017 & \\

55 & SDSS DR7 & $0.1$ & $0.376\pm 0.038$ & \cite{Shi:2017qpr} & 12 December 2017 & $(\Omega_{m},\Omega_b,\sigma_8)=(0.282,0.046,0.817)$ \\

56 & SDSS-IV & $1.52$ & $0.420 \pm 0.076$ &  \cite{Gil-Marin:2018cgo} & 8 January 2018  & $(\Omega_{m},\Omega_b h^2,\sigma_8)=(0.26479, 0.02258,0.8)$ \\ 

57 & SDSS-IV & $1.52$ & $0.396 \pm 0.079$ & \cite{Hou:2018yny} & 8 January 2018 & $(\Omega_{m},\Omega_b h^2,\sigma_8)=(0.31,0.022,0.8225)$ \\ 

\textbf{58} & SDSS-IV & $0.978$ & $0.379 \pm 0.176$ &  \cite{Zhao:2018jxv} & 9 January 2018 &$(\Omega_{m},\sigma_8)=(0.31,0.8)$\\

\textbf{59} & SDSS-IV & $1.23$ & $0.385 \pm 0.099$ &  \cite{Zhao:2018jxv} & 9 January 2018 & \\

\textbf{60} & SDSS-IV & $1.526$ & $0.342 \pm 0.070$ &  \cite{Zhao:2018jxv} & 9 January 2018 & \\

\textbf{61} & SDSS-IV & $1.944$ & $0.364 \pm 0.106$ &  \cite{Zhao:2018jxv} & 9 January 2018 & \\

\textbf{62} & VIPERS PDR2 & $0.60$ & $0.49 \pm 0.12$ &  \cite{Mohammad:2018mdy} & 6 June 2018 &$(\Omega_b,\Omega_m,h,\sigma_8)=(0.045,0.31,0.7,0.8)$\\

\textbf{63} & VIPERS PDR2 & $0.86$ & $0.46 \pm 0.09$ &  \cite{Mohammad:2018mdy} & 6 June 2018 & \\

\textbf{64} & BOSS DR12 voids & $0.57$ & $0.501 \pm 0.051$ &\cite{Nadathur:2019mct} & 1 April 2019 &$(\Omega_b,\Omega_m,h,\sigma_8)=(0.0482,0.307,0.6777,0.8228)$\\

\textbf{65} & 2MTF 6dFGSv & $0.03$ & $0.404 \pm 0.0815$ &\cite{Qin:2019axr} & 7 June 2019 &$(\Omega_b,\Omega_m,h,\sigma_8)=(0.0491,0.3121,0.6571,0.815)$\\

\textbf{66} & SDSS-IV & $0.72$ & $0.454 \pm 0.139$ &  \cite{Icaza-Lizaola:2019zgk} & 17 September 2019 &$(\Omega_{m},\Omega_b h^2,\sigma_8)=(0.31,0.022,0.8)$ \\

\hline
\end{longtable}

\begin{longtable}{ | c| c|c|c| c | c | c |c| }
\caption{The $E_G(z)$ data compilation  used in the present analysis. The subset of the datapoints with less correlation  is indicated with bold font in the index.} 
\label{tab:data-EG}\\
\hline
  
   Index &Dataset &$z$ & $E_G(z)$  & $\sigma_{E_G}$&Scale  [Mpc/h]  & Reference\\
    
\hline   
\hline  
 
\textbf{1}&KiDS GAMA&0.267 &0.43& 0.13 & $5<R<40$& \cite{Amon:2017lia}\\
2&SDSS BOSS LOWZ&0.27 &0.40& 0.05 & $25 < R < 150 $&\cite{Singh:2018flu}\\
3&CMB lens BOSS LOWZ&0.27 &0.46& 0.085 & $25 < R < 150 $&\cite{Singh:2018flu}\\
\textbf{4}&KiDS 2dFLenS BOSS LOWZ 2dFLOZ &0.305 &0.27 &0.08 & $5<R<60$& \cite{Amon:2017lia}\\
\textbf{5}&RCSLenS CFHTLenS WiggleZ BOSS WGZLoZ LOWZ&0.32& 0.40 &0.09 & $R>3 $&\cite{Blake:2015vea}\\
6&RCSLenS CFHTLenS WiggleZ BOSS WGZLoZ LOWZ&0.32 &0.48& 0.10&  $R>10$&\cite{Blake:2015vea}\\
7&SDSS &0.32&0.39&0.06& $10 < Rp < 50$&\cite{Reyes:2010tr}\\
\textbf{8}&KiDS 2dFLenS BOSS CMASS 2dFHIZ&0.554 &0.26& 0.07&  $5 < R < 60$ &\cite{Amon:2017lia}\\
\textbf{9}&RCSLenS CFHTLenS WiggleZ BOSS WGZHiZ CMASS&0.57 &0.31& 0.06&  $R>3$& \cite{Blake:2015vea}\\
\textbf{10}&RCSLenS CFHTLenS WiggleZ BOSS WGZHiZ CMASS&0.57 &0.30& 0.07& $ R>10$ &\cite{Blake:2015vea}\\
11&SDSS-III BOSS CMB lens CMASS&0.57&0.24&0.06& $ R>150$&\cite{Pullen:2015vtb}\\
12&CFHTLenS SDSS-III BOSS CMASS&0.57&0.42&0.056& $ 5<R<26$&\cite{Alam:2016qcl}\\
13&CMB lens BOSS CMASS&0.57 &0.39& 0.05 & $25 < R < 150 $&\cite{Singh:2018flu}\\
14&CFHTLenS BOSS CMASS&0.57 &0.43& 0.10 & $10 < R < 60 $&\cite{Jullo:2019lgq}\\
\textbf{15}&CFHTLenS VIPERS &0.60 &0.16& 0.09& $ 3 < R < 20$ &\cite{delaTorre:2016rxm}\\
\textbf{16}&CFHTLenS VIPERS &0.86 &0.09& 0.07 & $3 < R < 20 $&\cite{delaTorre:2016rxm}\\

\hline
\end{longtable}

\begin{longtable}{ | c|c|c| c | c | c |c| }
\caption{The $E_G(R)$ data compilation  in the range $0.15<z<0.43$ used in the present analysis.} 
\label{tab:data-EGRlow}\\
\hline
 
   Index &$R   [Mpc/h]$ & $E_G(R)$  & $\sigma_{E_G}$& z  & Reference\\
    
\hline   
\hline  
1&3.61 &	0.37 	&0.10 	&0.27&\cite{Singh:2018flu}\\
2&4.91 	&0.42 	&0.08 &0.27&\cite{Singh:2018flu}\\	
3&6.60 &	0.50 	&0.07 &0.27	&\cite{Singh:2018flu}\\
4&9.07 &	0.39 	&0.07 &0.27	&\cite{Singh:2018flu}\\
5&12.20 &	0.37 	&0.06 &0.27	&\cite{Singh:2018flu}\\
6&16.58 	&0.45 	&0.06 &0.27&\cite{Singh:2018flu}\\	
7&22.54 &	0.32 	&0.04 &0.27	&\cite{Singh:2018flu}\\
8&30.30 	&0.39 	&0.05 &0.27	&\cite{Singh:2018flu}\\
9&41.19 	&0.44 	&0.06 &0.27	&\cite{Singh:2018flu}\\
10&55.99 	&0.45 	&0.08 &0.27	&\cite{Singh:2018flu}\\
11&76.98 	&0.34 	&0.10 &0.27&\cite{Singh:2018flu}\\	
12&103.47 	&0.28 	&0.15 &0.27&\cite{Singh:2018flu}\\
13&2.45 &	0.28 &	0.23 & 0.32&	\cite{Reyes:2010tr}\\
14&3.41 &	0.49 &	0.16 	& 0.32&	\cite{Reyes:2010tr}\\
15&4.64 &	0.50 &	0.12 	& 0.32&	\cite{Reyes:2010tr}\\
16&6.62 &	0.32 &	0.09 	& 0.32&	\cite{Reyes:2010tr}\\
17&9.85 &	0.34 &	0.07 & 0.32&	\cite{Reyes:2010tr}\\
18&14.83 &	0.45 	&0.08 	& 0.32&	\cite{Reyes:2010tr}\\
19&22.10 &	0.43 	&0.09 	& 0.32&	\cite{Reyes:2010tr}\\
20&45.87 &	0.32 &0.10 & 0.32&	\cite{Reyes:2010tr}\\
21&1.76 &	0.74 &	0.21 &0.15-0.43&\cite{Blake:2015vea}\\	
22&2.23 &	0.71 	&0.15 &0.15-0.43&\cite{Blake:2015vea}\\	
23&2.85 &	0.35 	&0.14 &0.15-0.43&\cite{Blake:2015vea}\\	
24&3.56 &	0.30 &	0.11 &0.15-0.43&\cite{Blake:2015vea}\\
25&4.45 	&0.35 	&0.11 	&0.15-0.43&\cite{Blake:2015vea}\\
26&5.65 &	0.28 &	0.10 	&0.15-0.43&\cite{Blake:2015vea}\\
27&7.059 &	0.43 	&0.11 	&0.15-0.43&\cite{Blake:2015vea}\\
28&8.94 &	0.45 	&0.11 	&0.15-0.43&\cite{Blake:2015vea}\\
29&11.33 &	0.47 &	0.12 	&0.15-0.43&\cite{Blake:2015vea}\\
30&14.34 	&0.55 	&0.12 &0.15-0.43&\cite{Blake:2015vea}\\
31&17.98 &	0.40&	0.12 	&0.15-0.43&\cite{Blake:2015vea}\\
32&22.21 	&0.37 &	0.14 	&0.15-0.43&\cite{Blake:2015vea}\\
33&28.88 &	0.39 &	0.18 	&0.15-0.43&\cite{Blake:2015vea}\\
34&36.15 	&0.35 	&0.19 	&0.15-0.43&\cite{Blake:2015vea}\\
35&45.26 &0.30 &	0.30 &0.15-0.43&\cite{Blake:2015vea}\\
36&5.01 	&0.25 &	0.16 	&0.15-0.43&\cite{Amon:2017lia}\\
37&5.37 &	0.39 &	0.16 	&0.15-0.43&\cite{Amon:2017lia}\\
38&5.58 &	0.094 &	0.18 	&0.15-0.43&\cite{Amon:2017lia}\\
39&8.15 &	0.30 &	0.14 	&0.15-0.43&\cite{Amon:2017lia}\\
40&8.57 &	0.41 &	0.14 &0.15-0.43&\cite{Amon:2017lia}\\
41&9.02 &	0.41 &	0.24 &0.15-0.43&\cite{Amon:2017lia}\\
42&13.23 &	0.49 	&0.16 	&0.15-0.43&\cite{Amon:2017lia}\\
43&13.95 &	0.43 	&0.16 	&0.15-0.43&\cite{Amon:2017lia}\\
44&14.76 &	0.15 	&0.17 	&0.15-0.43&\cite{Amon:2017lia}\\
45&21.08 &	0.51 	&0.23 &0.15-0.43&\cite{Amon:2017lia}\\
46&22.75 &	0.33 &0.23	&0.15-0.43&\cite{Amon:2017lia}\\
47&23.96 &	0.32 	&0.32 	&0.15-0.43&\cite{Amon:2017lia}\\
48&35.52 &	0.33 &0.29 	&0.15-0.43&\cite{Amon:2017lia}\\
49&36.98 &	0.40 	&0.33 	&0.15-0.43&\cite{Amon:2017lia}\\
50&39.00 &	0.32 	&0.38 &0.15-0.43&\cite{Amon:2017lia}\\
51&56.60 &	0.37 &0.80 &0.15-0.43&\cite{Amon:2017lia}\\ 

\hline
\end{longtable}

\begin{longtable}{ | c|c|c| c | c | c |c| }
\caption{The $E_G(R)$ data compilation  in the range $0.43<z<1.2$ used in the present analysis.} 
\label{tab:data-EGRhigh}\\
\hline
 
   Index &$R [Mpc/h]$ & $E_G(R)$  & $\sigma_{E_G}$& z  & Reference\\
    
\hline   
\hline  
1&5.13 &	0.23 	&0.14 	&0.43-0.7&\cite{Amon:2017lia}\\
2&5.69 &	0.19 	&0.19 	&0.43-0.7&\cite{Amon:2017lia}\\
3&8.28 &	0.32 &	0.12 	&0.43-0.7&\cite{Amon:2017lia}\\
4&9.19 &	0.27 	&0.17 	&0.43-0.7&\cite{Amon:2017lia}\\
5&13.69 &	0.21 	&0.12 	&0.43-0.7&\cite{Amon:2017lia}\\
6&14.98 &	0.46 &0.25 	&0.43-0.7&\cite{Amon:2017lia}\\
7&24.43 &	0.95 	&0.47 	&0.43-0.7&\cite{Amon:2017lia}\\
8&22.02 &	0.22 	&0.13 &0.43-0.7&\cite{Amon:2017lia}\\
9&36.28 &	0.48 &	0.18 	&0.43-0.7&\cite{Amon:2017lia}\\
10&39.84 &	0.84 &0.57 	&0.43-0.7&\cite{Amon:2017lia}\\
11&59.78 &	0.54 	&0.45 &0.43-0.7&\cite{Amon:2017lia}\\
12&1.74	&0.34 &	0.29 &	0.43-0.7&\cite{Blake:2015vea}\\
13&2.25	&0.31 &	0.17 	&	0.43-0.7&\cite{Blake:2015vea}\\
14&2.74	&0.57 &	0.14	&	0.43-0.7&\cite{Blake:2015vea}\\
15&3.46	&0.43 	&0.11 	&	0.43-0.7&\cite{Blake:2015vea}\\
16&4.45	&0.35 &	0.10 	&	0.43-0.7&\cite{Blake:2015vea}\\
17&5.56	&0.30 &	0.09 &	0.43-0.7&\cite{Blake:2015vea}\\
18&6.92	&0.24 &	0.09&	0.43-0.7&\cite{Blake:2015vea}\\
19&8.84	&0.28 &	0.08 	&	0.43-0.7&\cite{Blake:2015vea}\\
20&11.15	&0.26 	&0.08 	&	0.43-0.7&\cite{Blake:2015vea}\\
21&13.90	&0.29 	&0.07 	&	0.43-0.7&\cite{Blake:2015vea}\\
22&17.74	&0.24 	&0.08	&	0.43-0.7&\cite{Blake:2015vea}\\
23&21.96	&0.25 	&0.09 	&	0.43-0.7&\cite{Blake:2015vea}\\
24&27.80	&0.32 	&0.10 	&	0.43-0.7&\cite{Blake:2015vea}\\
25&34.81	&0.41 	&0.14 	&	0.43-0.7&\cite{Blake:2015vea}\\
26&44.26	&0.62 	&0.19 	&	0.43-0.7&\cite{Blake:2015vea}\\
27&2.17 	&0.38 	&0.15 &0.57&	\cite{Alam:2016qcl}\\
28&3.30 	&0.26 	&0.11 &0.57&	\cite{Alam:2016qcl}\\
29&4.99 	&0.46 	&0.09 	&0.57&	\cite{Alam:2016qcl}\\
30&7.56 	&0.42 	&0.08 	&0.57&	\cite{Alam:2016qcl}\\
31&11.38 	&0.35 	&0.08 	&0.57&	\cite{Alam:2016qcl}\\
32&17.26 	&0.43 	&0.09 	&0.57&	\cite{Alam:2016qcl}\\
33&25.99 	&0.43 	&0.11 	&0.57&	\cite{Alam:2016qcl}\\
34&39.60 	&0.33 	&0.15 	&0.57&	\cite{Alam:2016qcl}\\
35&2.56 	&0.48 	&0.08 	&0.57&\cite{Singh:2018flu}\\
36&3.40 	&0.36 	&0.09 	&0.57&\cite{Singh:2018flu}\\
37&4.60 	&0.32 	&0.08 &0.57&\cite{Singh:2018flu}\\
38&6.12 	&0.40 	&0.08 	&0.57&\cite{Singh:2018flu}\\
39&8.20 	&0.33 	&0.07 	&0.57&\cite{Singh:2018flu}\\
40&11.00 	&0.43 	&0.09 	&0.57&\cite{Singh:2018flu}\\
41&14.87 	&0.32 	&0.07 	&0.57&\cite{Singh:2018flu}\\
42&19.75 	&0.34 	&0.12 &0.57&\cite{Singh:2018flu}\\
43&26.21 	&0.31 	&0.15 	&0.57&\cite{Singh:2018flu}\\
44&35.44 	&0.22 	&0.18 &0.57&\cite{Singh:2018flu}\\ 
45&2.22 	&0.27 	&0.17 	&0.57&\cite{Jullo:2019lgq}\\
46&3.56 	&0.13 	&0.23 &0.57&\cite{Jullo:2019lgq}\\
47&5.64 	&0.16 	&0.19 &0.57&\cite{Jullo:2019lgq}\\
48&8.86 	&0.49 	&0.28 &0.57&\cite{Jullo:2019lgq}\\
49&14.13 	&0.64 	&0.22 	&0.57&\cite{Jullo:2019lgq}\\
50&22.54 	&0.29 	&0.14 &0.57&\cite{Jullo:2019lgq}\\
51&35.37 	&0.40 	&0.21 &0.57&\cite{Jullo:2019lgq}\\
52&2.64 &0.31 	&0.14 	&0.5-0.7&\cite{delaTorre:2016rxm}\\
53&4.17  &0.20 	&0.15 	&0.5-0.7&\cite{delaTorre:2016rxm}\\
54&6.63  &0.22 	&0.17 	&0.5-0.7&\cite{delaTorre:2016rxm}\\
55&10.44  &0.01 	&0.20 	&0.5-0.7&\cite{delaTorre:2016rxm}\\
56&16.37  &0.09 	&0.26  &0.5-0.7&\cite{delaTorre:2016rxm}\\
57&2.61 	&0.34 	&0.12 	&0.7-1.2&\cite{delaTorre:2016rxm}\\
58&4.15 	&0.06 	&0.12 	&0.7-1.2&\cite{delaTorre:2016rxm}\\
59&6.60 	&0.11 	&0.13 	&0.7-1.2&\cite{delaTorre:2016rxm}\\
60&10.42 	&0.01	&0.16 	&0.7-1.2&\cite{delaTorre:2016rxm}\\
61&16.56 	&0.10 	&0.21 	&0.7-1.2&\cite{delaTorre:2016rxm}\\

\hline
\end{longtable} 

\end{widetext}

\raggedleft
\bibliography{Bibliography}

\begin{thebibliography}{178}%
\makeatletter
\providecommand \@ifxundefined [1]{%
 \@ifx{#1\undefined}
}%
\providecommand \@ifnum [1]{%
 \ifnum #1\expandafter \@firstoftwo
 \else \expandafter \@secondoftwo
 \fi
}%
\providecommand \@ifx [1]{%
 \ifx #1\expandafter \@firstoftwo
 \else \expandafter \@secondoftwo
 \fi
}%
\providecommand \natexlab [1]{#1}%
\providecommand \enquote  [1]{``#1''}%
\providecommand \bibnamefont  [1]{#1}%
\providecommand \bibfnamefont [1]{#1}%
\providecommand \citenamefont [1]{#1}%
\providecommand \href@noop [0]{\@secondoftwo}%
\providecommand \href [0]{\begingroup \@sanitize@url \@href}%
\providecommand \@href[1]{\@@startlink{#1}\@@href}%
\providecommand \@@href[1]{\endgroup#1\@@endlink}%
\providecommand \@sanitize@url [0]{\catcode `\\12\catcode `\$12\catcode
  `\&12\catcode `\#12\catcode `\^12\catcode `\_12\catcode `\%12\relax}%
\providecommand \@@startlink[1]{}%
\providecommand \@@endlink[0]{}%
\providecommand \url  [0]{\begingroup\@sanitize@url \@url }%
\providecommand \@url [1]{\endgroup\@href {#1}{\urlprefix }}%
\providecommand \urlprefix  [0]{URL }%
\providecommand \Eprint [0]{\href }%
\providecommand \doibase [0]{http://dx.doi.org/}%
\providecommand \selectlanguage [0]{\@gobble}%
\providecommand \bibinfo  [0]{\@secondoftwo}%
\providecommand \bibfield  [0]{\@secondoftwo}%
\providecommand \translation [1]{[#1]}%
\providecommand \BibitemOpen [0]{}%
\providecommand \bibitemStop [0]{}%
\providecommand \bibitemNoStop [0]{.\EOS\space}%
\providecommand \EOS [0]{\spacefactor3000\relax}%
\providecommand \BibitemShut  [1]{\csname bibitem#1\endcsname}%
\let\auto@bib@innerbib\@empty
\bibitem [{\citenamefont {Nesseris}\ \emph {et~al.}(2017)\citenamefont
  {Nesseris}, \citenamefont {Pantazis},\ and\ \citenamefont
  {Perivolaropoulos}}]{Nesseris:2017vor}%
  \BibitemOpen
  \bibfield  {author} {\bibinfo {author} {\bibfnamefont {Savvas}\ \bibnamefont
  {Nesseris}}, \bibinfo {author} {\bibfnamefont {George}\ \bibnamefont
  {Pantazis}}, \ and\ \bibinfo {author} {\bibfnamefont {Leandros}\ \bibnamefont
  {Perivolaropoulos}},\ }\bibfield  {title} {\enquote {\bibinfo {title}
  {{Tension and constraints on modified gravity parametrizations of
  $G_{\textrm{eff}}(z)$ from growth rate and Planck data}},}\ }\href {\doibase
  10.1103/PhysRevD.96.023542} {\bibfield  {journal} {\bibinfo  {journal} {Phys.
  Rev.}\ }\textbf {\bibinfo {volume} {D96}},\ \bibinfo {pages} {023542}
  (\bibinfo {year} {2017})},\ \Eprint {http://arxiv.org/abs/1703.10538}
  {arXiv:1703.10538 [astro-ph.CO]} \BibitemShut {NoStop}%
\bibitem [{\citenamefont {Kazantzidis}\ and\ \citenamefont
  {Perivolaropoulos}(2018)}]{Kazantzidis:2018rnb}%
  \BibitemOpen
  \bibfield  {author} {\bibinfo {author} {\bibfnamefont {Lavrentios}\
  \bibnamefont {Kazantzidis}}\ and\ \bibinfo {author} {\bibfnamefont
  {Leandros}\ \bibnamefont {Perivolaropoulos}},\ }\bibfield  {title} {\enquote
  {\bibinfo {title} {{Evolution of the $f\sigma_8$ tension with the
  Planck15/$\Lambda$CDM determination and implications for modified gravity
  theories}},}\ }\href {\doibase 10.1103/PhysRevD.97.103503} {\bibfield
  {journal} {\bibinfo  {journal} {Phys. Rev.}\ }\textbf {\bibinfo {volume}
  {D97}},\ \bibinfo {pages} {103503} (\bibinfo {year} {2018})},\ \Eprint
  {http://arxiv.org/abs/1803.01337} {arXiv:1803.01337 [astro-ph.CO]}
  \BibitemShut {NoStop}%
\bibitem [{\citenamefont {Perivolaropoulos}\ and\ \citenamefont
  {Kazantzidis}(2019)}]{Perivolaropoulos:2019vkb}%
  \BibitemOpen
  \bibfield  {author} {\bibinfo {author} {\bibfnamefont {Leandros}\
  \bibnamefont {Perivolaropoulos}}\ and\ \bibinfo {author} {\bibfnamefont
  {Lavrentios}\ \bibnamefont {Kazantzidis}},\ }\bibfield  {title} {\enquote
  {\bibinfo {title} {{Hints of modified gravity in cosmos and in the lab?}}}\
  }\href {\doibase 10.1142/S021827181942001X} {\bibfield  {journal} {\bibinfo
  {journal} {Int. J. Mod. Phys.}\ }\textbf {\bibinfo {volume} {D28}},\ \bibinfo
  {pages} {1942001} (\bibinfo {year} {2019})},\ \Eprint
  {http://arxiv.org/abs/1904.09462} {arXiv:1904.09462 [gr-qc]} \BibitemShut
  {NoStop}%
\bibitem [{\citenamefont {Kazantzidis}\ and\ \citenamefont
  {Perivolaropoulos}(2019)}]{Kazantzidis:2019dvk}%
  \BibitemOpen
  \bibfield  {author} {\bibinfo {author} {\bibfnamefont {Lavrentios}\
  \bibnamefont {Kazantzidis}}\ and\ \bibinfo {author} {\bibfnamefont
  {Leandros}\ \bibnamefont {Perivolaropoulos}},\ }\bibfield  {title} {\enquote
  {\bibinfo {title} {{Is gravity getting weaker at low z? Observational
  evidence and theoretical implications}},}\ }\href@noop {} {\  (\bibinfo
  {year} {2019})},\ \Eprint {http://arxiv.org/abs/1907.03176} {arXiv:1907.03176
  [astro-ph.CO]} \BibitemShut {NoStop}%
\bibitem [{\citenamefont {Carroll}(2001)}]{Carroll:2000fy}%
  \BibitemOpen
  \bibfield  {author} {\bibinfo {author} {\bibfnamefont {Sean~M.}\ \bibnamefont
  {Carroll}},\ }\bibfield  {title} {\enquote {\bibinfo {title} {{The
  Cosmological constant}},}\ }\href {\doibase 10.12942/lrr-2001-1} {\bibfield
  {journal} {\bibinfo  {journal} {Living Rev. Rel.}\ }\textbf {\bibinfo
  {volume} {4}},\ \bibinfo {pages} {1} (\bibinfo {year} {2001})},\ \Eprint
  {http://arxiv.org/abs/astro-ph/0004075} {arXiv:astro-ph/0004075 [astro-ph]}
  \BibitemShut {NoStop}%
\bibitem [{\citenamefont {Will}(2014)}]{Will:2014kxa}%
  \BibitemOpen
  \bibfield  {author} {\bibinfo {author} {\bibfnamefont {Clifford~M.}\
  \bibnamefont {Will}},\ }\bibfield  {title} {\enquote {\bibinfo {title} {{The
  Confrontation between General Relativity and Experiment}},}\ }\href {\doibase
  10.12942/lrr-2014-4} {\bibfield  {journal} {\bibinfo  {journal} {Living Rev.
  Rel.}\ }\textbf {\bibinfo {volume} {17}},\ \bibinfo {pages} {4} (\bibinfo
  {year} {2014})},\ \Eprint {http://arxiv.org/abs/1403.7377} {arXiv:1403.7377
  [gr-qc]} \BibitemShut {NoStop}%
\bibitem [{\citenamefont {Riess}\ \emph {et~al.}(1998)\citenamefont {Riess}
  \emph {et~al.}}]{Riess:1998cb}%
  \BibitemOpen
  \bibfield  {author} {\bibinfo {author} {\bibfnamefont {Adam~G.}\ \bibnamefont
  {Riess}} \emph {et~al.} (\bibinfo {collaboration} {Supernova Search Team}),\
  }\bibfield  {title} {\enquote {\bibinfo {title} {{Observational evidence from
  supernovae for an accelerating universe and a cosmological constant}},}\
  }\href {\doibase 10.1086/300499} {\bibfield  {journal} {\bibinfo  {journal}
  {Astron. J.}\ }\textbf {\bibinfo {volume} {116}},\ \bibinfo {pages}
  {1009--1038} (\bibinfo {year} {1998})},\ \Eprint
  {http://arxiv.org/abs/astro-ph/9805201} {arXiv:astro-ph/9805201 [astro-ph]}
  \BibitemShut {NoStop}%
\bibitem [{\citenamefont {Perlmutter}\ \emph {et~al.}(1999)\citenamefont
  {Perlmutter} \emph {et~al.}}]{Perlmutter:1998np}%
  \BibitemOpen
  \bibfield  {author} {\bibinfo {author} {\bibfnamefont {S.}~\bibnamefont
  {Perlmutter}} \emph {et~al.} (\bibinfo {collaboration} {Supernova Cosmology
  Project}),\ }\bibfield  {title} {\enquote {\bibinfo {title} {{Measurements of
  $\Omega$ and $\Lambda$ from 42 high redshift supernovae}},}\ }\href {\doibase
  10.1086/307221} {\bibfield  {journal} {\bibinfo  {journal} {Astrophys. J.}\
  }\textbf {\bibinfo {volume} {517}},\ \bibinfo {pages} {565--586} (\bibinfo
  {year} {1999})},\ \Eprint {http://arxiv.org/abs/astro-ph/9812133}
  {arXiv:astro-ph/9812133 [astro-ph]} \BibitemShut {NoStop}%
\bibitem [{\citenamefont {Weinberg}(1989)}]{Weinberg:1988cp}%
  \BibitemOpen
  \bibfield  {author} {\bibinfo {author} {\bibfnamefont {Steven}\ \bibnamefont
  {Weinberg}},\ }\bibfield  {title} {\enquote {\bibinfo {title} {{The
  Cosmological Constant Problem}},}\ }\href {\doibase 10.1103/RevModPhys.61.1}
  {\bibfield  {journal} {\bibinfo  {journal} {Rev. Mod. Phys.}\ }\textbf
  {\bibinfo {volume} {61}},\ \bibinfo {pages} {1--23} (\bibinfo {year}
  {1989})}\BibitemShut {NoStop}%
\bibitem [{\citenamefont {Martin}(2012)}]{Martin:2012bt}%
  \BibitemOpen
  \bibfield  {author} {\bibinfo {author} {\bibfnamefont {Jerome}\ \bibnamefont
  {Martin}},\ }\bibfield  {title} {\enquote {\bibinfo {title} {{Everything You
  Always Wanted To Know About The Cosmological Constant Problem (But Were
  Afraid To Ask)}},}\ }\href {\doibase 10.1016/j.crhy.2012.04.008} {\bibfield
  {journal} {\bibinfo  {journal} {Comptes Rendus Physique}\ }\textbf {\bibinfo
  {volume} {13}},\ \bibinfo {pages} {566--665} (\bibinfo {year} {2012})},\
  \Eprint {http://arxiv.org/abs/1205.3365} {arXiv:1205.3365 [astro-ph.CO]}
  \BibitemShut {NoStop}%
\bibitem [{\citenamefont {Burgess}(2015)}]{Burgess:2013ara}%
  \BibitemOpen
  \bibfield  {author} {\bibinfo {author} {\bibfnamefont {C.~P.}\ \bibnamefont
  {Burgess}},\ }\bibfield  {title} {\enquote {\bibinfo {title} {{The
  Cosmological Constant Problem: Why it's hard to get Dark Energy from
  Micro-physics}},}\ }in\ \href {\doibase
  10.1093/acprof:oso/9780198728856.003.0004} {\emph {\bibinfo {booktitle}
  {{Proceedings, 100th Les Houches Summer School: Post-Planck Cosmology: Les
  Houches, France, July 8 - August 2, 2013}}}}\ (\bibinfo {year} {2015})\ pp.\
  \bibinfo {pages} {149--197},\ \Eprint {http://arxiv.org/abs/1309.4133}
  {arXiv:1309.4133 [hep-th]} \BibitemShut {NoStop}%
\bibitem [{\citenamefont {{Steinhardt}}(1997)}]{1997cpp..conf..123S}%
  \BibitemOpen
  \bibfield  {author} {\bibinfo {author} {\bibfnamefont {P.~J.}\ \bibnamefont
  {{Steinhardt}}},\ }\bibfield  {title} {\enquote {\bibinfo {title}
  {{Cosmological Challenges for the 21st Century}},}\ }in\ \href@noop {} {\emph
  {\bibinfo {booktitle} {Critical Problems in Physics}}},\ \bibinfo {editor}
  {edited by\ \bibinfo {editor} {\bibfnamefont {V.~L.}\ \bibnamefont
  {{Fitch}}}, \bibinfo {editor} {\bibfnamefont {D.~R.}\ \bibnamefont
  {{Marlow}}}, \ and\ \bibinfo {editor} {\bibfnamefont {M.~A.~E.}\ \bibnamefont
  {{Dementi}}}}\ (\bibinfo {year} {1997})\ p.\ \bibinfo {pages}
  {123}\BibitemShut {NoStop}%
\bibitem [{\citenamefont {Velten}\ \emph {et~al.}(2014)\citenamefont {Velten},
  \citenamefont {vom Marttens},\ and\ \citenamefont
  {Zimdahl}}]{Velten:2014nra}%
  \BibitemOpen
  \bibfield  {author} {\bibinfo {author} {\bibfnamefont {H.~E.~S.}\
  \bibnamefont {Velten}}, \bibinfo {author} {\bibfnamefont {R.~F.}\
  \bibnamefont {vom Marttens}}, \ and\ \bibinfo {author} {\bibfnamefont
  {W.}~\bibnamefont {Zimdahl}},\ }\bibfield  {title} {\enquote {\bibinfo
  {title} {{Aspects of the cosmological “coincidence problem”}},}\ }\href
  {\doibase 10.1140/epjc/s10052-014-3160-4} {\bibfield  {journal} {\bibinfo
  {journal} {Eur. Phys. J.}\ }\textbf {\bibinfo {volume} {C74}},\ \bibinfo
  {pages} {3160} (\bibinfo {year} {2014})},\ \Eprint
  {http://arxiv.org/abs/1410.2509} {arXiv:1410.2509 [astro-ph.CO]} \BibitemShut
  {NoStop}%
\bibitem [{\citenamefont {Ade}\ \emph {et~al.}(2016{\natexlab{a}})\citenamefont
  {Ade} \emph {et~al.}}]{Ade:2015xua}%
  \BibitemOpen
  \bibfield  {author} {\bibinfo {author} {\bibfnamefont {P.~A.~R.}\
  \bibnamefont {Ade}} \emph {et~al.} (\bibinfo {collaboration} {Planck}),\
  }\bibfield  {title} {\enquote {\bibinfo {title} {{Planck 2015 results. XIII.
  Cosmological parameters}},}\ }\href {\doibase 10.1051/0004-6361/201525830}
  {\bibfield  {journal} {\bibinfo  {journal} {Astron. Astrophys.}\ }\textbf
  {\bibinfo {volume} {594}},\ \bibinfo {pages} {A13} (\bibinfo {year}
  {2016}{\natexlab{a}})},\ \Eprint {http://arxiv.org/abs/1502.01589}
  {arXiv:1502.01589 [astro-ph.CO]} \BibitemShut {NoStop}%
\bibitem [{\citenamefont {Aghanim}\ \emph {et~al.}(2018)\citenamefont {Aghanim}
  \emph {et~al.}}]{planck18}%
  \BibitemOpen
  \bibfield  {author} {\bibinfo {author} {\bibfnamefont {N.}~\bibnamefont
  {Aghanim}} \emph {et~al.} (\bibinfo {collaboration} {Planck}),\ }\bibfield
  {title} {\enquote {\bibinfo {title} {{Planck 2018 results. VI. Cosmological
  parameters}},}\ }\href@noop {} {\  (\bibinfo {year} {2018})},\ \Eprint
  {http://arxiv.org/abs/1807.06209} {arXiv:1807.06209 [astro-ph.CO]}
  \BibitemShut {NoStop}%
\bibitem [{\citenamefont {Riess}\ \emph {et~al.}(2016)\citenamefont {Riess}
  \emph {et~al.}}]{Riess:2016jrr}%
  \BibitemOpen
  \bibfield  {author} {\bibinfo {author} {\bibfnamefont {Adam~G.}\ \bibnamefont
  {Riess}} \emph {et~al.},\ }\bibfield  {title} {\enquote {\bibinfo {title} {{A
  2.4\% Determination of the Local Value of the Hubble Constant}},}\ }\href
  {\doibase 10.3847/0004-637X/826/1/56} {\bibfield  {journal} {\bibinfo
  {journal} {Astrophys. J.}\ }\textbf {\bibinfo {volume} {826}},\ \bibinfo
  {pages} {56} (\bibinfo {year} {2016})},\ \Eprint
  {http://arxiv.org/abs/1604.01424} {arXiv:1604.01424 [astro-ph.CO]}
  \BibitemShut {NoStop}%
\bibitem [{\citenamefont {Riess}\ \emph {et~al.}(2018)\citenamefont {Riess}
  \emph {et~al.}}]{Riess:2018byc}%
  \BibitemOpen
  \bibfield  {author} {\bibinfo {author} {\bibfnamefont {Adam~G.}\ \bibnamefont
  {Riess}} \emph {et~al.},\ }\bibfield  {title} {\enquote {\bibinfo {title}
  {{Milky Way Cepheid Standards for Measuring Cosmic Distances and Application
  to Gaia DR2: Implications for the Hubble Constant}},}\ }\href {\doibase
  10.3847/1538-4357/aac82e} {\bibfield  {journal} {\bibinfo  {journal}
  {Astrophys. J.}\ }\textbf {\bibinfo {volume} {861}},\ \bibinfo {pages} {126}
  (\bibinfo {year} {2018})},\ \Eprint {http://arxiv.org/abs/1804.10655}
  {arXiv:1804.10655 [astro-ph.CO]} \BibitemShut {NoStop}%
\bibitem [{\citenamefont {Birrer}\ \emph {et~al.}(2019)\citenamefont {Birrer}
  \emph {et~al.}}]{Birrer:2018vtm}%
  \BibitemOpen
  \bibfield  {author} {\bibinfo {author} {\bibfnamefont {S.}~\bibnamefont
  {Birrer}} \emph {et~al.},\ }\bibfield  {title} {\enquote {\bibinfo {title}
  {{H0LiCOW - IX. Cosmographic analysis of the doubly imaged quasar SDSS
  1206+4332 and a new measurement of the Hubble constant}},}\ }\href {\doibase
  10.1093/mnras/stz200} {\bibfield  {journal} {\bibinfo  {journal} {Mon. Not.
  Roy. Astron. Soc.}\ }\textbf {\bibinfo {volume} {484}},\ \bibinfo {pages}
  {4726} (\bibinfo {year} {2019})},\ \Eprint {http://arxiv.org/abs/1809.01274}
  {arXiv:1809.01274 [astro-ph.CO]} \BibitemShut {NoStop}%
\bibitem [{\citenamefont {Zhao}\ \emph {et~al.}(2009)\citenamefont {Zhao},
  \citenamefont {Pogosian}, \citenamefont {Silvestri},\ and\ \citenamefont
  {Zylberberg}}]{Zhao:2008bn}%
  \BibitemOpen
  \bibfield  {author} {\bibinfo {author} {\bibfnamefont {Gong-Bo}\ \bibnamefont
  {Zhao}}, \bibinfo {author} {\bibfnamefont {Levon}\ \bibnamefont {Pogosian}},
  \bibinfo {author} {\bibfnamefont {Alessandra}\ \bibnamefont {Silvestri}}, \
  and\ \bibinfo {author} {\bibfnamefont {Joel}\ \bibnamefont {Zylberberg}},\
  }\bibfield  {title} {\enquote {\bibinfo {title} {{Searching for modified
  growth patterns with tomographic surveys}},}\ }\href {\doibase
  10.1103/PhysRevD.79.083513} {\bibfield  {journal} {\bibinfo  {journal} {Phys.
  Rev.}\ }\textbf {\bibinfo {volume} {D79}},\ \bibinfo {pages} {083513}
  (\bibinfo {year} {2009})},\ \Eprint {http://arxiv.org/abs/0809.3791}
  {arXiv:0809.3791 [astro-ph]} \BibitemShut {NoStop}%
\bibitem [{\citenamefont {Cai}\ and\ \citenamefont
  {Bernstein}(2012)}]{Cai:2011wj}%
  \BibitemOpen
  \bibfield  {author} {\bibinfo {author} {\bibfnamefont {Yan-Chuan}\
  \bibnamefont {Cai}}\ and\ \bibinfo {author} {\bibfnamefont {Gary}\
  \bibnamefont {Bernstein}},\ }\bibfield  {title} {\enquote {\bibinfo {title}
  {{Combining weak lensing tomography and spectroscopic redshift surveys}},}\
  }\href {\doibase 10.1111/j.1365-2966.2012.20676.x} {\bibfield  {journal}
  {\bibinfo  {journal} {Mon. Not. Roy. Astron. Soc.}\ }\textbf {\bibinfo
  {volume} {422}},\ \bibinfo {pages} {1045--1056} (\bibinfo {year} {2012})},\
  \Eprint {http://arxiv.org/abs/1112.4478} {arXiv:1112.4478 [astro-ph.CO]}
  \BibitemShut {NoStop}%
\bibitem [{\citenamefont {Joudaki}\ and\ \citenamefont
  {Kaplinghat}(2012)}]{Joudaki:2011nw}%
  \BibitemOpen
  \bibfield  {author} {\bibinfo {author} {\bibfnamefont {Shahab}\ \bibnamefont
  {Joudaki}}\ and\ \bibinfo {author} {\bibfnamefont {Manoj}\ \bibnamefont
  {Kaplinghat}},\ }\bibfield  {title} {\enquote {\bibinfo {title} {{Dark Energy
  and Neutrino Masses from Future Measurements of the Expansion History and
  Growth of Structure}},}\ }\href {\doibase 10.1103/PhysRevD.86.023526}
  {\bibfield  {journal} {\bibinfo  {journal} {Phys. Rev.}\ }\textbf {\bibinfo
  {volume} {D86}},\ \bibinfo {pages} {023526} (\bibinfo {year} {2012})},\
  \Eprint {http://arxiv.org/abs/1106.0299} {arXiv:1106.0299 [astro-ph.CO]}
  \BibitemShut {NoStop}%
\bibitem [{\citenamefont {Tereno}\ \emph {et~al.}(2011)\citenamefont {Tereno},
  \citenamefont {Semboloni},\ and\ \citenamefont {Schrabback}}]{Tereno:2010dt}%
  \BibitemOpen
  \bibfield  {author} {\bibinfo {author} {\bibfnamefont {Ismael}\ \bibnamefont
  {Tereno}}, \bibinfo {author} {\bibfnamefont {Elisabetta}\ \bibnamefont
  {Semboloni}}, \ and\ \bibinfo {author} {\bibfnamefont {Tim}\ \bibnamefont
  {Schrabback}},\ }\bibfield  {title} {\enquote {\bibinfo {title} {{COSMOS
  weak-lensing constraints on modified gravity}},}\ }\href {\doibase
  10.1051/0004-6361/201016273} {\bibfield  {journal} {\bibinfo  {journal}
  {Astron. Astrophys.}\ }\textbf {\bibinfo {volume} {530}},\ \bibinfo {pages}
  {A68} (\bibinfo {year} {2011})},\ \Eprint {http://arxiv.org/abs/1012.5854}
  {arXiv:1012.5854 [astro-ph.CO]} \BibitemShut {NoStop}%
\bibitem [{\citenamefont {Simpson}\ \emph {et~al.}(2013)\citenamefont {Simpson}
  \emph {et~al.}}]{Simpson:2012ra}%
  \BibitemOpen
  \bibfield  {author} {\bibinfo {author} {\bibfnamefont {Fergus}\ \bibnamefont
  {Simpson}} \emph {et~al.},\ }\bibfield  {title} {\enquote {\bibinfo {title}
  {{CFHTLenS: Testing the Laws of Gravity with Tomographic Weak Lensing and
  Redshift Space Distortions}},}\ }\href {\doibase 10.1093/mnras/sts493}
  {\bibfield  {journal} {\bibinfo  {journal} {Mon. Not. Roy. Astron. Soc.}\
  }\textbf {\bibinfo {volume} {429}},\ \bibinfo {pages} {2249} (\bibinfo {year}
  {2013})},\ \Eprint {http://arxiv.org/abs/1212.3339} {arXiv:1212.3339
  [astro-ph.CO]} \BibitemShut {NoStop}%
\bibitem [{\citenamefont {Zhao}\ \emph {et~al.}(2015)\citenamefont {Zhao},
  \citenamefont {Bacon}, \citenamefont {Maartens}, \citenamefont {Santos},\
  and\ \citenamefont {Raccanelli}}]{Zhao:2015vta}%
  \BibitemOpen
  \bibfield  {author} {\bibinfo {author} {\bibfnamefont {Gongbo}\ \bibnamefont
  {Zhao}}, \bibinfo {author} {\bibfnamefont {David}\ \bibnamefont {Bacon}},
  \bibinfo {author} {\bibfnamefont {Roy}\ \bibnamefont {Maartens}}, \bibinfo
  {author} {\bibfnamefont {Mario}\ \bibnamefont {Santos}}, \ and\ \bibinfo
  {author} {\bibfnamefont {Alvise}\ \bibnamefont {Raccanelli}},\ }\bibfield
  {title} {\enquote {\bibinfo {title} {{Model-independent constraints on dark
  energy and modified gravity with the SKA}},}\ }\bibfield  {booktitle} {\emph
  {\bibinfo {booktitle} {{Proceedings, Advancing Astrophysics with the Square
  Kilometre Array (AASKA14): Giardini Naxos, Italy, June 9-13, 2014}}},\ }\href
  {\doibase 10.22323/1.215.0165} {\bibfield  {journal} {\bibinfo  {journal}
  {PoS}\ }\textbf {\bibinfo {volume} {AASKA14}},\ \bibinfo {pages} {165}
  (\bibinfo {year} {2015})}\BibitemShut {NoStop}%
\bibitem [{\citenamefont {Joudaki}\ \emph {et~al.}(2018)\citenamefont {Joudaki}
  \emph {et~al.}}]{Joudaki:2017zdt}%
  \BibitemOpen
  \bibfield  {author} {\bibinfo {author} {\bibfnamefont {Shahab}\ \bibnamefont
  {Joudaki}} \emph {et~al.},\ }\bibfield  {title} {\enquote {\bibinfo {title}
  {{KiDS-450 + 2dFLenS: Cosmological parameter constraints from weak
  gravitational lensing tomography and overlapping redshift-space galaxy
  clustering}},}\ }\href {\doibase 10.1093/mnras/stx2820} {\bibfield  {journal}
  {\bibinfo  {journal} {Mon. Not. Roy. Astron. Soc.}\ }\textbf {\bibinfo
  {volume} {474}},\ \bibinfo {pages} {4894--4924} (\bibinfo {year} {2018})},\
  \Eprint {http://arxiv.org/abs/1707.06627} {arXiv:1707.06627 [astro-ph.CO]}
  \BibitemShut {NoStop}%
\bibitem [{\citenamefont {Bertschinger}(2006)}]{Bertschinger:2006aw}%
  \BibitemOpen
  \bibfield  {author} {\bibinfo {author} {\bibfnamefont {Edmund}\ \bibnamefont
  {Bertschinger}},\ }\bibfield  {title} {\enquote {\bibinfo {title} {{On the
  Growth of Perturbations as a Test of Dark Energy}},}\ }\href {\doibase
  10.1086/506021} {\bibfield  {journal} {\bibinfo  {journal} {Astrophys. J.}\
  }\textbf {\bibinfo {volume} {648}},\ \bibinfo {pages} {797--806} (\bibinfo
  {year} {2006})},\ \Eprint {http://arxiv.org/abs/astro-ph/0604485}
  {arXiv:astro-ph/0604485 [astro-ph]} \BibitemShut {NoStop}%
\bibitem [{\citenamefont {Nesseris}\ and\ \citenamefont
  {Perivolaropoulos}(2007{\natexlab{a}})}]{Nesseris:2006er}%
  \BibitemOpen
  \bibfield  {author} {\bibinfo {author} {\bibfnamefont {S.}~\bibnamefont
  {Nesseris}}\ and\ \bibinfo {author} {\bibfnamefont {Leandros}\ \bibnamefont
  {Perivolaropoulos}},\ }\bibfield  {title} {\enquote {\bibinfo {title}
  {{Crossing the Phantom Divide: Theoretical Implications and Observational
  Status}},}\ }\href {\doibase 10.1088/1475-7516/2007/01/018} {\bibfield
  {journal} {\bibinfo  {journal} {JCAP}\ }\textbf {\bibinfo {volume} {0701}},\
  \bibinfo {pages} {018} (\bibinfo {year} {2007}{\natexlab{a}})},\ \Eprint
  {http://arxiv.org/abs/astro-ph/0610092} {arXiv:astro-ph/0610092 [astro-ph]}
  \BibitemShut {NoStop}%
\bibitem [{\citenamefont {Basilakos}\ \emph {et~al.}(2013)\citenamefont
  {Basilakos}, \citenamefont {Nesseris},\ and\ \citenamefont
  {Perivolaropoulos}}]{Basilakos:2013nfa}%
  \BibitemOpen
  \bibfield  {author} {\bibinfo {author} {\bibfnamefont {Spyros}\ \bibnamefont
  {Basilakos}}, \bibinfo {author} {\bibfnamefont {Savvas}\ \bibnamefont
  {Nesseris}}, \ and\ \bibinfo {author} {\bibfnamefont {Leandros}\ \bibnamefont
  {Perivolaropoulos}},\ }\bibfield  {title} {\enquote {\bibinfo {title}
  {{Observational constraints on viable f(R) parametrizations with geometrical
  and dynamical probes}},}\ }\href {\doibase 10.1103/PhysRevD.87.123529}
  {\bibfield  {journal} {\bibinfo  {journal} {Phys. Rev.}\ }\textbf {\bibinfo
  {volume} {D87}},\ \bibinfo {pages} {123529} (\bibinfo {year} {2013})},\
  \Eprint {http://arxiv.org/abs/1302.6051} {arXiv:1302.6051 [astro-ph.CO]}
  \BibitemShut {NoStop}%
\bibitem [{\citenamefont {Ruiz}\ and\ \citenamefont
  {Huterer}(2015)}]{Ruiz:2014hma}%
  \BibitemOpen
  \bibfield  {author} {\bibinfo {author} {\bibfnamefont {Eduardo~J.}\
  \bibnamefont {Ruiz}}\ and\ \bibinfo {author} {\bibfnamefont {Dragan}\
  \bibnamefont {Huterer}},\ }\bibfield  {title} {\enquote {\bibinfo {title}
  {{Testing the dark energy consistency with geometry and growth}},}\ }\href
  {\doibase 10.1103/PhysRevD.91.063009} {\bibfield  {journal} {\bibinfo
  {journal} {Phys. Rev.}\ }\textbf {\bibinfo {volume} {D91}},\ \bibinfo {pages}
  {063009} (\bibinfo {year} {2015})},\ \Eprint {http://arxiv.org/abs/1410.5832}
  {arXiv:1410.5832 [astro-ph.CO]} \BibitemShut {NoStop}%
\bibitem [{\citenamefont {Rozo}\ \emph {et~al.}(2010)\citenamefont {Rozo} \emph
  {et~al.}}]{Rozo:2009jj}%
  \BibitemOpen
  \bibfield  {author} {\bibinfo {author} {\bibfnamefont {Eduardo}\ \bibnamefont
  {Rozo}} \emph {et~al.} (\bibinfo {collaboration} {DSDD}),\ }\bibfield
  {title} {\enquote {\bibinfo {title} {{Cosmological Constraints from the SDSS
  maxBCG Cluster Catalog}},}\ }\href {\doibase 10.1088/0004-637X/708/1/645}
  {\bibfield  {journal} {\bibinfo  {journal} {Astrophys. J.}\ }\textbf
  {\bibinfo {volume} {708}},\ \bibinfo {pages} {645--660} (\bibinfo {year}
  {2010})},\ \Eprint {http://arxiv.org/abs/0902.3702} {arXiv:0902.3702
  [astro-ph.CO]} \BibitemShut {NoStop}%
\bibitem [{\citenamefont {Rapetti}\ \emph {et~al.}(2009)\citenamefont
  {Rapetti}, \citenamefont {Allen}, \citenamefont {Mantz},\ and\ \citenamefont
  {Ebeling}}]{Rapetti:2008rm}%
  \BibitemOpen
  \bibfield  {author} {\bibinfo {author} {\bibfnamefont {David}\ \bibnamefont
  {Rapetti}}, \bibinfo {author} {\bibfnamefont {Steven~W.}\ \bibnamefont
  {Allen}}, \bibinfo {author} {\bibfnamefont {Adam}\ \bibnamefont {Mantz}}, \
  and\ \bibinfo {author} {\bibfnamefont {Harald}\ \bibnamefont {Ebeling}},\
  }\bibfield  {title} {\enquote {\bibinfo {title} {{Constraints on modified
  gravity from the observed X-ray luminosity function of galaxy clusters}},}\
  }\href {\doibase 10.1111/j.1365-2966.2009.15510.x} {\bibfield  {journal}
  {\bibinfo  {journal} {Mon. Not. Roy. Astron. Soc.}\ }\textbf {\bibinfo
  {volume} {400}},\ \bibinfo {pages} {699} (\bibinfo {year} {2009})},\ \Eprint
  {http://arxiv.org/abs/0812.2259} {arXiv:0812.2259 [astro-ph]} \BibitemShut
  {NoStop}%
\bibitem [{\citenamefont {Bocquet}\ \emph {et~al.}(2015)\citenamefont {Bocquet}
  \emph {et~al.}}]{Bocquet:2014lmj}%
  \BibitemOpen
  \bibfield  {author} {\bibinfo {author} {\bibfnamefont {S.}~\bibnamefont
  {Bocquet}} \emph {et~al.} (\bibinfo {collaboration} {SPT}),\ }\bibfield
  {title} {\enquote {\bibinfo {title} {{Mass Calibration and Cosmological
  Analysis of the SPT-SZ Galaxy Cluster Sample Using Velocity Dispersion
  $\sigma_v$ and X-ray $Y_\textrm{X}$ Measurements}},}\ }\href {\doibase
  10.1088/0004-637X/799/2/214} {\bibfield  {journal} {\bibinfo  {journal}
  {Astrophys. J.}\ }\textbf {\bibinfo {volume} {799}},\ \bibinfo {pages} {214}
  (\bibinfo {year} {2015})},\ \Eprint {http://arxiv.org/abs/1407.2942}
  {arXiv:1407.2942 [astro-ph.CO]} \BibitemShut {NoStop}%
\bibitem [{\citenamefont {Schmidt}(2008)}]{Schmidt:2008hc}%
  \BibitemOpen
  \bibfield  {author} {\bibinfo {author} {\bibfnamefont {Fabian}\ \bibnamefont
  {Schmidt}},\ }\bibfield  {title} {\enquote {\bibinfo {title} {{Weak Lensing
  Probes of Modified Gravity}},}\ }\href {\doibase 10.1103/PhysRevD.78.043002}
  {\bibfield  {journal} {\bibinfo  {journal} {Phys. Rev.}\ }\textbf {\bibinfo
  {volume} {D78}},\ \bibinfo {pages} {043002} (\bibinfo {year} {2008})},\
  \Eprint {http://arxiv.org/abs/0805.4812} {arXiv:0805.4812 [astro-ph]}
  \BibitemShut {NoStop}%
\bibitem [{\citenamefont {Hildebrandt}\ \emph {et~al.}(2017)\citenamefont
  {Hildebrandt} \emph {et~al.}}]{kids1}%
  \BibitemOpen
  \bibfield  {author} {\bibinfo {author} {\bibfnamefont {H.}~\bibnamefont
  {Hildebrandt}} \emph {et~al.},\ }\bibfield  {title} {\enquote {\bibinfo
  {title} {{KiDS-450: Cosmological parameter constraints from tomographic weak
  gravitational lensing}},}\ }\href {\doibase 10.1093/mnras/stw2805} {\bibfield
   {journal} {\bibinfo  {journal} {Mon. Not. Roy. Astron. Soc.}\ }\textbf
  {\bibinfo {volume} {465}},\ \bibinfo {pages} {1454} (\bibinfo {year}
  {2017})},\ \Eprint {http://arxiv.org/abs/1606.05338} {arXiv:1606.05338
  [astro-ph.CO]} \BibitemShut {NoStop}%
\bibitem [{\citenamefont {Heymans}\ \emph {et~al.}(2012)\citenamefont {Heymans}
  \emph {et~al.}}]{cfhtlens}%
  \BibitemOpen
  \bibfield  {author} {\bibinfo {author} {\bibfnamefont {Catherine}\
  \bibnamefont {Heymans}} \emph {et~al.},\ }\bibfield  {title} {\enquote
  {\bibinfo {title} {{CFHTLenS: The Canada-France-Hawaii Telescope Lensing
  Survey}},}\ }\href {\doibase 10.1111/j.1365-2966.2012.21952.x} {\bibfield
  {journal} {\bibinfo  {journal} {Mon. Not. Roy. Astron. Soc.}\ }\textbf
  {\bibinfo {volume} {427}},\ \bibinfo {pages} {146} (\bibinfo {year}
  {2012})},\ \Eprint {http://arxiv.org/abs/1210.0032} {arXiv:1210.0032
  [astro-ph.CO]} \BibitemShut {NoStop}%
\bibitem [{\citenamefont {Troxel}\ \emph
  {et~al.}(2018{\natexlab{a}})\citenamefont {Troxel} \emph
  {et~al.}}]{Troxel:2017xyo}%
  \BibitemOpen
  \bibfield  {author} {\bibinfo {author} {\bibfnamefont {M.~A.}\ \bibnamefont
  {Troxel}} \emph {et~al.} (\bibinfo {collaboration} {DES}),\ }\bibfield
  {title} {\enquote {\bibinfo {title} {{Dark Energy Survey Year 1 results:
  Cosmological constraints from cosmic shear}},}\ }\href {\doibase
  10.1103/PhysRevD.98.043528} {\bibfield  {journal} {\bibinfo  {journal} {Phys.
  Rev.}\ }\textbf {\bibinfo {volume} {D98}},\ \bibinfo {pages} {043528}
  (\bibinfo {year} {2018}{\natexlab{a}})},\ \Eprint
  {http://arxiv.org/abs/1708.01538} {arXiv:1708.01538 [astro-ph.CO]}
  \BibitemShut {NoStop}%
\bibitem [{\citenamefont {Köhlinger}\ \emph {et~al.}(2017)\citenamefont
  {Köhlinger} \emph {et~al.}}]{kids2}%
  \BibitemOpen
  \bibfield  {author} {\bibinfo {author} {\bibfnamefont {F.}~\bibnamefont
  {Köhlinger}} \emph {et~al.},\ }\bibfield  {title} {\enquote {\bibinfo
  {title} {{KiDS-450: The tomographic weak lensing power spectrum and
  constraints on cosmological parameters}},}\ }\href {\doibase
  10.1093/mnras/stx1820} {\bibfield  {journal} {\bibinfo  {journal} {Mon. Not.
  Roy. Astron. Soc.}\ }\textbf {\bibinfo {volume} {471}},\ \bibinfo {pages}
  {4412--4435} (\bibinfo {year} {2017})},\ \Eprint
  {http://arxiv.org/abs/1706.02892} {arXiv:1706.02892 [astro-ph.CO]}
  \BibitemShut {NoStop}%
\bibitem [{\citenamefont {Abbott}\ \emph {et~al.}(2018)\citenamefont {Abbott}
  \emph {et~al.}}]{des3}%
  \BibitemOpen
  \bibfield  {author} {\bibinfo {author} {\bibfnamefont {T.~M.~C.}\
  \bibnamefont {Abbott}} \emph {et~al.} (\bibinfo {collaboration} {DES}),\
  }\bibfield  {title} {\enquote {\bibinfo {title} {{Dark Energy Survey year 1
  results: Cosmological constraints from galaxy clustering and weak
  lensing}},}\ }\href {\doibase 10.1103/PhysRevD.98.043526} {\bibfield
  {journal} {\bibinfo  {journal} {Phys. Rev.}\ }\textbf {\bibinfo {volume}
  {D98}},\ \bibinfo {pages} {043526} (\bibinfo {year} {2018})},\ \Eprint
  {http://arxiv.org/abs/1708.01530} {arXiv:1708.01530 [astro-ph.CO]}
  \BibitemShut {NoStop}%
\bibitem [{\citenamefont {Abbott}\ \emph {et~al.}(2019)\citenamefont {Abbott}
  \emph {et~al.}}]{Abbott:2018xao}%
  \BibitemOpen
  \bibfield  {author} {\bibinfo {author} {\bibfnamefont {T.~M.~C.}\
  \bibnamefont {Abbott}} \emph {et~al.} (\bibinfo {collaboration} {DES}),\
  }\bibfield  {title} {\enquote {\bibinfo {title} {{Dark Energy Survey Year 1
  Results: Constraints on Extended Cosmological Models from Galaxy Clustering
  and Weak Lensing}},}\ }\href {\doibase 10.1103/PhysRevD.99.123505} {\bibfield
   {journal} {\bibinfo  {journal} {Phys. Rev.}\ }\textbf {\bibinfo {volume}
  {D99}},\ \bibinfo {pages} {123505} (\bibinfo {year} {2019})},\ \Eprint
  {http://arxiv.org/abs/1810.02499} {arXiv:1810.02499 [astro-ph.CO]}
  \BibitemShut {NoStop}%
\bibitem [{\citenamefont {Samushia}\ \emph {et~al.}(2013)\citenamefont
  {Samushia} \emph {et~al.}}]{Samushia:2012iq}%
  \BibitemOpen
  \bibfield  {author} {\bibinfo {author} {\bibfnamefont {Lado}\ \bibnamefont
  {Samushia}} \emph {et~al.},\ }\bibfield  {title} {\enquote {\bibinfo {title}
  {{The Clustering of Galaxies in the SDSS-III DR9 Baryon Oscillation
  Spectroscopic Survey: Testing Deviations from $\Lambda$ and General
  Relativity using anisotropic clustering of galaxies}},}\ }\href {\doibase
  10.1093/mnras/sts443} {\bibfield  {journal} {\bibinfo  {journal} {Mon. Not.
  Roy. Astron. Soc.}\ }\textbf {\bibinfo {volume} {429}},\ \bibinfo {pages}
  {1514--1528} (\bibinfo {year} {2013})},\ \Eprint
  {http://arxiv.org/abs/1206.5309} {arXiv:1206.5309 [astro-ph.CO]} \BibitemShut
  {NoStop}%
\bibitem [{\citenamefont {Macaulay}\ \emph {et~al.}(2013)\citenamefont
  {Macaulay}, \citenamefont {Wehus},\ and\ \citenamefont
  {Eriksen}}]{Macaulay:2013swa}%
  \BibitemOpen
  \bibfield  {author} {\bibinfo {author} {\bibfnamefont {Edward}\ \bibnamefont
  {Macaulay}}, \bibinfo {author} {\bibfnamefont {Ingunn~Kathrine}\ \bibnamefont
  {Wehus}}, \ and\ \bibinfo {author} {\bibfnamefont {Hans~Kristian}\
  \bibnamefont {Eriksen}},\ }\bibfield  {title} {\enquote {\bibinfo {title}
  {{Lower Growth Rate from Recent Redshift Space Distortion Measurements than
  Expected from Planck}},}\ }\href {\doibase 10.1103/PhysRevLett.111.161301}
  {\bibfield  {journal} {\bibinfo  {journal} {Phys. Rev. Lett.}\ }\textbf
  {\bibinfo {volume} {111}},\ \bibinfo {pages} {161301} (\bibinfo {year}
  {2013})},\ \Eprint {http://arxiv.org/abs/1303.6583} {arXiv:1303.6583
  [astro-ph.CO]} \BibitemShut {NoStop}%
\bibitem [{\citenamefont {Johnson}\ \emph {et~al.}(2016)\citenamefont
  {Johnson}, \citenamefont {Blake}, \citenamefont {Dossett}, \citenamefont
  {Koda}, \citenamefont {Parkinson},\ and\ \citenamefont
  {Joudaki}}]{Johnson:2015aaa}%
  \BibitemOpen
  \bibfield  {author} {\bibinfo {author} {\bibfnamefont {Andrew}\ \bibnamefont
  {Johnson}}, \bibinfo {author} {\bibfnamefont {Chris}\ \bibnamefont {Blake}},
  \bibinfo {author} {\bibfnamefont {Jason}\ \bibnamefont {Dossett}}, \bibinfo
  {author} {\bibfnamefont {Jun}\ \bibnamefont {Koda}}, \bibinfo {author}
  {\bibfnamefont {David}\ \bibnamefont {Parkinson}}, \ and\ \bibinfo {author}
  {\bibfnamefont {Shahab}\ \bibnamefont {Joudaki}},\ }\bibfield  {title}
  {\enquote {\bibinfo {title} {{Searching for Modified Gravity: Scale and
  Redshift Dependent Constraints from Galaxy Peculiar Velocities}},}\ }\href
  {\doibase 10.1093/mnras/stw447} {\bibfield  {journal} {\bibinfo  {journal}
  {Mon. Not. Roy. Astron. Soc.}\ }\textbf {\bibinfo {volume} {458}},\ \bibinfo
  {pages} {2725--2744} (\bibinfo {year} {2016})},\ \Eprint
  {http://arxiv.org/abs/1504.06885} {arXiv:1504.06885 [astro-ph.CO]}
  \BibitemShut {NoStop}%
\bibitem [{\citenamefont {Kumar}\ and\ \citenamefont
  {Nunes}(2016)}]{Kumar:2016zpg}%
  \BibitemOpen
  \bibfield  {author} {\bibinfo {author} {\bibfnamefont {Suresh}\ \bibnamefont
  {Kumar}}\ and\ \bibinfo {author} {\bibfnamefont {Rafael~C.}\ \bibnamefont
  {Nunes}},\ }\bibfield  {title} {\enquote {\bibinfo {title} {{Probing the
  interaction between dark matter and dark energy in the presence of massive
  neutrinos}},}\ }\href {\doibase 10.1103/PhysRevD.94.123511} {\bibfield
  {journal} {\bibinfo  {journal} {Phys. Rev.}\ }\textbf {\bibinfo {volume}
  {D94}},\ \bibinfo {pages} {123511} (\bibinfo {year} {2016})},\ \Eprint
  {http://arxiv.org/abs/1608.02454} {arXiv:1608.02454 [astro-ph.CO]}
  \BibitemShut {NoStop}%
\bibitem [{\citenamefont {Pourtsidou}\ and\ \citenamefont
  {Tram}(2016)}]{Pourtsidou:2016ico}%
  \BibitemOpen
  \bibfield  {author} {\bibinfo {author} {\bibfnamefont {Alkistis}\
  \bibnamefont {Pourtsidou}}\ and\ \bibinfo {author} {\bibfnamefont {Thomas}\
  \bibnamefont {Tram}},\ }\bibfield  {title} {\enquote {\bibinfo {title}
  {{Reconciling CMB and structure growth measurements with dark energy
  interactions}},}\ }\href {\doibase 10.1103/PhysRevD.94.043518} {\bibfield
  {journal} {\bibinfo  {journal} {Phys. Rev.}\ }\textbf {\bibinfo {volume}
  {D94}},\ \bibinfo {pages} {043518} (\bibinfo {year} {2016})},\ \Eprint
  {http://arxiv.org/abs/1604.04222} {arXiv:1604.04222 [astro-ph.CO]}
  \BibitemShut {NoStop}%
\bibitem [{\citenamefont {Barros}\ \emph {et~al.}(2019)\citenamefont {Barros},
  \citenamefont {Amendola}, \citenamefont {Barreiro},\ and\ \citenamefont
  {Nunes}}]{Barros:2018efl}%
  \BibitemOpen
  \bibfield  {author} {\bibinfo {author} {\bibfnamefont {Bruno~J.}\
  \bibnamefont {Barros}}, \bibinfo {author} {\bibfnamefont {Luca}\ \bibnamefont
  {Amendola}}, \bibinfo {author} {\bibfnamefont {Tiago}\ \bibnamefont
  {Barreiro}}, \ and\ \bibinfo {author} {\bibfnamefont {Nelson~J.}\
  \bibnamefont {Nunes}},\ }\bibfield  {title} {\enquote {\bibinfo {title}
  {{Coupled quintessence with a $\Lambda$CDM background: removing the
  $\sigma_8$ tension}},}\ }\href {\doibase 10.1088/1475-7516/2019/01/007}
  {\bibfield  {journal} {\bibinfo  {journal} {JCAP}\ }\textbf {\bibinfo
  {volume} {1901}},\ \bibinfo {pages} {007} (\bibinfo {year} {2019})},\ \Eprint
  {http://arxiv.org/abs/1802.09216} {arXiv:1802.09216 [astro-ph.CO]}
  \BibitemShut {NoStop}%
\bibitem [{\citenamefont {Camera}\ \emph {et~al.}(2019)\citenamefont {Camera},
  \citenamefont {Martinelli},\ and\ \citenamefont {Bertacca}}]{Camera:2019vbp}%
  \BibitemOpen
  \bibfield  {author} {\bibinfo {author} {\bibfnamefont {Stefano}\ \bibnamefont
  {Camera}}, \bibinfo {author} {\bibfnamefont {Matteo}\ \bibnamefont
  {Martinelli}}, \ and\ \bibinfo {author} {\bibfnamefont {Daniele}\
  \bibnamefont {Bertacca}},\ }\bibfield  {title} {\enquote {\bibinfo {title}
  {{Does quartessence ease cosmic tensions?}}}\ }\href {\doibase
  10.1016/j.dark.2018.11.008} {\bibfield  {journal} {\bibinfo  {journal} {Phys.
  Dark Univ.}\ }\textbf {\bibinfo {volume} {23}},\ \bibinfo {pages} {100247}
  (\bibinfo {year} {2019})},\ \Eprint {http://arxiv.org/abs/1704.06277}
  {arXiv:1704.06277 [astro-ph.CO]} \BibitemShut {NoStop}%
\bibitem [{\citenamefont {Yang}\ \emph {et~al.}(2019)\citenamefont {Yang},
  \citenamefont {Pan}, \citenamefont {Di~Valentino}, \citenamefont
  {Saridakis},\ and\ \citenamefont {Chakraborty}}]{Yang:2018qmz}%
  \BibitemOpen
  \bibfield  {author} {\bibinfo {author} {\bibfnamefont {Weiqiang}\
  \bibnamefont {Yang}}, \bibinfo {author} {\bibfnamefont {Supriya}\
  \bibnamefont {Pan}}, \bibinfo {author} {\bibfnamefont {Eleonora}\
  \bibnamefont {Di~Valentino}}, \bibinfo {author} {\bibfnamefont {Emmanuel~N.}\
  \bibnamefont {Saridakis}}, \ and\ \bibinfo {author} {\bibfnamefont {Subenoy}\
  \bibnamefont {Chakraborty}},\ }\bibfield  {title} {\enquote {\bibinfo {title}
  {{Observational constraints on one-parameter dynamical dark-energy
  parametrizations and the $H_0$ tension}},}\ }\href {\doibase
  10.1103/PhysRevD.99.043543} {\bibfield  {journal} {\bibinfo  {journal} {Phys.
  Rev.}\ }\textbf {\bibinfo {volume} {D99}},\ \bibinfo {pages} {043543}
  (\bibinfo {year} {2019})},\ \Eprint {http://arxiv.org/abs/1810.05141}
  {arXiv:1810.05141 [astro-ph.CO]} \BibitemShut {NoStop}%
\bibitem [{\citenamefont {Lambiase}\ \emph {et~al.}(2019)\citenamefont
  {Lambiase}, \citenamefont {Mohanty}, \citenamefont {Narang},\ and\
  \citenamefont {Parashari}}]{Lambiase:2018ows}%
  \BibitemOpen
  \bibfield  {author} {\bibinfo {author} {\bibfnamefont {Gaetano}\ \bibnamefont
  {Lambiase}}, \bibinfo {author} {\bibfnamefont {Subhendra}\ \bibnamefont
  {Mohanty}}, \bibinfo {author} {\bibfnamefont {Ashish}\ \bibnamefont
  {Narang}}, \ and\ \bibinfo {author} {\bibfnamefont {Priyank}\ \bibnamefont
  {Parashari}},\ }\bibfield  {title} {\enquote {\bibinfo {title} {{Testing dark
  energy models in the light of $\sigma _8$ tension}},}\ }\href {\doibase
  10.1140/epjc/s10052-019-6634-6} {\bibfield  {journal} {\bibinfo  {journal}
  {Eur. Phys. J.}\ }\textbf {\bibinfo {volume} {C79}},\ \bibinfo {pages} {141}
  (\bibinfo {year} {2019})},\ \Eprint {http://arxiv.org/abs/1804.07154}
  {arXiv:1804.07154 [astro-ph.CO]} \BibitemShut {NoStop}%
\bibitem [{\citenamefont {Gomez-Valent}\ and\ \citenamefont
  {Sola}(2017)}]{Gomez-Valent:2017idt}%
  \BibitemOpen
  \bibfield  {author} {\bibinfo {author} {\bibfnamefont {Adria}\ \bibnamefont
  {Gomez-Valent}}\ and\ \bibinfo {author} {\bibfnamefont {Joan}\ \bibnamefont
  {Sola}},\ }\bibfield  {title} {\enquote {\bibinfo {title} {{Relaxing the
  $\sigma_8$-tension through running vacuum in the Universe}},}\ }\href
  {\doibase 10.1209/0295-5075/120/39001} {\bibfield  {journal} {\bibinfo
  {journal} {EPL}\ }\textbf {\bibinfo {volume} {120}},\ \bibinfo {pages}
  {39001} (\bibinfo {year} {2017})},\ \Eprint {http://arxiv.org/abs/1711.00692}
  {arXiv:1711.00692 [astro-ph.CO]} \BibitemShut {NoStop}%
\bibitem [{\citenamefont {Gómez-Valent}\ and\ \citenamefont
  {Solà}(2018)}]{Gomez-Valent:2018nib}%
  \BibitemOpen
  \bibfield  {author} {\bibinfo {author} {\bibfnamefont {Adrià}\ \bibnamefont
  {Gómez-Valent}}\ and\ \bibinfo {author} {\bibfnamefont {Joan}\ \bibnamefont
  {Solà}},\ }\bibfield  {title} {\enquote {\bibinfo {title} {{Density
  perturbations for running vacuum: a successful approach to structure
  formation and to the $\sigma_8$-tension}},}\ }\href@noop {} {\  (\bibinfo
  {year} {2018})},\ \Eprint {http://arxiv.org/abs/1801.08501} {arXiv:1801.08501
  [astro-ph.CO]} \BibitemShut {NoStop}%
\bibitem [{\citenamefont {Diaz~Rivero}\ \emph {et~al.}(2019)\citenamefont
  {Diaz~Rivero}, \citenamefont {Miranda},\ and\ \citenamefont
  {Dvorkin}}]{DiazRivero:2019ukx}%
  \BibitemOpen
  \bibfield  {author} {\bibinfo {author} {\bibfnamefont {Ana}\ \bibnamefont
  {Diaz~Rivero}}, \bibinfo {author} {\bibfnamefont {V.}~\bibnamefont
  {Miranda}}, \ and\ \bibinfo {author} {\bibfnamefont {Cora}\ \bibnamefont
  {Dvorkin}},\ }\bibfield  {title} {\enquote {\bibinfo {title} {{Observable
  Predictions for Massive-Neutrino Cosmologies with Model-Independent Dark
  Energy}},}\ }\href {\doibase 10.1103/PhysRevD.100.063504} {\bibfield
  {journal} {\bibinfo  {journal} {Phys. Rev.}\ }\textbf {\bibinfo {volume}
  {D100}},\ \bibinfo {pages} {063504} (\bibinfo {year} {2019})},\ \Eprint
  {http://arxiv.org/abs/1903.03125} {arXiv:1903.03125 [astro-ph.CO]}
  \BibitemShut {NoStop}%
\bibitem [{\citenamefont {Tsujikawa}(2007)}]{Tsujikawa:2007gd}%
  \BibitemOpen
  \bibfield  {author} {\bibinfo {author} {\bibfnamefont {Shinji}\ \bibnamefont
  {Tsujikawa}},\ }\bibfield  {title} {\enquote {\bibinfo {title} {{Matter
  density perturbations and effective gravitational constant in modified
  gravity models of dark energy}},}\ }\href {\doibase
  10.1103/PhysRevD.76.023514} {\bibfield  {journal} {\bibinfo  {journal} {Phys.
  Rev.}\ }\textbf {\bibinfo {volume} {D76}},\ \bibinfo {pages} {023514}
  (\bibinfo {year} {2007})},\ \Eprint {http://arxiv.org/abs/0705.1032}
  {arXiv:0705.1032 [astro-ph]} \BibitemShut {NoStop}%
\bibitem [{\citenamefont {Hu}\ and\ \citenamefont {Sawicki}(2007)}]{Hu:2007nk}%
  \BibitemOpen
  \bibfield  {author} {\bibinfo {author} {\bibfnamefont {Wayne}\ \bibnamefont
  {Hu}}\ and\ \bibinfo {author} {\bibfnamefont {Ignacy}\ \bibnamefont
  {Sawicki}},\ }\bibfield  {title} {\enquote {\bibinfo {title} {{Models of f(R)
  Cosmic Acceleration that Evade Solar-System Tests}},}\ }\href {\doibase
  10.1103/PhysRevD.76.064004} {\bibfield  {journal} {\bibinfo  {journal} {Phys.
  Rev.}\ }\textbf {\bibinfo {volume} {D76}},\ \bibinfo {pages} {064004}
  (\bibinfo {year} {2007})},\ \Eprint {http://arxiv.org/abs/0705.1158}
  {arXiv:0705.1158 [astro-ph]} \BibitemShut {NoStop}%
\bibitem [{\citenamefont {Gannouji}\ \emph {et~al.}(2009)\citenamefont
  {Gannouji}, \citenamefont {Moraes},\ and\ \citenamefont
  {Polarski}}]{Gannouji:2008wt}%
  \BibitemOpen
  \bibfield  {author} {\bibinfo {author} {\bibfnamefont {R.}~\bibnamefont
  {Gannouji}}, \bibinfo {author} {\bibfnamefont {B.}~\bibnamefont {Moraes}}, \
  and\ \bibinfo {author} {\bibfnamefont {D.}~\bibnamefont {Polarski}},\
  }\bibfield  {title} {\enquote {\bibinfo {title} {{The growth of matter
  perturbations in f(R) models}},}\ }\href {\doibase
  10.1088/1475-7516/2009/02/034} {\bibfield  {journal} {\bibinfo  {journal}
  {JCAP}\ }\textbf {\bibinfo {volume} {0902}},\ \bibinfo {pages} {034}
  (\bibinfo {year} {2009})},\ \Eprint {http://arxiv.org/abs/0809.3374}
  {arXiv:0809.3374 [astro-ph]} \BibitemShut {NoStop}%
\bibitem [{\citenamefont {Capozziello}\ and\ \citenamefont
  {De~Laurentis}(2011)}]{Capozziello:2011et}%
  \BibitemOpen
  \bibfield  {author} {\bibinfo {author} {\bibfnamefont {Salvatore}\
  \bibnamefont {Capozziello}}\ and\ \bibinfo {author} {\bibfnamefont
  {Mariafelicia}\ \bibnamefont {De~Laurentis}},\ }\bibfield  {title} {\enquote
  {\bibinfo {title} {{Extended Theories of Gravity}},}\ }\href {\doibase
  10.1016/j.physrep.2011.09.003} {\bibfield  {journal} {\bibinfo  {journal}
  {Phys. Rept.}\ }\textbf {\bibinfo {volume} {509}},\ \bibinfo {pages}
  {167--321} (\bibinfo {year} {2011})},\ \Eprint
  {http://arxiv.org/abs/1108.6266} {arXiv:1108.6266 [gr-qc]} \BibitemShut
  {NoStop}%
\bibitem [{\citenamefont {Clifton}\ \emph {et~al.}(2012)\citenamefont
  {Clifton}, \citenamefont {Ferreira}, \citenamefont {Padilla},\ and\
  \citenamefont {Skordis}}]{Clifton:2011jh}%
  \BibitemOpen
  \bibfield  {author} {\bibinfo {author} {\bibfnamefont {Timothy}\ \bibnamefont
  {Clifton}}, \bibinfo {author} {\bibfnamefont {Pedro~G.}\ \bibnamefont
  {Ferreira}}, \bibinfo {author} {\bibfnamefont {Antonio}\ \bibnamefont
  {Padilla}}, \ and\ \bibinfo {author} {\bibfnamefont {Constantinos}\
  \bibnamefont {Skordis}},\ }\bibfield  {title} {\enquote {\bibinfo {title}
  {{Modified Gravity and Cosmology}},}\ }\href {\doibase
  10.1016/j.physrep.2012.01.001} {\bibfield  {journal} {\bibinfo  {journal}
  {Phys. Rept.}\ }\textbf {\bibinfo {volume} {513}},\ \bibinfo {pages} {1--189}
  (\bibinfo {year} {2012})},\ \Eprint {http://arxiv.org/abs/1106.2476}
  {arXiv:1106.2476 [astro-ph.CO]} \BibitemShut {NoStop}%
\bibitem [{\citenamefont {Nojiri}\ and\ \citenamefont
  {Odintsov}(2011)}]{Nojiri:2010wj}%
  \BibitemOpen
  \bibfield  {author} {\bibinfo {author} {\bibfnamefont {Shin'ichi}\
  \bibnamefont {Nojiri}}\ and\ \bibinfo {author} {\bibfnamefont {Sergei~D.}\
  \bibnamefont {Odintsov}},\ }\bibfield  {title} {\enquote {\bibinfo {title}
  {{Unified cosmic history in modified gravity: from F(R) theory to Lorentz
  non-invariant models}},}\ }\href {\doibase 10.1016/j.physrep.2011.04.001}
  {\bibfield  {journal} {\bibinfo  {journal} {Phys. Rept.}\ }\textbf {\bibinfo
  {volume} {505}},\ \bibinfo {pages} {59--144} (\bibinfo {year} {2011})},\
  \Eprint {http://arxiv.org/abs/1011.0544} {arXiv:1011.0544 [gr-qc]}
  \BibitemShut {NoStop}%
\bibitem [{\citenamefont {Boubekeur}\ \emph {et~al.}(2014)\citenamefont
  {Boubekeur}, \citenamefont {Giusarma}, \citenamefont {Mena},\ and\
  \citenamefont {Ramírez}}]{Boubekeur:2014uaa}%
  \BibitemOpen
  \bibfield  {author} {\bibinfo {author} {\bibfnamefont {Lotfi}\ \bibnamefont
  {Boubekeur}}, \bibinfo {author} {\bibfnamefont {Elena}\ \bibnamefont
  {Giusarma}}, \bibinfo {author} {\bibfnamefont {Olga}\ \bibnamefont {Mena}}, \
  and\ \bibinfo {author} {\bibfnamefont {Héctor}\ \bibnamefont {Ramírez}},\
  }\bibfield  {title} {\enquote {\bibinfo {title} {{Current status of modified
  gravity}},}\ }\href {\doibase 10.1103/PhysRevD.90.103512} {\bibfield
  {journal} {\bibinfo  {journal} {Phys. Rev.}\ }\textbf {\bibinfo {volume}
  {D90}},\ \bibinfo {pages} {103512} (\bibinfo {year} {2014})},\ \Eprint
  {http://arxiv.org/abs/1407.6837} {arXiv:1407.6837 [astro-ph.CO]} \BibitemShut
  {NoStop}%
\bibitem [{\citenamefont {Cai}\ \emph {et~al.}(2018)\citenamefont {Cai},
  \citenamefont {Li}, \citenamefont {Saridakis},\ and\ \citenamefont
  {Xue}}]{Cai:2018rzd}%
  \BibitemOpen
  \bibfield  {author} {\bibinfo {author} {\bibfnamefont {Yi-Fu}\ \bibnamefont
  {Cai}}, \bibinfo {author} {\bibfnamefont {Chunlong}\ \bibnamefont {Li}},
  \bibinfo {author} {\bibfnamefont {Emmanuel~N.}\ \bibnamefont {Saridakis}}, \
  and\ \bibinfo {author} {\bibfnamefont {Lingqin}\ \bibnamefont {Xue}},\
  }\bibfield  {title} {\enquote {\bibinfo {title} {{$f(T)$ gravity after
  GW170817 and GRB170817A}},}\ }\href {\doibase 10.1103/PhysRevD.97.103513}
  {\bibfield  {journal} {\bibinfo  {journal} {Phys. Rev.}\ }\textbf {\bibinfo
  {volume} {D97}},\ \bibinfo {pages} {103513} (\bibinfo {year} {2018})},\
  \Eprint {http://arxiv.org/abs/1801.05827} {arXiv:1801.05827 [gr-qc]}
  \BibitemShut {NoStop}%
\bibitem [{\citenamefont {Pérez-Romero}\ and\ \citenamefont
  {Nesseris}(2018)}]{Perez-Romero:2017njc}%
  \BibitemOpen
  \bibfield  {author} {\bibinfo {author} {\bibfnamefont {Judit}\ \bibnamefont
  {Pérez-Romero}}\ and\ \bibinfo {author} {\bibfnamefont {Savvas}\
  \bibnamefont {Nesseris}},\ }\bibfield  {title} {\enquote {\bibinfo {title}
  {{Cosmological constraints and comparison of viable $f(R)$ models}},}\ }\href
  {\doibase 10.1103/PhysRevD.97.023525} {\bibfield  {journal} {\bibinfo
  {journal} {Phys. Rev.}\ }\textbf {\bibinfo {volume} {D97}},\ \bibinfo {pages}
  {023525} (\bibinfo {year} {2018})},\ \Eprint
  {http://arxiv.org/abs/1710.05634} {arXiv:1710.05634 [astro-ph.CO]}
  \BibitemShut {NoStop}%
\bibitem [{\citenamefont {Nojiri}\ \emph {et~al.}(2017)\citenamefont {Nojiri},
  \citenamefont {Odintsov},\ and\ \citenamefont {Oikonomou}}]{Nojiri:2017ncd}%
  \BibitemOpen
  \bibfield  {author} {\bibinfo {author} {\bibfnamefont {S.}~\bibnamefont
  {Nojiri}}, \bibinfo {author} {\bibfnamefont {S.~D.}\ \bibnamefont
  {Odintsov}}, \ and\ \bibinfo {author} {\bibfnamefont {V.~K.}\ \bibnamefont
  {Oikonomou}},\ }\bibfield  {title} {\enquote {\bibinfo {title} {{Modified
  Gravity Theories on a Nutshell: Inflation, Bounce and Late-time
  Evolution}},}\ }\href {\doibase 10.1016/j.physrep.2017.06.001} {\bibfield
  {journal} {\bibinfo  {journal} {Phys. Rept.}\ }\textbf {\bibinfo {volume}
  {692}},\ \bibinfo {pages} {1--104} (\bibinfo {year} {2017})},\ \Eprint
  {http://arxiv.org/abs/1705.11098} {arXiv:1705.11098 [gr-qc]} \BibitemShut
  {NoStop}%
\bibitem [{\citenamefont {Song}\ \emph {et~al.}(2007)\citenamefont {Song},
  \citenamefont {Hu},\ and\ \citenamefont {Sawicki}}]{Song:2006ej}%
  \BibitemOpen
  \bibfield  {author} {\bibinfo {author} {\bibfnamefont {Yong-Seon}\
  \bibnamefont {Song}}, \bibinfo {author} {\bibfnamefont {Wayne}\ \bibnamefont
  {Hu}}, \ and\ \bibinfo {author} {\bibfnamefont {Ignacy}\ \bibnamefont
  {Sawicki}},\ }\bibfield  {title} {\enquote {\bibinfo {title} {{The Large
  Scale Structure of f(R) Gravity}},}\ }\href {\doibase
  10.1103/PhysRevD.75.044004} {\bibfield  {journal} {\bibinfo  {journal} {Phys.
  Rev.}\ }\textbf {\bibinfo {volume} {D75}},\ \bibinfo {pages} {044004}
  (\bibinfo {year} {2007})},\ \Eprint {http://arxiv.org/abs/astro-ph/0610532}
  {arXiv:astro-ph/0610532 [astro-ph]} \BibitemShut {NoStop}%
\bibitem [{\citenamefont {Brax}\ \emph {et~al.}(2008)\citenamefont {Brax},
  \citenamefont {van~de Bruck}, \citenamefont {Davis},\ and\ \citenamefont
  {Shaw}}]{Brax:2008hh}%
  \BibitemOpen
  \bibfield  {author} {\bibinfo {author} {\bibfnamefont {Philippe}\
  \bibnamefont {Brax}}, \bibinfo {author} {\bibfnamefont {Carsten}\
  \bibnamefont {van~de Bruck}}, \bibinfo {author} {\bibfnamefont
  {Anne-Christine}\ \bibnamefont {Davis}}, \ and\ \bibinfo {author}
  {\bibfnamefont {Douglas~J.}\ \bibnamefont {Shaw}},\ }\bibfield  {title}
  {\enquote {\bibinfo {title} {{f(R) Gravity and Chameleon Theories}},}\ }\href
  {\doibase 10.1103/PhysRevD.78.104021} {\bibfield  {journal} {\bibinfo
  {journal} {Phys. Rev.}\ }\textbf {\bibinfo {volume} {D78}},\ \bibinfo {pages}
  {104021} (\bibinfo {year} {2008})},\ \Eprint {http://arxiv.org/abs/0806.3415}
  {arXiv:0806.3415 [astro-ph]} \BibitemShut {NoStop}%
\bibitem [{\citenamefont {Koyama}(2016)}]{Koyama:2015vza}%
  \BibitemOpen
  \bibfield  {author} {\bibinfo {author} {\bibfnamefont {Kazuya}\ \bibnamefont
  {Koyama}},\ }\bibfield  {title} {\enquote {\bibinfo {title} {{Cosmological
  Tests of Modified Gravity}},}\ }\href {\doibase
  10.1088/0034-4885/79/4/046902} {\bibfield  {journal} {\bibinfo  {journal}
  {Rept. Prog. Phys.}\ }\textbf {\bibinfo {volume} {79}},\ \bibinfo {pages}
  {046902} (\bibinfo {year} {2016})},\ \Eprint
  {http://arxiv.org/abs/1504.04623} {arXiv:1504.04623 [astro-ph.CO]}
  \BibitemShut {NoStop}%
\bibitem [{\citenamefont {Ishak}(2019)}]{Ishak:2018his}%
  \BibitemOpen
  \bibfield  {author} {\bibinfo {author} {\bibfnamefont {Mustapha}\
  \bibnamefont {Ishak}},\ }\bibfield  {title} {\enquote {\bibinfo {title}
  {{Testing General Relativity in Cosmology}},}\ }\href {\doibase
  10.1007/s41114-018-0017-4} {\bibfield  {journal} {\bibinfo  {journal} {Living
  Rev. Rel.}\ }\textbf {\bibinfo {volume} {22}},\ \bibinfo {pages} {1}
  (\bibinfo {year} {2019})},\ \Eprint {http://arxiv.org/abs/1806.10122}
  {arXiv:1806.10122 [astro-ph.CO]} \BibitemShut {NoStop}%
\bibitem [{\citenamefont {Zhang}\ \emph {et~al.}(2007)\citenamefont {Zhang},
  \citenamefont {Liguori}, \citenamefont {Bean},\ and\ \citenamefont
  {Dodelson}}]{Zhang:2007nk}%
  \BibitemOpen
  \bibfield  {author} {\bibinfo {author} {\bibfnamefont {Pengjie}\ \bibnamefont
  {Zhang}}, \bibinfo {author} {\bibfnamefont {Michele}\ \bibnamefont
  {Liguori}}, \bibinfo {author} {\bibfnamefont {Rachel}\ \bibnamefont {Bean}},
  \ and\ \bibinfo {author} {\bibfnamefont {Scott}\ \bibnamefont {Dodelson}},\
  }\bibfield  {title} {\enquote {\bibinfo {title} {{Probing Gravity at
  Cosmological Scales by Measurements which Test the Relationship between
  Gravitational Lensing and Matter Overdensity}},}\ }\href {\doibase
  10.1103/PhysRevLett.99.141302} {\bibfield  {journal} {\bibinfo  {journal}
  {Phys. Rev. Lett.}\ }\textbf {\bibinfo {volume} {99}},\ \bibinfo {pages}
  {141302} (\bibinfo {year} {2007})},\ \Eprint {http://arxiv.org/abs/0704.1932}
  {arXiv:0704.1932 [astro-ph]} \BibitemShut {NoStop}%
\bibitem [{\citenamefont {Leonard}\ \emph {et~al.}(2015)\citenamefont
  {Leonard}, \citenamefont {Ferreira},\ and\ \citenamefont
  {Heymans}}]{Leonard:2015cba}%
  \BibitemOpen
  \bibfield  {author} {\bibinfo {author} {\bibfnamefont {C.~Danielle}\
  \bibnamefont {Leonard}}, \bibinfo {author} {\bibfnamefont {Pedro~G.}\
  \bibnamefont {Ferreira}}, \ and\ \bibinfo {author} {\bibfnamefont
  {Catherine}\ \bibnamefont {Heymans}},\ }\bibfield  {title} {\enquote
  {\bibinfo {title} {{Testing gravity with $E_G$: mapping theory onto
  observations}},}\ }\href {\doibase 10.1088/1475-7516/2015/12/051} {\bibfield
  {journal} {\bibinfo  {journal} {JCAP}\ }\textbf {\bibinfo {volume} {1512}},\
  \bibinfo {pages} {051} (\bibinfo {year} {2015})},\ \Eprint
  {http://arxiv.org/abs/1510.04287} {arXiv:1510.04287 [astro-ph.CO]}
  \BibitemShut {NoStop}%
\bibitem [{\citenamefont {de~la Torre}\ \emph {et~al.}(2017)\citenamefont
  {de~la Torre} \emph {et~al.}}]{delaTorre:2016rxm}%
  \BibitemOpen
  \bibfield  {author} {\bibinfo {author} {\bibfnamefont {S.}~\bibnamefont
  {de~la Torre}} \emph {et~al.},\ }\bibfield  {title} {\enquote {\bibinfo
  {title} {{The VIMOS Public Extragalactic Redshift Survey (VIPERS). Gravity
  test from the combination of redshift-space distortions and galaxy-galaxy
  lensing at $0.5 < z < 1.2$}},}\ }\href {\doibase 10.1051/0004-6361/201630276}
  {\bibfield  {journal} {\bibinfo  {journal} {Astron. Astrophys.}\ }\textbf
  {\bibinfo {volume} {608}},\ \bibinfo {pages} {A44} (\bibinfo {year}
  {2017})},\ \Eprint {http://arxiv.org/abs/1612.05647} {arXiv:1612.05647
  [astro-ph.CO]} \BibitemShut {NoStop}%
\bibitem [{\citenamefont {Bardeen}(1980)}]{Bardeen:1980kt}%
  \BibitemOpen
  \bibfield  {author} {\bibinfo {author} {\bibfnamefont {James~M.}\
  \bibnamefont {Bardeen}},\ }\bibfield  {title} {\enquote {\bibinfo {title}
  {{Gauge Invariant Cosmological Perturbations}},}\ }\href {\doibase
  10.1103/PhysRevD.22.1882} {\bibfield  {journal} {\bibinfo  {journal} {Phys.
  Rev.}\ }\textbf {\bibinfo {volume} {D22}},\ \bibinfo {pages} {1882--1905}
  (\bibinfo {year} {1980})}\BibitemShut {NoStop}%
\bibitem [{\citenamefont {Reyes}\ \emph {et~al.}(2010)\citenamefont {Reyes},
  \citenamefont {Mandelbaum}, \citenamefont {Seljak}, \citenamefont {Baldauf},
  \citenamefont {Gunn}, \citenamefont {Lombriser},\ and\ \citenamefont
  {Smith}}]{Reyes:2010tr}%
  \BibitemOpen
  \bibfield  {author} {\bibinfo {author} {\bibfnamefont {Reinabelle}\
  \bibnamefont {Reyes}}, \bibinfo {author} {\bibfnamefont {Rachel}\
  \bibnamefont {Mandelbaum}}, \bibinfo {author} {\bibfnamefont {Uros}\
  \bibnamefont {Seljak}}, \bibinfo {author} {\bibfnamefont {Tobias}\
  \bibnamefont {Baldauf}}, \bibinfo {author} {\bibfnamefont {James~E.}\
  \bibnamefont {Gunn}}, \bibinfo {author} {\bibfnamefont {Lucas}\ \bibnamefont
  {Lombriser}}, \ and\ \bibinfo {author} {\bibfnamefont {Robert~E.}\
  \bibnamefont {Smith}},\ }\bibfield  {title} {\enquote {\bibinfo {title}
  {{Confirmation of general relativity on large scales from weak lensing and
  galaxy velocities}},}\ }\href {\doibase 10.1038/nature08857} {\bibfield
  {journal} {\bibinfo  {journal} {Nature}\ }\textbf {\bibinfo {volume} {464}},\
  \bibinfo {pages} {256--258} (\bibinfo {year} {2010})},\ \Eprint
  {http://arxiv.org/abs/1003.2185} {arXiv:1003.2185 [astro-ph.CO]} \BibitemShut
  {NoStop}%
\bibitem [{\citenamefont {Ade}\ \emph {et~al.}(2016{\natexlab{b}})\citenamefont
  {Ade} \emph {et~al.}}]{Ade:2015zua}%
  \BibitemOpen
  \bibfield  {author} {\bibinfo {author} {\bibfnamefont {P.~A.~R.}\
  \bibnamefont {Ade}} \emph {et~al.} (\bibinfo {collaboration} {Planck}),\
  }\bibfield  {title} {\enquote {\bibinfo {title} {{Planck 2015 results. XV.
  Gravitational lensing}},}\ }\href {\doibase 10.1051/0004-6361/201525941}
  {\bibfield  {journal} {\bibinfo  {journal} {Astron. Astrophys.}\ }\textbf
  {\bibinfo {volume} {594}},\ \bibinfo {pages} {A15} (\bibinfo {year}
  {2016}{\natexlab{b}})},\ \Eprint {http://arxiv.org/abs/1502.01591}
  {arXiv:1502.01591 [astro-ph.CO]} \BibitemShut {NoStop}%
\bibitem [{\citenamefont {Pullen}\ \emph {et~al.}(2015)\citenamefont {Pullen},
  \citenamefont {Alam},\ and\ \citenamefont {Ho}}]{Pullen:2014fva}%
  \BibitemOpen
  \bibfield  {author} {\bibinfo {author} {\bibfnamefont {Anthony~R.}\
  \bibnamefont {Pullen}}, \bibinfo {author} {\bibfnamefont {Shadab}\
  \bibnamefont {Alam}}, \ and\ \bibinfo {author} {\bibfnamefont {Shirley}\
  \bibnamefont {Ho}},\ }\bibfield  {title} {\enquote {\bibinfo {title}
  {{Probing gravity at large scales through CMB lensing}},}\ }\href {\doibase
  10.1093/mnras/stv554} {\bibfield  {journal} {\bibinfo  {journal} {Mon. Not.
  Roy. Astron. Soc.}\ }\textbf {\bibinfo {volume} {449}},\ \bibinfo {pages}
  {4326--4335} (\bibinfo {year} {2015})},\ \Eprint
  {http://arxiv.org/abs/1412.4454} {arXiv:1412.4454 [astro-ph.CO]} \BibitemShut
  {NoStop}%
\bibitem [{\citenamefont {Pullen}\ \emph {et~al.}(2016)\citenamefont {Pullen},
  \citenamefont {Alam}, \citenamefont {He},\ and\ \citenamefont
  {Ho}}]{Pullen:2015vtb}%
  \BibitemOpen
  \bibfield  {author} {\bibinfo {author} {\bibfnamefont {Anthony~R.}\
  \bibnamefont {Pullen}}, \bibinfo {author} {\bibfnamefont {Shadab}\
  \bibnamefont {Alam}}, \bibinfo {author} {\bibfnamefont {Siyu}\ \bibnamefont
  {He}}, \ and\ \bibinfo {author} {\bibfnamefont {Shirley}\ \bibnamefont
  {Ho}},\ }\bibfield  {title} {\enquote {\bibinfo {title} {{Constraining
  Gravity at the Largest Scales through CMB Lensing and Galaxy Velocities}},}\
  }\href {\doibase 10.1093/mnras/stw1249} {\bibfield  {journal} {\bibinfo
  {journal} {Mon. Not. Roy. Astron. Soc.}\ }\textbf {\bibinfo {volume} {460}},\
  \bibinfo {pages} {4098--4108} (\bibinfo {year} {2016})},\ \Eprint
  {http://arxiv.org/abs/1511.04457} {arXiv:1511.04457 [astro-ph.CO]}
  \BibitemShut {NoStop}%
\bibitem [{\citenamefont {Bartelmann}\ and\ \citenamefont
  {Schneider}(2001)}]{Bartelmann:1999yn}%
  \BibitemOpen
  \bibfield  {author} {\bibinfo {author} {\bibfnamefont {Matthias}\
  \bibnamefont {Bartelmann}}\ and\ \bibinfo {author} {\bibfnamefont {Peter}\
  \bibnamefont {Schneider}},\ }\bibfield  {title} {\enquote {\bibinfo {title}
  {{Weak gravitational lensing}},}\ }\href {\doibase
  10.1016/S0370-1573(00)00082-X} {\bibfield  {journal} {\bibinfo  {journal}
  {Phys. Rept.}\ }\textbf {\bibinfo {volume} {340}},\ \bibinfo {pages}
  {291--472} (\bibinfo {year} {2001})},\ \Eprint
  {http://arxiv.org/abs/astro-ph/9912508} {arXiv:astro-ph/9912508 [astro-ph]}
  \BibitemShut {NoStop}%
\bibitem [{\citenamefont {Hoekstra}\ \emph {et~al.}(2004)\citenamefont
  {Hoekstra}, \citenamefont {Yee},\ and\ \citenamefont
  {Gladders}}]{Hoekstra:2003pn}%
  \BibitemOpen
  \bibfield  {author} {\bibinfo {author} {\bibfnamefont {Henk}\ \bibnamefont
  {Hoekstra}}, \bibinfo {author} {\bibfnamefont {Howard K.~C.}\ \bibnamefont
  {Yee}}, \ and\ \bibinfo {author} {\bibfnamefont {Michael~D.}\ \bibnamefont
  {Gladders}},\ }\bibfield  {title} {\enquote {\bibinfo {title} {{Properties of
  galaxy dark matter halos from weak lensing}},}\ }\href {\doibase
  10.1086/382726} {\bibfield  {journal} {\bibinfo  {journal} {Astrophys. J.}\
  }\textbf {\bibinfo {volume} {606}},\ \bibinfo {pages} {67--77} (\bibinfo
  {year} {2004})},\ \Eprint {http://arxiv.org/abs/astro-ph/0306515}
  {arXiv:astro-ph/0306515 [astro-ph]} \BibitemShut {NoStop}%
\bibitem [{\citenamefont {Mandelbaum}\ \emph {et~al.}(2005)\citenamefont
  {Mandelbaum}, \citenamefont {Hirata}, \citenamefont {Seljak}, \citenamefont
  {Guzik}, \citenamefont {Padmanabhan}, \citenamefont {Blake}, \citenamefont
  {Blanton}, \citenamefont {Lupton},\ and\ \citenamefont
  {Brinkmann}}]{Mandelbaum:2005wv}%
  \BibitemOpen
  \bibfield  {author} {\bibinfo {author} {\bibfnamefont {Rachel}\ \bibnamefont
  {Mandelbaum}}, \bibinfo {author} {\bibfnamefont {Christopher~M.}\
  \bibnamefont {Hirata}}, \bibinfo {author} {\bibfnamefont {Uros}\ \bibnamefont
  {Seljak}}, \bibinfo {author} {\bibfnamefont {Jacek}\ \bibnamefont {Guzik}},
  \bibinfo {author} {\bibfnamefont {Nikhil}\ \bibnamefont {Padmanabhan}},
  \bibinfo {author} {\bibfnamefont {Cullen}\ \bibnamefont {Blake}}, \bibinfo
  {author} {\bibfnamefont {Michael~R.}\ \bibnamefont {Blanton}}, \bibinfo
  {author} {\bibfnamefont {Robert}\ \bibnamefont {Lupton}}, \ and\ \bibinfo
  {author} {\bibfnamefont {Jonathan}\ \bibnamefont {Brinkmann}},\ }\bibfield
  {title} {\enquote {\bibinfo {title} {{Systematic errors in weak lensing:
  Application to SDSS galaxy-galaxy weak lensing}},}\ }\href {\doibase
  10.1111/j.1365-2966.2005.09282.x} {\bibfield  {journal} {\bibinfo  {journal}
  {Mon. Not. Roy. Astron. Soc.}\ }\textbf {\bibinfo {volume} {361}},\ \bibinfo
  {pages} {1287--1322} (\bibinfo {year} {2005})},\ \Eprint
  {http://arxiv.org/abs/astro-ph/0501201} {arXiv:astro-ph/0501201 [astro-ph]}
  \BibitemShut {NoStop}%
\bibitem [{\citenamefont {Kilbinger}(2015)}]{Kilbinger:2014cea}%
  \BibitemOpen
  \bibfield  {author} {\bibinfo {author} {\bibfnamefont {Martin}\ \bibnamefont
  {Kilbinger}},\ }\bibfield  {title} {\enquote {\bibinfo {title} {{Cosmology
  with cosmic shear observations: a review}},}\ }\href {\doibase
  10.1088/0034-4885/78/8/086901} {\bibfield  {journal} {\bibinfo  {journal}
  {Rept. Prog. Phys.}\ }\textbf {\bibinfo {volume} {78}},\ \bibinfo {pages}
  {086901} (\bibinfo {year} {2015})},\ \Eprint {http://arxiv.org/abs/1411.0115}
  {arXiv:1411.0115 [astro-ph.CO]} \BibitemShut {NoStop}%
\bibitem [{\citenamefont {Mukhanov}\ \emph {et~al.}(1992)\citenamefont
  {Mukhanov}, \citenamefont {Feldman},\ and\ \citenamefont
  {Brandenberger}}]{Mukhanov:1990me}%
  \BibitemOpen
  \bibfield  {author} {\bibinfo {author} {\bibfnamefont {Viatcheslav~F.}\
  \bibnamefont {Mukhanov}}, \bibinfo {author} {\bibfnamefont {H.~A.}\
  \bibnamefont {Feldman}}, \ and\ \bibinfo {author} {\bibfnamefont {Robert~H.}\
  \bibnamefont {Brandenberger}},\ }\bibfield  {title} {\enquote {\bibinfo
  {title} {{Theory of cosmological perturbations. Part 1. Classical
  perturbations. Part 2. Quantum theory of perturbations. Part 3.
  Extensions}},}\ }\href {\doibase 10.1016/0370-1573(92)90044-Z} {\bibfield
  {journal} {\bibinfo  {journal} {Phys. Rept.}\ }\textbf {\bibinfo {volume}
  {215}},\ \bibinfo {pages} {203--333} (\bibinfo {year} {1992})}\BibitemShut
  {NoStop}%
\bibitem [{\citenamefont {Ma}\ and\ \citenamefont
  {Bertschinger}(1995)}]{Ma:1995ey}%
  \BibitemOpen
  \bibfield  {author} {\bibinfo {author} {\bibfnamefont {Chung-Pei}\
  \bibnamefont {Ma}}\ and\ \bibinfo {author} {\bibfnamefont {Edmund}\
  \bibnamefont {Bertschinger}},\ }\bibfield  {title} {\enquote {\bibinfo
  {title} {{Cosmological perturbation theory in the synchronous and conformal
  Newtonian gauges}},}\ }\href {\doibase 10.1086/176550} {\bibfield  {journal}
  {\bibinfo  {journal} {Astrophys. J.}\ }\textbf {\bibinfo {volume} {455}},\
  \bibinfo {pages} {7--25} (\bibinfo {year} {1995})},\ \Eprint
  {http://arxiv.org/abs/astro-ph/9506072} {arXiv:astro-ph/9506072 [astro-ph]}
  \BibitemShut {NoStop}%
\bibitem [{\citenamefont {Esposito-Farese}\ and\ \citenamefont
  {Polarski}(2001)}]{EspositoFarese:2000ij}%
  \BibitemOpen
  \bibfield  {author} {\bibinfo {author} {\bibfnamefont {Gilles}\ \bibnamefont
  {Esposito-Farese}}\ and\ \bibinfo {author} {\bibfnamefont {D.}~\bibnamefont
  {Polarski}},\ }\bibfield  {title} {\enquote {\bibinfo {title} {{Scalar tensor
  gravity in an accelerating universe}},}\ }\href {\doibase
  10.1103/PhysRevD.63.063504} {\bibfield  {journal} {\bibinfo  {journal} {Phys.
  Rev.}\ }\textbf {\bibinfo {volume} {D63}},\ \bibinfo {pages} {063504}
  (\bibinfo {year} {2001})},\ \Eprint {http://arxiv.org/abs/gr-qc/0009034}
  {arXiv:gr-qc/0009034 [gr-qc]} \BibitemShut {NoStop}%
\bibitem [{\citenamefont {Kaiser}(1987)}]{Kaiser:1987qv}%
  \BibitemOpen
  \bibfield  {author} {\bibinfo {author} {\bibfnamefont {N.}~\bibnamefont
  {Kaiser}},\ }\bibfield  {title} {\enquote {\bibinfo {title} {{Clustering in
  real space and in redshift space}},}\ }\href@noop {} {\bibfield  {journal}
  {\bibinfo  {journal} {Mon. Not. Roy. Astron. Soc.}\ }\textbf {\bibinfo
  {volume} {227}},\ \bibinfo {pages} {1--27} (\bibinfo {year}
  {1987})}\BibitemShut {NoStop}%
\bibitem [{\citenamefont {Hamilton}(1997)}]{Hamilton:1997zq}%
  \BibitemOpen
  \bibfield  {author} {\bibinfo {author} {\bibfnamefont {A.~J.~S.}\
  \bibnamefont {Hamilton}},\ }\bibfield  {title} {\enquote {\bibinfo {title}
  {{Linear redshift distortions: A Review}},}\ }in\ \href {\doibase
  10.1007/978-94-011-4960-0_17} {\emph {\bibinfo {booktitle} {{Ringberg
  Workshop on Large Scale Structure Ringberg, Germany, September 23-28,
  1996}}}}\ (\bibinfo {year} {1997})\ \Eprint
  {http://arxiv.org/abs/astro-ph/9708102} {arXiv:astro-ph/9708102 [astro-ph]}
  \BibitemShut {NoStop}%
\bibitem [{\citenamefont {Samushia}\ \emph {et~al.}(2014)\citenamefont
  {Samushia} \emph {et~al.}}]{boss}%
  \BibitemOpen
  \bibfield  {author} {\bibinfo {author} {\bibfnamefont {Lado}\ \bibnamefont
  {Samushia}} \emph {et~al.},\ }\bibfield  {title} {\enquote {\bibinfo {title}
  {{The clustering of galaxies in the SDSS-III Baryon Oscillation Spectroscopic
  Survey: measuring growth rate and geometry with anisotropic clustering}},}\
  }\href {\doibase 10.1093/mnras/stu197} {\bibfield  {journal} {\bibinfo
  {journal} {Mon. Not. Roy. Astron. Soc.}\ }\textbf {\bibinfo {volume} {439}},\
  \bibinfo {pages} {3504--3519} (\bibinfo {year} {2014})},\ \Eprint
  {http://arxiv.org/abs/1312.4899} {arXiv:1312.4899 [astro-ph.CO]} \BibitemShut
  {NoStop}%
\bibitem [{\citenamefont {Beutler}\ \emph {et~al.}(2012)\citenamefont
  {Beutler}, \citenamefont {Blake}, \citenamefont {Colless}, \citenamefont
  {Jones}, \citenamefont {Staveley-Smith}, \citenamefont {Poole}, \citenamefont
  {Campbell}, \citenamefont {Parker}, \citenamefont {Saunders},\ and\
  \citenamefont {Watson}}]{Beutler:2012px}%
  \BibitemOpen
  \bibfield  {author} {\bibinfo {author} {\bibfnamefont {Florian}\ \bibnamefont
  {Beutler}}, \bibinfo {author} {\bibfnamefont {Chris}\ \bibnamefont {Blake}},
  \bibinfo {author} {\bibfnamefont {Matthew}\ \bibnamefont {Colless}}, \bibinfo
  {author} {\bibfnamefont {D.~Heath}\ \bibnamefont {Jones}}, \bibinfo {author}
  {\bibfnamefont {Lister}\ \bibnamefont {Staveley-Smith}}, \bibinfo {author}
  {\bibfnamefont {Gregory~B.}\ \bibnamefont {Poole}}, \bibinfo {author}
  {\bibfnamefont {Lachlan}\ \bibnamefont {Campbell}}, \bibinfo {author}
  {\bibfnamefont {Quentin}\ \bibnamefont {Parker}}, \bibinfo {author}
  {\bibfnamefont {Will}\ \bibnamefont {Saunders}}, \ and\ \bibinfo {author}
  {\bibfnamefont {Fred}\ \bibnamefont {Watson}},\ }\bibfield  {title} {\enquote
  {\bibinfo {title} {{The 6dF Galaxy Survey: $z \approx 0$ measurement of the
  growth rate and $\sigma_8$}},}\ }\href {\doibase
  10.1111/j.1365-2966.2012.21136.x} {\bibfield  {journal} {\bibinfo  {journal}
  {Mon. Not. Roy. Astron. Soc.}\ }\textbf {\bibinfo {volume} {423}},\ \bibinfo
  {pages} {3430--3444} (\bibinfo {year} {2012})},\ \Eprint
  {http://arxiv.org/abs/1204.4725} {arXiv:1204.4725 [astro-ph.CO]} \BibitemShut
  {NoStop}%
\bibitem [{\citenamefont {Bertschinger}(2011)}]{Bertschinger:2011kk}%
  \BibitemOpen
  \bibfield  {author} {\bibinfo {author} {\bibfnamefont {Edmund}\ \bibnamefont
  {Bertschinger}},\ }\bibfield  {title} {\enquote {\bibinfo {title} {{One
  Gravitational Potential or Two? Forecasts and Tests}},}\ }\href {\doibase
  10.1098/rsta.2011.0369} {\bibfield  {journal} {\bibinfo  {journal} {Phil.
  Trans. Roy. Soc. Lond.}\ }\textbf {\bibinfo {volume} {A369}},\ \bibinfo
  {pages} {4947--4961} (\bibinfo {year} {2011})},\ \Eprint
  {http://arxiv.org/abs/1111.4659} {arXiv:1111.4659 [astro-ph.CO]} \BibitemShut
  {NoStop}%
\bibitem [{\citenamefont {Bertschinger}\ and\ \citenamefont
  {Zukin}(2008)}]{Bertschinger:2008zb}%
  \BibitemOpen
  \bibfield  {author} {\bibinfo {author} {\bibfnamefont {Edmund}\ \bibnamefont
  {Bertschinger}}\ and\ \bibinfo {author} {\bibfnamefont {Phillip}\
  \bibnamefont {Zukin}},\ }\bibfield  {title} {\enquote {\bibinfo {title}
  {{Distinguishing Modified Gravity from Dark Energy}},}\ }\href {\doibase
  10.1103/PhysRevD.78.024015} {\bibfield  {journal} {\bibinfo  {journal} {Phys.
  Rev.}\ }\textbf {\bibinfo {volume} {D78}},\ \bibinfo {pages} {024015}
  (\bibinfo {year} {2008})},\ \Eprint {http://arxiv.org/abs/0801.2431}
  {arXiv:0801.2431 [astro-ph]} \BibitemShut {NoStop}%
\bibitem [{\citenamefont {Zhao}\ \emph {et~al.}(2010)\citenamefont {Zhao},
  \citenamefont {Giannantonio}, \citenamefont {Pogosian}, \citenamefont
  {Silvestri}, \citenamefont {Bacon}, \citenamefont {Koyama}, \citenamefont
  {Nichol},\ and\ \citenamefont {Song}}]{Zhao:2010dz}%
  \BibitemOpen
  \bibfield  {author} {\bibinfo {author} {\bibfnamefont {Gong-Bo}\ \bibnamefont
  {Zhao}}, \bibinfo {author} {\bibfnamefont {Tommaso}\ \bibnamefont
  {Giannantonio}}, \bibinfo {author} {\bibfnamefont {Levon}\ \bibnamefont
  {Pogosian}}, \bibinfo {author} {\bibfnamefont {Alessandra}\ \bibnamefont
  {Silvestri}}, \bibinfo {author} {\bibfnamefont {David~J.}\ \bibnamefont
  {Bacon}}, \bibinfo {author} {\bibfnamefont {Kazuya}\ \bibnamefont {Koyama}},
  \bibinfo {author} {\bibfnamefont {Robert~C.}\ \bibnamefont {Nichol}}, \ and\
  \bibinfo {author} {\bibfnamefont {Yong-Seon}\ \bibnamefont {Song}},\
  }\bibfield  {title} {\enquote {\bibinfo {title} {{Probing modifications of
  General Relativity using current cosmological observations}},}\ }\href
  {\doibase 10.1103/PhysRevD.81.103510} {\bibfield  {journal} {\bibinfo
  {journal} {Phys. Rev.}\ }\textbf {\bibinfo {volume} {D81}},\ \bibinfo {pages}
  {103510} (\bibinfo {year} {2010})},\ \Eprint {http://arxiv.org/abs/1003.0001}
  {arXiv:1003.0001 [astro-ph.CO]} \BibitemShut {NoStop}%
\bibitem [{\citenamefont {Daniel}\ \emph {et~al.}(2010)\citenamefont {Daniel},
  \citenamefont {Linder}, \citenamefont {Smith}, \citenamefont {Caldwell},
  \citenamefont {Cooray}, \citenamefont {Leauthaud},\ and\ \citenamefont
  {Lombriser}}]{Daniel:2010ky}%
  \BibitemOpen
  \bibfield  {author} {\bibinfo {author} {\bibfnamefont {Scott~F.}\
  \bibnamefont {Daniel}}, \bibinfo {author} {\bibfnamefont {Eric~V.}\
  \bibnamefont {Linder}}, \bibinfo {author} {\bibfnamefont {Tristan~L.}\
  \bibnamefont {Smith}}, \bibinfo {author} {\bibfnamefont {Robert~R.}\
  \bibnamefont {Caldwell}}, \bibinfo {author} {\bibfnamefont {Asantha}\
  \bibnamefont {Cooray}}, \bibinfo {author} {\bibfnamefont {Alexie}\
  \bibnamefont {Leauthaud}}, \ and\ \bibinfo {author} {\bibfnamefont {Lucas}\
  \bibnamefont {Lombriser}},\ }\bibfield  {title} {\enquote {\bibinfo {title}
  {{Testing General Relativity with Current Cosmological Data}},}\ }\href
  {\doibase 10.1103/PhysRevD.81.123508} {\bibfield  {journal} {\bibinfo
  {journal} {Phys. Rev.}\ }\textbf {\bibinfo {volume} {D81}},\ \bibinfo {pages}
  {123508} (\bibinfo {year} {2010})},\ \Eprint {http://arxiv.org/abs/1002.1962}
  {arXiv:1002.1962 [astro-ph.CO]} \BibitemShut {NoStop}%
\bibitem [{\citenamefont {Song}\ \emph {et~al.}(2011)\citenamefont {Song},
  \citenamefont {Zhao}, \citenamefont {Bacon}, \citenamefont {Koyama},
  \citenamefont {Nichol},\ and\ \citenamefont {Pogosian}}]{Song:2010fg}%
  \BibitemOpen
  \bibfield  {author} {\bibinfo {author} {\bibfnamefont {Yong-Seon}\
  \bibnamefont {Song}}, \bibinfo {author} {\bibfnamefont {Gong-Bo}\
  \bibnamefont {Zhao}}, \bibinfo {author} {\bibfnamefont {David}\ \bibnamefont
  {Bacon}}, \bibinfo {author} {\bibfnamefont {Kazuya}\ \bibnamefont {Koyama}},
  \bibinfo {author} {\bibfnamefont {Robert~C.}\ \bibnamefont {Nichol}}, \ and\
  \bibinfo {author} {\bibfnamefont {Levon}\ \bibnamefont {Pogosian}},\
  }\bibfield  {title} {\enquote {\bibinfo {title} {{Complementarity of Weak
  Lensing and Peculiar Velocity Measurements in Testing General Relativity}},}\
  }\href {\doibase 10.1103/PhysRevD.84.083523} {\bibfield  {journal} {\bibinfo
  {journal} {Phys. Rev.}\ }\textbf {\bibinfo {volume} {D84}},\ \bibinfo {pages}
  {083523} (\bibinfo {year} {2011})},\ \Eprint {http://arxiv.org/abs/1011.2106}
  {arXiv:1011.2106 [astro-ph.CO]} \BibitemShut {NoStop}%
\bibitem [{\citenamefont {Daniel}\ and\ \citenamefont
  {Linder}(2010)}]{Daniel:2010yt}%
  \BibitemOpen
  \bibfield  {author} {\bibinfo {author} {\bibfnamefont {Scott~F.}\
  \bibnamefont {Daniel}}\ and\ \bibinfo {author} {\bibfnamefont {Eric~V.}\
  \bibnamefont {Linder}},\ }\bibfield  {title} {\enquote {\bibinfo {title}
  {{Confronting General Relativity with Further Cosmological Data}},}\ }\href
  {\doibase 10.1103/PhysRevD.82.103523} {\bibfield  {journal} {\bibinfo
  {journal} {Phys. Rev.}\ }\textbf {\bibinfo {volume} {D82}},\ \bibinfo {pages}
  {103523} (\bibinfo {year} {2010})},\ \Eprint {http://arxiv.org/abs/1008.0397}
  {arXiv:1008.0397 [astro-ph.CO]} \BibitemShut {NoStop}%
\bibitem [{\citenamefont {Hu}\ \emph {et~al.}(2013)\citenamefont {Hu},
  \citenamefont {Liguori}, \citenamefont {Bartolo},\ and\ \citenamefont
  {Matarrese}}]{Hu:2013aqa}%
  \BibitemOpen
  \bibfield  {author} {\bibinfo {author} {\bibfnamefont {Bin}\ \bibnamefont
  {Hu}}, \bibinfo {author} {\bibfnamefont {Michele}\ \bibnamefont {Liguori}},
  \bibinfo {author} {\bibfnamefont {Nicola}\ \bibnamefont {Bartolo}}, \ and\
  \bibinfo {author} {\bibfnamefont {Sabino}\ \bibnamefont {Matarrese}},\
  }\bibfield  {title} {\enquote {\bibinfo {title} {{Parametrized modified
  gravity constraints after Planck}},}\ }\href {\doibase
  10.1103/PhysRevD.88.123514} {\bibfield  {journal} {\bibinfo  {journal} {Phys.
  Rev.}\ }\textbf {\bibinfo {volume} {D88}},\ \bibinfo {pages} {123514}
  (\bibinfo {year} {2013})},\ \Eprint {http://arxiv.org/abs/1307.5276}
  {arXiv:1307.5276 [astro-ph.CO]} \BibitemShut {NoStop}%
\bibitem [{\citenamefont {Ferté}\ \emph {et~al.}(2019)\citenamefont {Ferté},
  \citenamefont {Kirk}, \citenamefont {Liddle},\ and\ \citenamefont
  {Zuntz}}]{Ferte:2017bpf}%
  \BibitemOpen
  \bibfield  {author} {\bibinfo {author} {\bibfnamefont {Agnès}\ \bibnamefont
  {Ferté}}, \bibinfo {author} {\bibfnamefont {Donnacha}\ \bibnamefont {Kirk}},
  \bibinfo {author} {\bibfnamefont {Andrew~R.}\ \bibnamefont {Liddle}}, \ and\
  \bibinfo {author} {\bibfnamefont {Joe}\ \bibnamefont {Zuntz}},\ }\bibfield
  {title} {\enquote {\bibinfo {title} {{Testing gravity on cosmological scales
  with cosmic shear, cosmic microwave background anisotropies, and
  redshift-space distortions}},}\ }\href {\doibase 10.1103/PhysRevD.99.083512}
  {\bibfield  {journal} {\bibinfo  {journal} {Phys. Rev.}\ }\textbf {\bibinfo
  {volume} {D99}},\ \bibinfo {pages} {083512} (\bibinfo {year} {2019})},\
  \Eprint {http://arxiv.org/abs/1712.01846} {arXiv:1712.01846 [astro-ph.CO]}
  \BibitemShut {NoStop}%
\bibitem [{\citenamefont {Bernardeau}\ \emph {et~al.}(2002)\citenamefont
  {Bernardeau}, \citenamefont {Colombi}, \citenamefont {Gaztanaga},\ and\
  \citenamefont {Scoccimarro}}]{Bernardeau:2001qr}%
  \BibitemOpen
  \bibfield  {author} {\bibinfo {author} {\bibfnamefont {F.}~\bibnamefont
  {Bernardeau}}, \bibinfo {author} {\bibfnamefont {S.}~\bibnamefont {Colombi}},
  \bibinfo {author} {\bibfnamefont {E.}~\bibnamefont {Gaztanaga}}, \ and\
  \bibinfo {author} {\bibfnamefont {R.}~\bibnamefont {Scoccimarro}},\
  }\bibfield  {title} {\enquote {\bibinfo {title} {{Large scale structure of
  the universe and cosmological perturbation theory}},}\ }\href {\doibase
  10.1016/S0370-1573(02)00135-7} {\bibfield  {journal} {\bibinfo  {journal}
  {Phys. Rept.}\ }\textbf {\bibinfo {volume} {367}},\ \bibinfo {pages} {1--248}
  (\bibinfo {year} {2002})},\ \Eprint {http://arxiv.org/abs/astro-ph/0112551}
  {arXiv:astro-ph/0112551 [astro-ph]} \BibitemShut {NoStop}%
\bibitem [{\citenamefont {{Peebles}}(1980)}]{1980lssu.book.....P}%
  \BibitemOpen
  \bibfield  {author} {\bibinfo {author} {\bibfnamefont {P.~J.~E.}\
  \bibnamefont {{Peebles}}},\ }\href@noop {} {\emph {\bibinfo {title} {Research
  supported by the National Science Foundation.~Princeton, N.J., Princeton
  University Press, 1980.~435 p.}}}\ (\bibinfo {year} {1980})\BibitemShut
  {NoStop}%
\bibitem [{\citenamefont {Lahav}\ \emph {et~al.}(1991)\citenamefont {Lahav},
  \citenamefont {Lilje}, \citenamefont {Primack},\ and\ \citenamefont
  {Rees}}]{Lahav:1991wc}%
  \BibitemOpen
  \bibfield  {author} {\bibinfo {author} {\bibfnamefont {Ofer}\ \bibnamefont
  {Lahav}}, \bibinfo {author} {\bibfnamefont {Per~B.}\ \bibnamefont {Lilje}},
  \bibinfo {author} {\bibfnamefont {Joel~R.}\ \bibnamefont {Primack}}, \ and\
  \bibinfo {author} {\bibfnamefont {Martin~J.}\ \bibnamefont {Rees}},\
  }\bibfield  {title} {\enquote {\bibinfo {title} {{Dynamical effects of the
  cosmological constant}},}\ }\href@noop {} {\bibfield  {journal} {\bibinfo
  {journal} {Mon. Not. Roy. Astron. Soc.}\ }\textbf {\bibinfo {volume} {251}},\
  \bibinfo {pages} {128--136} (\bibinfo {year} {1991})}\BibitemShut {NoStop}%
\bibitem [{\citenamefont {Wang}\ and\ \citenamefont
  {Steinhardt}(1998)}]{Wang:1998gt}%
  \BibitemOpen
  \bibfield  {author} {\bibinfo {author} {\bibfnamefont {Li-Min}\ \bibnamefont
  {Wang}}\ and\ \bibinfo {author} {\bibfnamefont {Paul~J.}\ \bibnamefont
  {Steinhardt}},\ }\bibfield  {title} {\enquote {\bibinfo {title} {{Cluster
  abundance constraints on quintessence models}},}\ }\href {\doibase
  10.1086/306436} {\bibfield  {journal} {\bibinfo  {journal} {Astrophys. J.}\
  }\textbf {\bibinfo {volume} {508}},\ \bibinfo {pages} {483--490} (\bibinfo
  {year} {1998})},\ \Eprint {http://arxiv.org/abs/astro-ph/9804015}
  {arXiv:astro-ph/9804015 [astro-ph]} \BibitemShut {NoStop}%
\bibitem [{\citenamefont {Linder}(2005)}]{Linder:2005in}%
  \BibitemOpen
  \bibfield  {author} {\bibinfo {author} {\bibfnamefont {Eric~V.}\ \bibnamefont
  {Linder}},\ }\bibfield  {title} {\enquote {\bibinfo {title} {{Cosmic growth
  history and expansion history}},}\ }\href {\doibase
  10.1103/PhysRevD.72.043529} {\bibfield  {journal} {\bibinfo  {journal} {Phys.
  Rev.}\ }\textbf {\bibinfo {volume} {D72}},\ \bibinfo {pages} {043529}
  (\bibinfo {year} {2005})},\ \Eprint {http://arxiv.org/abs/astro-ph/0507263}
  {arXiv:astro-ph/0507263 [astro-ph]} \BibitemShut {NoStop}%
\bibitem [{\citenamefont {Polarski}\ and\ \citenamefont
  {Gannouji}(2008)}]{Polarski:2007rr}%
  \BibitemOpen
  \bibfield  {author} {\bibinfo {author} {\bibfnamefont {David}\ \bibnamefont
  {Polarski}}\ and\ \bibinfo {author} {\bibfnamefont {Radouane}\ \bibnamefont
  {Gannouji}},\ }\bibfield  {title} {\enquote {\bibinfo {title} {{On the growth
  of linear perturbations}},}\ }\href {\doibase 10.1016/j.physletb.2008.01.032}
  {\bibfield  {journal} {\bibinfo  {journal} {Phys. Lett.}\ }\textbf {\bibinfo
  {volume} {B660}},\ \bibinfo {pages} {439--443} (\bibinfo {year} {2008})},\
  \Eprint {http://arxiv.org/abs/0710.1510} {arXiv:0710.1510 [astro-ph]}
  \BibitemShut {NoStop}%
\bibitem [{\citenamefont {Linder}\ and\ \citenamefont
  {Cahn}(2007)}]{Linder:2007hg}%
  \BibitemOpen
  \bibfield  {author} {\bibinfo {author} {\bibfnamefont {Eric~V.}\ \bibnamefont
  {Linder}}\ and\ \bibinfo {author} {\bibfnamefont {Robert~N.}\ \bibnamefont
  {Cahn}},\ }\bibfield  {title} {\enquote {\bibinfo {title} {{Parameterized
  Beyond-Einstein Growth}},}\ }\href {\doibase
  10.1016/j.astropartphys.2007.09.003} {\bibfield  {journal} {\bibinfo
  {journal} {Astropart. Phys.}\ }\textbf {\bibinfo {volume} {28}},\ \bibinfo
  {pages} {481--488} (\bibinfo {year} {2007})},\ \Eprint
  {http://arxiv.org/abs/astro-ph/0701317} {arXiv:astro-ph/0701317 [astro-ph]}
  \BibitemShut {NoStop}%
\bibitem [{\citenamefont {Gannouji}\ and\ \citenamefont
  {Polarski}(2008)}]{Gannouji:2008jr}%
  \BibitemOpen
  \bibfield  {author} {\bibinfo {author} {\bibfnamefont {Radouane}\
  \bibnamefont {Gannouji}}\ and\ \bibinfo {author} {\bibfnamefont {David}\
  \bibnamefont {Polarski}},\ }\bibfield  {title} {\enquote {\bibinfo {title}
  {{The growth of matter perturbations in some scalar-tensor DE models}},}\
  }\href {\doibase 10.1088/1475-7516/2008/05/018} {\bibfield  {journal}
  {\bibinfo  {journal} {JCAP}\ }\textbf {\bibinfo {volume} {0805}},\ \bibinfo
  {pages} {018} (\bibinfo {year} {2008})},\ \Eprint
  {http://arxiv.org/abs/0802.4196} {arXiv:0802.4196 [astro-ph]} \BibitemShut
  {NoStop}%
\bibitem [{\citenamefont {Nesseris}\ and\ \citenamefont
  {Sapone}(2015)}]{Nesseris:2015fqa}%
  \BibitemOpen
  \bibfield  {author} {\bibinfo {author} {\bibfnamefont {Savvas}\ \bibnamefont
  {Nesseris}}\ and\ \bibinfo {author} {\bibfnamefont {Domenico}\ \bibnamefont
  {Sapone}},\ }\bibfield  {title} {\enquote {\bibinfo {title} {{Accuracy of the
  growth index in the presence of dark energy perturbations}},}\ }\href
  {\doibase 10.1103/PhysRevD.92.023013} {\bibfield  {journal} {\bibinfo
  {journal} {Phys. Rev.}\ }\textbf {\bibinfo {volume} {D92}},\ \bibinfo {pages}
  {023013} (\bibinfo {year} {2015})},\ \Eprint
  {http://arxiv.org/abs/1505.06601} {arXiv:1505.06601 [astro-ph.CO]}
  \BibitemShut {NoStop}%
\bibitem [{\citenamefont {Polarski}\ \emph {et~al.}(2016)\citenamefont
  {Polarski}, \citenamefont {Starobinsky},\ and\ \citenamefont
  {Giacomini}}]{Polarski:2016ieb}%
  \BibitemOpen
  \bibfield  {author} {\bibinfo {author} {\bibfnamefont {David}\ \bibnamefont
  {Polarski}}, \bibinfo {author} {\bibfnamefont {Alexei~A.}\ \bibnamefont
  {Starobinsky}}, \ and\ \bibinfo {author} {\bibfnamefont {Hector}\
  \bibnamefont {Giacomini}},\ }\bibfield  {title} {\enquote {\bibinfo {title}
  {{When is the growth index constant?}}}\ }\href {\doibase
  10.1088/1475-7516/2016/12/037} {\bibfield  {journal} {\bibinfo  {journal}
  {JCAP}\ }\textbf {\bibinfo {volume} {1612}},\ \bibinfo {pages} {037}
  (\bibinfo {year} {2016})},\ \Eprint {http://arxiv.org/abs/1610.00363}
  {arXiv:1610.00363 [astro-ph.CO]} \BibitemShut {NoStop}%
\bibitem [{\citenamefont {Pogosian}\ and\ \citenamefont
  {Silvestri}(2008)}]{Pogosian:2007sw}%
  \BibitemOpen
  \bibfield  {author} {\bibinfo {author} {\bibfnamefont {Levon}\ \bibnamefont
  {Pogosian}}\ and\ \bibinfo {author} {\bibfnamefont {Alessandra}\ \bibnamefont
  {Silvestri}},\ }\bibfield  {title} {\enquote {\bibinfo {title} {{The pattern
  of growth in viable f(R) cosmologies}},}\ }\href {\doibase
  10.1103/PhysRevD.77.023503, 10.1103/PhysRevD.81.049901} {\bibfield  {journal}
  {\bibinfo  {journal} {Phys. Rev.}\ }\textbf {\bibinfo {volume} {D77}},\
  \bibinfo {pages} {023503} (\bibinfo {year} {2008})},\ \bibinfo {note}
  {[Erratum: Phys. Rev.D81,049901(2010)]},\ \Eprint
  {http://arxiv.org/abs/0709.0296} {arXiv:0709.0296 [astro-ph]} \BibitemShut
  {NoStop}%
\bibitem [{\citenamefont {Jain}\ and\ \citenamefont
  {Khoury}(2010)}]{Jain:2010ka}%
  \BibitemOpen
  \bibfield  {author} {\bibinfo {author} {\bibfnamefont {Bhuvnesh}\
  \bibnamefont {Jain}}\ and\ \bibinfo {author} {\bibfnamefont {Justin}\
  \bibnamefont {Khoury}},\ }\bibfield  {title} {\enquote {\bibinfo {title}
  {{Cosmological Tests of Gravity}},}\ }\href {\doibase
  10.1016/j.aop.2010.04.002} {\bibfield  {journal} {\bibinfo  {journal} {Annals
  Phys.}\ }\textbf {\bibinfo {volume} {325}},\ \bibinfo {pages} {1479--1516}
  (\bibinfo {year} {2010})},\ \Eprint {http://arxiv.org/abs/1004.3294}
  {arXiv:1004.3294 [astro-ph.CO]} \BibitemShut {NoStop}%
\bibitem [{\citenamefont {Amendola}\ \emph {et~al.}(2008)\citenamefont
  {Amendola}, \citenamefont {Kunz},\ and\ \citenamefont
  {Sapone}}]{Amendola:2007rr}%
  \BibitemOpen
  \bibfield  {author} {\bibinfo {author} {\bibfnamefont {Luca}\ \bibnamefont
  {Amendola}}, \bibinfo {author} {\bibfnamefont {Martin}\ \bibnamefont {Kunz}},
  \ and\ \bibinfo {author} {\bibfnamefont {Domenico}\ \bibnamefont {Sapone}},\
  }\bibfield  {title} {\enquote {\bibinfo {title} {{Measuring the dark side
  (with weak lensing)}},}\ }\href {\doibase 10.1088/1475-7516/2008/04/013}
  {\bibfield  {journal} {\bibinfo  {journal} {JCAP}\ }\textbf {\bibinfo
  {volume} {0804}},\ \bibinfo {pages} {013} (\bibinfo {year} {2008})},\ \Eprint
  {http://arxiv.org/abs/0704.2421} {arXiv:0704.2421 [astro-ph]} \BibitemShut
  {NoStop}%
\bibitem [{\citenamefont {Pogosian}\ \emph {et~al.}(2010)\citenamefont
  {Pogosian}, \citenamefont {Silvestri}, \citenamefont {Koyama},\ and\
  \citenamefont {Zhao}}]{Pogosian:2010tj}%
  \BibitemOpen
  \bibfield  {author} {\bibinfo {author} {\bibfnamefont {Levon}\ \bibnamefont
  {Pogosian}}, \bibinfo {author} {\bibfnamefont {Alessandra}\ \bibnamefont
  {Silvestri}}, \bibinfo {author} {\bibfnamefont {Kazuya}\ \bibnamefont
  {Koyama}}, \ and\ \bibinfo {author} {\bibfnamefont {Gong-Bo}\ \bibnamefont
  {Zhao}},\ }\bibfield  {title} {\enquote {\bibinfo {title} {{How to optimally
  parametrize deviations from General Relativity in the evolution of
  cosmological perturbations?}}}\ }\href {\doibase 10.1103/PhysRevD.81.104023}
  {\bibfield  {journal} {\bibinfo  {journal} {Phys. Rev.}\ }\textbf {\bibinfo
  {volume} {D81}},\ \bibinfo {pages} {104023} (\bibinfo {year} {2010})},\
  \Eprint {http://arxiv.org/abs/1002.2382} {arXiv:1002.2382 [astro-ph.CO]}
  \BibitemShut {NoStop}%
\bibitem [{\citenamefont {Daniel}\ and\ \citenamefont
  {Linder}(2013)}]{Daniel:2012kn}%
  \BibitemOpen
  \bibfield  {author} {\bibinfo {author} {\bibfnamefont {Scott~F.}\
  \bibnamefont {Daniel}}\ and\ \bibinfo {author} {\bibfnamefont {Eric~V.}\
  \bibnamefont {Linder}},\ }\bibfield  {title} {\enquote {\bibinfo {title}
  {{Constraining Cosmic Expansion and Gravity with Galaxy Redshift Surveys}},}\
  }\href {\doibase 10.1088/1475-7516/2013/02/007} {\bibfield  {journal}
  {\bibinfo  {journal} {JCAP}\ }\textbf {\bibinfo {volume} {1302}},\ \bibinfo
  {pages} {007} (\bibinfo {year} {2013})},\ \Eprint
  {http://arxiv.org/abs/1212.0009} {arXiv:1212.0009 [astro-ph.CO]} \BibitemShut
  {NoStop}%
\bibitem [{\citenamefont {Huterer}\ \emph {et~al.}(2015)\citenamefont {Huterer}
  \emph {et~al.}}]{Huterer:2013xky}%
  \BibitemOpen
  \bibfield  {author} {\bibinfo {author} {\bibfnamefont {Dragan}\ \bibnamefont
  {Huterer}} \emph {et~al.},\ }\bibfield  {title} {\enquote {\bibinfo {title}
  {{Growth of Cosmic Structure: Probing Dark Energy Beyond Expansion}},}\
  }\href {\doibase 10.1016/j.astropartphys.2014.07.004} {\bibfield  {journal}
  {\bibinfo  {journal} {Astropart. Phys.}\ }\textbf {\bibinfo {volume} {63}},\
  \bibinfo {pages} {23--41} (\bibinfo {year} {2015})},\ \Eprint
  {http://arxiv.org/abs/1309.5385} {arXiv:1309.5385 [astro-ph.CO]} \BibitemShut
  {NoStop}%
\bibitem [{\citenamefont {Mueller}\ \emph {et~al.}(2018)\citenamefont
  {Mueller}, \citenamefont {Percival}, \citenamefont {Linder}, \citenamefont
  {Alam}, \citenamefont {Zhao}, \citenamefont {Sánchez}, \citenamefont
  {Beutler},\ and\ \citenamefont {Brinkmann}}]{Mueller:2016kpu}%
  \BibitemOpen
  \bibfield  {author} {\bibinfo {author} {\bibfnamefont {Eva-Maria}\
  \bibnamefont {Mueller}}, \bibinfo {author} {\bibfnamefont {Will}\
  \bibnamefont {Percival}}, \bibinfo {author} {\bibfnamefont {Eric}\
  \bibnamefont {Linder}}, \bibinfo {author} {\bibfnamefont {Shadab}\
  \bibnamefont {Alam}}, \bibinfo {author} {\bibfnamefont {Gong-Bo}\
  \bibnamefont {Zhao}}, \bibinfo {author} {\bibfnamefont {Ariel~G.}\
  \bibnamefont {Sánchez}}, \bibinfo {author} {\bibfnamefont {Florian}\
  \bibnamefont {Beutler}}, \ and\ \bibinfo {author} {\bibfnamefont {Jon}\
  \bibnamefont {Brinkmann}},\ }\bibfield  {title} {\enquote {\bibinfo {title}
  {{The clustering of galaxies in the completed SDSS-III Baryon Oscillation
  Spectroscopic Survey: constraining modified gravity}},}\ }\href {\doibase
  10.1093/mnras/stx3232} {\bibfield  {journal} {\bibinfo  {journal} {Mon. Not.
  Roy. Astron. Soc.}\ }\textbf {\bibinfo {volume} {475}},\ \bibinfo {pages}
  {2122--2131} (\bibinfo {year} {2018})},\ \Eprint
  {http://arxiv.org/abs/1612.00812} {arXiv:1612.00812 [astro-ph.CO]}
  \BibitemShut {NoStop}%
\bibitem [{\citenamefont {Perenon}\ \emph {et~al.}(2019)\citenamefont
  {Perenon}, \citenamefont {Bel}, \citenamefont {Maartens},\ and\ \citenamefont
  {de~la Cruz-Dombriz}}]{Perenon:2019dpc}%
  \BibitemOpen
  \bibfield  {author} {\bibinfo {author} {\bibfnamefont {Louis}\ \bibnamefont
  {Perenon}}, \bibinfo {author} {\bibfnamefont {Julien}\ \bibnamefont {Bel}},
  \bibinfo {author} {\bibfnamefont {Roy}\ \bibnamefont {Maartens}}, \ and\
  \bibinfo {author} {\bibfnamefont {Alvaro}\ \bibnamefont {de~la
  Cruz-Dombriz}},\ }\bibfield  {title} {\enquote {\bibinfo {title} {{Optimising
  growth of structure constraints on modified gravity}},}\ }\href {\doibase
  10.1088/1475-7516/2019/06/020} {\bibfield  {journal} {\bibinfo  {journal}
  {JCAP}\ }\textbf {\bibinfo {volume} {1906}},\ \bibinfo {pages} {020}
  (\bibinfo {year} {2019})},\ \Eprint {http://arxiv.org/abs/1901.11063}
  {arXiv:1901.11063 [astro-ph.CO]} \BibitemShut {NoStop}%
\bibitem [{\citenamefont {Arjona}\ \emph {et~al.}(2018)\citenamefont {Arjona},
  \citenamefont {Cardona},\ and\ \citenamefont {Nesseris}}]{Arjona:2018jhh}%
  \BibitemOpen
  \bibfield  {author} {\bibinfo {author} {\bibfnamefont {Rubén}\ \bibnamefont
  {Arjona}}, \bibinfo {author} {\bibfnamefont {Wilmar}\ \bibnamefont
  {Cardona}}, \ and\ \bibinfo {author} {\bibfnamefont {Savvas}\ \bibnamefont
  {Nesseris}},\ }\bibfield  {title} {\enquote {\bibinfo {title} {{Unraveling
  the effective fluid approach for $f(R)$ models in the sub-horizon
  approximation}},}\ }\href@noop {} {\  (\bibinfo {year} {2018})},\ \Eprint
  {http://arxiv.org/abs/1811.02469} {arXiv:1811.02469 [astro-ph.CO]}
  \BibitemShut {NoStop}%
\bibitem [{\citenamefont {Amendola}\ \emph {et~al.}(2013)\citenamefont
  {Amendola}, \citenamefont {Kunz}, \citenamefont {Motta}, \citenamefont
  {Saltas},\ and\ \citenamefont {Sawicki}}]{Amendola:2012ky}%
  \BibitemOpen
  \bibfield  {author} {\bibinfo {author} {\bibfnamefont {Luca}\ \bibnamefont
  {Amendola}}, \bibinfo {author} {\bibfnamefont {Martin}\ \bibnamefont {Kunz}},
  \bibinfo {author} {\bibfnamefont {Mariele}\ \bibnamefont {Motta}}, \bibinfo
  {author} {\bibfnamefont {Ippocratis~D.}\ \bibnamefont {Saltas}}, \ and\
  \bibinfo {author} {\bibfnamefont {Ignacy}\ \bibnamefont {Sawicki}},\
  }\bibfield  {title} {\enquote {\bibinfo {title} {{Observables and
  unobservables in dark energy cosmologies}},}\ }\href {\doibase
  10.1103/PhysRevD.87.023501} {\bibfield  {journal} {\bibinfo  {journal} {Phys.
  Rev.}\ }\textbf {\bibinfo {volume} {D87}},\ \bibinfo {pages} {023501}
  (\bibinfo {year} {2013})},\ \Eprint {http://arxiv.org/abs/1210.0439}
  {arXiv:1210.0439 [astro-ph.CO]} \BibitemShut {NoStop}%
\bibitem [{\citenamefont {Motta}\ \emph {et~al.}(2013)\citenamefont {Motta},
  \citenamefont {Sawicki}, \citenamefont {Saltas}, \citenamefont {Amendola},\
  and\ \citenamefont {Kunz}}]{Motta:2013cwa}%
  \BibitemOpen
  \bibfield  {author} {\bibinfo {author} {\bibfnamefont {Mariele}\ \bibnamefont
  {Motta}}, \bibinfo {author} {\bibfnamefont {Ignacy}\ \bibnamefont {Sawicki}},
  \bibinfo {author} {\bibfnamefont {Ippocratis~D.}\ \bibnamefont {Saltas}},
  \bibinfo {author} {\bibfnamefont {Luca}\ \bibnamefont {Amendola}}, \ and\
  \bibinfo {author} {\bibfnamefont {Martin}\ \bibnamefont {Kunz}},\ }\bibfield
  {title} {\enquote {\bibinfo {title} {{Probing Dark Energy through Scale
  Dependence}},}\ }\href {\doibase 10.1103/PhysRevD.88.124035} {\bibfield
  {journal} {\bibinfo  {journal} {Phys. Rev.}\ }\textbf {\bibinfo {volume}
  {D88}},\ \bibinfo {pages} {124035} (\bibinfo {year} {2013})},\ \Eprint
  {http://arxiv.org/abs/1305.0008} {arXiv:1305.0008 [astro-ph.CO]} \BibitemShut
  {NoStop}%
\bibitem [{\citenamefont {Pinho}\ \emph {et~al.}(2018)\citenamefont {Pinho},
  \citenamefont {Casas},\ and\ \citenamefont {Amendola}}]{Pinho:2018unz}%
  \BibitemOpen
  \bibfield  {author} {\bibinfo {author} {\bibfnamefont {Ana~Marta}\
  \bibnamefont {Pinho}}, \bibinfo {author} {\bibfnamefont {Santiago}\
  \bibnamefont {Casas}}, \ and\ \bibinfo {author} {\bibfnamefont {Luca}\
  \bibnamefont {Amendola}},\ }\bibfield  {title} {\enquote {\bibinfo {title}
  {{Model-independent reconstruction of the linear anisotropic stress
  $\eta$}},}\ }\href {\doibase 10.1088/1475-7516/2018/11/027} {\bibfield
  {journal} {\bibinfo  {journal} {JCAP}\ }\textbf {\bibinfo {volume} {1811}},\
  \bibinfo {pages} {027} (\bibinfo {year} {2018})},\ \Eprint
  {http://arxiv.org/abs/1805.00027} {arXiv:1805.00027 [astro-ph.CO]}
  \BibitemShut {NoStop}%
\bibitem [{\citenamefont {Sanchez}\ and\ \citenamefont
  {Perivolaropoulos}(2010)}]{Sanchez:2010ng}%
  \BibitemOpen
  \bibfield  {author} {\bibinfo {author} {\bibfnamefont {J.~C.~Bueno}\
  \bibnamefont {Sanchez}}\ and\ \bibinfo {author} {\bibfnamefont
  {L.}~\bibnamefont {Perivolaropoulos}},\ }\bibfield  {title} {\enquote
  {\bibinfo {title} {{Evolution of Dark Energy Perturbations in Scalar-Tensor
  Cosmologies}},}\ }\href {\doibase 10.1103/PhysRevD.81.103505} {\bibfield
  {journal} {\bibinfo  {journal} {Phys. Rev.}\ }\textbf {\bibinfo {volume}
  {D81}},\ \bibinfo {pages} {103505} (\bibinfo {year} {2010})},\ \Eprint
  {http://arxiv.org/abs/1002.2042} {arXiv:1002.2042 [astro-ph.CO]} \BibitemShut
  {NoStop}%
\bibitem [{\citenamefont {Muller}\ \emph {et~al.}(2008)\citenamefont {Muller},
  \citenamefont {Williams},\ and\ \citenamefont {Turyshev}}]{Muller:2005sr}%
  \BibitemOpen
  \bibfield  {author} {\bibinfo {author} {\bibfnamefont {Jurgen}\ \bibnamefont
  {Muller}}, \bibinfo {author} {\bibfnamefont {James~G.}\ \bibnamefont
  {Williams}}, \ and\ \bibinfo {author} {\bibfnamefont {Slava~G.}\ \bibnamefont
  {Turyshev}},\ }\bibfield  {title} {\enquote {\bibinfo {title} {{Lunar laser
  ranging contributions to relativity and geodesy}},}\ }\bibfield  {booktitle}
  {\emph {\bibinfo {booktitle} {{Proceedings, 359th WE-Heraeus Seminar :
  Lasers, Clocks, and Drag-Free: Technologies for Future Exploration in Space
  and tests of Gravity: Bremen, Germany, May 30-June 1, 2005}}},\ }\href
  {\doibase 10.1007/978-3-540-34377-6_21} {\bibfield  {journal} {\bibinfo
  {journal} {Astrophys. Space Sci. Libr.}\ }\textbf {\bibinfo {volume} {349}},\
  \bibinfo {pages} {457--472} (\bibinfo {year} {2008})},\ \Eprint
  {http://arxiv.org/abs/gr-qc/0509114} {arXiv:gr-qc/0509114 [gr-qc]}
  \BibitemShut {NoStop}%
\bibitem [{\citenamefont {Pitjeva}\ and\ \citenamefont
  {Pitjev}(2013)}]{Pitjeva:2013chs}%
  \BibitemOpen
  \bibfield  {author} {\bibinfo {author} {\bibfnamefont {E.~V.}\ \bibnamefont
  {Pitjeva}}\ and\ \bibinfo {author} {\bibfnamefont {N.~P.}\ \bibnamefont
  {Pitjev}},\ }\bibfield  {title} {\enquote {\bibinfo {title} {{Relativistic
  effects and dark matter in the Solar system from observations of planets and
  spacecraft}},}\ }\href {\doibase 10.1093/mnras/stt695} {\bibfield  {journal}
  {\bibinfo  {journal} {Mon. Not. Roy. Astron. Soc.}\ }\textbf {\bibinfo
  {volume} {432}},\ \bibinfo {pages} {3431} (\bibinfo {year} {2013})},\ \Eprint
  {http://arxiv.org/abs/1306.3043} {arXiv:1306.3043 [astro-ph.EP]} \BibitemShut
  {NoStop}%
\bibitem [{\citenamefont {Gannouji}\ \emph {et~al.}(2006)\citenamefont
  {Gannouji}, \citenamefont {Polarski}, \citenamefont {Ranquet},\ and\
  \citenamefont {Starobinsky}}]{Gannouji:2006jm}%
  \BibitemOpen
  \bibfield  {author} {\bibinfo {author} {\bibfnamefont {Radouane}\
  \bibnamefont {Gannouji}}, \bibinfo {author} {\bibfnamefont {David}\
  \bibnamefont {Polarski}}, \bibinfo {author} {\bibfnamefont {Andre}\
  \bibnamefont {Ranquet}}, \ and\ \bibinfo {author} {\bibfnamefont {Alexei~A.}\
  \bibnamefont {Starobinsky}},\ }\bibfield  {title} {\enquote {\bibinfo {title}
  {{Scalar-Tensor Models of Normal and Phantom Dark Energy}},}\ }\href
  {\doibase 10.1088/1475-7516/2006/09/016} {\bibfield  {journal} {\bibinfo
  {journal} {JCAP}\ }\textbf {\bibinfo {volume} {0609}},\ \bibinfo {pages}
  {016} (\bibinfo {year} {2006})},\ \Eprint
  {http://arxiv.org/abs/astro-ph/0606287} {arXiv:astro-ph/0606287 [astro-ph]}
  \BibitemShut {NoStop}%
\bibitem [{\citenamefont {Nesseris}\ and\ \citenamefont
  {Perivolaropoulos}(2007{\natexlab{b}})}]{Nesseris:2006hp}%
  \BibitemOpen
  \bibfield  {author} {\bibinfo {author} {\bibfnamefont {S.}~\bibnamefont
  {Nesseris}}\ and\ \bibinfo {author} {\bibfnamefont {Leandros}\ \bibnamefont
  {Perivolaropoulos}},\ }\bibfield  {title} {\enquote {\bibinfo {title} {{The
  Limits of Extended Quintessence}},}\ }\href {\doibase
  10.1103/PhysRevD.75.023517} {\bibfield  {journal} {\bibinfo  {journal} {Phys.
  Rev.}\ }\textbf {\bibinfo {volume} {D75}},\ \bibinfo {pages} {023517}
  (\bibinfo {year} {2007}{\natexlab{b}})},\ \Eprint
  {http://arxiv.org/abs/astro-ph/0611238} {arXiv:astro-ph/0611238 [astro-ph]}
  \BibitemShut {NoStop}%
\bibitem [{\citenamefont {Ade}\ \emph {et~al.}(2016{\natexlab{c}})\citenamefont
  {Ade} \emph {et~al.}}]{Ade:2015rim}%
  \BibitemOpen
  \bibfield  {author} {\bibinfo {author} {\bibfnamefont {P.~A.~R.}\
  \bibnamefont {Ade}} \emph {et~al.} (\bibinfo {collaboration} {Planck}),\
  }\bibfield  {title} {\enquote {\bibinfo {title} {{Planck 2015 results. XIV.
  Dark energy and modified gravity}},}\ }\href {\doibase
  10.1051/0004-6361/201525814} {\bibfield  {journal} {\bibinfo  {journal}
  {Astron. Astrophys.}\ }\textbf {\bibinfo {volume} {594}},\ \bibinfo {pages}
  {A14} (\bibinfo {year} {2016}{\natexlab{c}})},\ \Eprint
  {http://arxiv.org/abs/1502.01590} {arXiv:1502.01590 [astro-ph.CO]}
  \BibitemShut {NoStop}%
\bibitem [{\citenamefont {Di~Valentino}\ \emph {et~al.}(2016)\citenamefont
  {Di~Valentino}, \citenamefont {Melchiorri},\ and\ \citenamefont
  {Silk}}]{DiValentino:2015bja}%
  \BibitemOpen
  \bibfield  {author} {\bibinfo {author} {\bibfnamefont {Eleonora}\
  \bibnamefont {Di~Valentino}}, \bibinfo {author} {\bibfnamefont {Alessandro}\
  \bibnamefont {Melchiorri}}, \ and\ \bibinfo {author} {\bibfnamefont {Joseph}\
  \bibnamefont {Silk}},\ }\bibfield  {title} {\enquote {\bibinfo {title}
  {{Cosmological hints of modified gravity?}}}\ }\href {\doibase
  10.1103/PhysRevD.93.023513} {\bibfield  {journal} {\bibinfo  {journal} {Phys.
  Rev.}\ }\textbf {\bibinfo {volume} {D93}},\ \bibinfo {pages} {023513}
  (\bibinfo {year} {2016})},\ \Eprint {http://arxiv.org/abs/1509.07501}
  {arXiv:1509.07501 [astro-ph.CO]} \BibitemShut {NoStop}%
\bibitem [{\citenamefont {Gannouji}\ \emph {et~al.}(2018)\citenamefont
  {Gannouji}, \citenamefont {Kazantzidis}, \citenamefont {Perivolaropoulos},\
  and\ \citenamefont {Polarski}}]{Gannouji:2018ncm}%
  \BibitemOpen
  \bibfield  {author} {\bibinfo {author} {\bibfnamefont {Radouane}\
  \bibnamefont {Gannouji}}, \bibinfo {author} {\bibfnamefont {Lavrentios}\
  \bibnamefont {Kazantzidis}}, \bibinfo {author} {\bibfnamefont {Leandros}\
  \bibnamefont {Perivolaropoulos}}, \ and\ \bibinfo {author} {\bibfnamefont
  {David}\ \bibnamefont {Polarski}},\ }\bibfield  {title} {\enquote {\bibinfo
  {title} {{Consistency of modified gravity with a decreasing $G_{\rm eff}(z)$
  in a $\Lambda$CDM background}},}\ }\href {\doibase
  10.1103/PhysRevD.98.104044} {\bibfield  {journal} {\bibinfo  {journal} {Phys.
  Rev.}\ }\textbf {\bibinfo {volume} {D98}},\ \bibinfo {pages} {104044}
  (\bibinfo {year} {2018})},\ \Eprint {http://arxiv.org/abs/1809.07034}
  {arXiv:1809.07034 [gr-qc]} \BibitemShut {NoStop}%
\bibitem [{\citenamefont {Linder}(2018)}]{Linder:2018jil}%
  \BibitemOpen
  \bibfield  {author} {\bibinfo {author} {\bibfnamefont {Eric~V.}\ \bibnamefont
  {Linder}},\ }\bibfield  {title} {\enquote {\bibinfo {title} {{No Slip
  Gravity}},}\ }\href {\doibase 10.1088/1475-7516/2018/03/005} {\bibfield
  {journal} {\bibinfo  {journal} {JCAP}\ }\textbf {\bibinfo {volume} {1803}},\
  \bibinfo {pages} {005} (\bibinfo {year} {2018})},\ \Eprint
  {http://arxiv.org/abs/1801.01503} {arXiv:1801.01503 [astro-ph.CO]}
  \BibitemShut {NoStop}%
\bibitem [{\citenamefont {Arjona}\ \emph {et~al.}(2019)\citenamefont {Arjona},
  \citenamefont {Cardona},\ and\ \citenamefont {Nesseris}}]{Arjona:2019rfn}%
  \BibitemOpen
  \bibfield  {author} {\bibinfo {author} {\bibfnamefont {Rubén}\ \bibnamefont
  {Arjona}}, \bibinfo {author} {\bibfnamefont {Wilmar}\ \bibnamefont
  {Cardona}}, \ and\ \bibinfo {author} {\bibfnamefont {Savvas}\ \bibnamefont
  {Nesseris}},\ }\bibfield  {title} {\enquote {\bibinfo {title} {{Designing
  Horndeski and the effective fluid approach}},}\ }\href {\doibase
  10.1103/PhysRevD.100.063526} {\bibfield  {journal} {\bibinfo  {journal}
  {Phys. Rev.}\ }\textbf {\bibinfo {volume} {D100}},\ \bibinfo {pages} {063526}
  (\bibinfo {year} {2019})},\ \Eprint {http://arxiv.org/abs/1904.06294}
  {arXiv:1904.06294 [astro-ph.CO]} \BibitemShut {NoStop}%
\bibitem [{\citenamefont {Tsujikawa}(2015)}]{Tsujikawa:2015mga}%
  \BibitemOpen
  \bibfield  {author} {\bibinfo {author} {\bibfnamefont {Shinji}\ \bibnamefont
  {Tsujikawa}},\ }\bibfield  {title} {\enquote {\bibinfo {title} {{Possibility
  of realizing weak gravity in redshift space distortion measurements}},}\
  }\href {\doibase 10.1103/PhysRevD.92.044029} {\bibfield  {journal} {\bibinfo
  {journal} {Phys. Rev.}\ }\textbf {\bibinfo {volume} {D92}},\ \bibinfo {pages}
  {044029} (\bibinfo {year} {2015})},\ \Eprint
  {http://arxiv.org/abs/1505.02459} {arXiv:1505.02459 [astro-ph.CO]}
  \BibitemShut {NoStop}%
\bibitem [{\citenamefont {Amendola}\ \emph {et~al.}(2020)\citenamefont
  {Amendola}, \citenamefont {Bettoni}, \citenamefont {Pinho},\ and\
  \citenamefont {Casas}}]{Amendola:2019laa}%
  \BibitemOpen
  \bibfield  {author} {\bibinfo {author} {\bibfnamefont {Luca}\ \bibnamefont
  {Amendola}}, \bibinfo {author} {\bibfnamefont {Dario}\ \bibnamefont
  {Bettoni}}, \bibinfo {author} {\bibfnamefont {Ana~Marta}\ \bibnamefont
  {Pinho}}, \ and\ \bibinfo {author} {\bibfnamefont {Santiago}\ \bibnamefont
  {Casas}},\ }\bibfield  {title} {\enquote {\bibinfo {title} {{Measuring
  gravity at cosmological scales}},}\ }\href {\doibase 10.3390/universe6020020}
  {\bibfield  {journal} {\bibinfo  {journal} {Universe}\ }\textbf {\bibinfo
  {volume} {6}},\ \bibinfo {pages} {20} (\bibinfo {year} {2020})},\ \Eprint
  {http://arxiv.org/abs/1902.06978} {arXiv:1902.06978 [astro-ph.CO]}
  \BibitemShut {NoStop}%
\bibitem [{\citenamefont {Moradinezhad~Dizgah}\ and\ \citenamefont
  {Durrer}(2016)}]{Dizgah:2016bgm}%
  \BibitemOpen
  \bibfield  {author} {\bibinfo {author} {\bibfnamefont {Azadeh}\ \bibnamefont
  {Moradinezhad~Dizgah}}\ and\ \bibinfo {author} {\bibfnamefont {Ruth}\
  \bibnamefont {Durrer}},\ }\bibfield  {title} {\enquote {\bibinfo {title}
  {{Lensing corrections to the $E_g(z)$ statistics from large scale
  structure}},}\ }\href {\doibase 10.1088/1475-7516/2016/09/035} {\bibfield
  {journal} {\bibinfo  {journal} {JCAP}\ }\textbf {\bibinfo {volume} {1609}},\
  \bibinfo {pages} {035} (\bibinfo {year} {2016})},\ \Eprint
  {http://arxiv.org/abs/1604.08914} {arXiv:1604.08914 [astro-ph.CO]}
  \BibitemShut {NoStop}%
\bibitem [{\citenamefont {Ghosh}\ and\ \citenamefont
  {Durrer}(2019)}]{Ghosh:2018ijm}%
  \BibitemOpen
  \bibfield  {author} {\bibinfo {author} {\bibfnamefont {Basundhara}\
  \bibnamefont {Ghosh}}\ and\ \bibinfo {author} {\bibfnamefont {Ruth}\
  \bibnamefont {Durrer}},\ }\bibfield  {title} {\enquote {\bibinfo {title}
  {{The observable $E_g$ statistics}},}\ }\href {\doibase
  10.1088/1475-7516/2019/06/010} {\bibfield  {journal} {\bibinfo  {journal}
  {JCAP}\ }\textbf {\bibinfo {volume} {1906}},\ \bibinfo {pages} {010}
  (\bibinfo {year} {2019})},\ \Eprint {http://arxiv.org/abs/1812.09546}
  {arXiv:1812.09546 [astro-ph.CO]} \BibitemShut {NoStop}%
\bibitem [{\citenamefont {Troxel}\ \emph
  {et~al.}(2018{\natexlab{b}})\citenamefont {Troxel} \emph {et~al.}}]{des2}%
  \BibitemOpen
  \bibfield  {author} {\bibinfo {author} {\bibfnamefont {M.~A.}\ \bibnamefont
  {Troxel}} \emph {et~al.} (\bibinfo {collaboration} {DES}),\ }\bibfield
  {title} {\enquote {\bibinfo {title} {{Dark Energy Survey Year 1 results:
  Cosmological constraints from cosmic shear}},}\ }\href {\doibase
  10.1103/PhysRevD.98.043528} {\bibfield  {journal} {\bibinfo  {journal} {Phys.
  Rev.}\ }\textbf {\bibinfo {volume} {D98}},\ \bibinfo {pages} {043528}
  (\bibinfo {year} {2018}{\natexlab{b}})},\ \Eprint
  {http://arxiv.org/abs/1708.01538} {arXiv:1708.01538 [astro-ph.CO]}
  \BibitemShut {NoStop}%
\bibitem [{\citenamefont {Drlica-Wagner}\ \emph {et~al.}(2018)\citenamefont
  {Drlica-Wagner} \emph {et~al.}}]{Drlica-Wagner:2017tkk}%
  \BibitemOpen
  \bibfield  {author} {\bibinfo {author} {\bibfnamefont {A.}~\bibnamefont
  {Drlica-Wagner}} \emph {et~al.} (\bibinfo {collaboration} {DES}),\ }\bibfield
   {title} {\enquote {\bibinfo {title} {{Dark Energy Survey Year 1 Results:
  Photometric Data Set for Cosmology}},}\ }\href {\doibase
  10.3847/1538-4365/aab4f5} {\bibfield  {journal} {\bibinfo  {journal}
  {Astrophys. J. Suppl.}\ }\textbf {\bibinfo {volume} {235}},\ \bibinfo {pages}
  {33} (\bibinfo {year} {2018})},\ \Eprint {http://arxiv.org/abs/1708.01531}
  {arXiv:1708.01531 [astro-ph.CO]} \BibitemShut {NoStop}%
\bibitem [{\citenamefont {Hoyle}\ \emph {et~al.}(2018)\citenamefont {Hoyle}
  \emph {et~al.}}]{Hoyle:2017mee}%
  \BibitemOpen
  \bibfield  {author} {\bibinfo {author} {\bibfnamefont {B.}~\bibnamefont
  {Hoyle}} \emph {et~al.} (\bibinfo {collaboration} {DES}),\ }\bibfield
  {title} {\enquote {\bibinfo {title} {{Dark Energy Survey Year 1 Results:
  Redshift distributions of the weak lensing source galaxies}},}\ }\href
  {\doibase 10.1093/mnras/sty957} {\bibfield  {journal} {\bibinfo  {journal}
  {Mon. Not. Roy. Astron. Soc.}\ }\textbf {\bibinfo {volume} {478}},\ \bibinfo
  {pages} {592--610} (\bibinfo {year} {2018})},\ \Eprint
  {http://arxiv.org/abs/1708.01532} {arXiv:1708.01532 [astro-ph.CO]}
  \BibitemShut {NoStop}%
\bibitem [{\citenamefont {Elvin-Poole}\ \emph {et~al.}(2018)\citenamefont
  {Elvin-Poole} \emph {et~al.}}]{Elvin-Poole:2017xsf}%
  \BibitemOpen
  \bibfield  {author} {\bibinfo {author} {\bibfnamefont {J.}~\bibnamefont
  {Elvin-Poole}} \emph {et~al.} (\bibinfo {collaboration} {DES}),\ }\bibfield
  {title} {\enquote {\bibinfo {title} {{Dark Energy Survey year 1 results:
  Galaxy clustering for combined probes}},}\ }\href {\doibase
  10.1103/PhysRevD.98.042006} {\bibfield  {journal} {\bibinfo  {journal} {Phys.
  Rev.}\ }\textbf {\bibinfo {volume} {D98}},\ \bibinfo {pages} {042006}
  (\bibinfo {year} {2018})},\ \Eprint {http://arxiv.org/abs/1708.01536}
  {arXiv:1708.01536 [astro-ph.CO]} \BibitemShut {NoStop}%
\bibitem [{\citenamefont {Amendola}\ \emph {et~al.}(2018)\citenamefont
  {Amendola} \emph {et~al.}}]{euclid2}%
  \BibitemOpen
  \bibfield  {author} {\bibinfo {author} {\bibfnamefont {Luca}\ \bibnamefont
  {Amendola}} \emph {et~al.},\ }\bibfield  {title} {\enquote {\bibinfo {title}
  {{Cosmology and fundamental physics with the Euclid satellite}},}\ }\href
  {\doibase 10.1007/s41114-017-0010-3} {\bibfield  {journal} {\bibinfo
  {journal} {Living Rev. Rel.}\ }\textbf {\bibinfo {volume} {21}},\ \bibinfo
  {pages} {2} (\bibinfo {year} {2018})},\ \Eprint
  {http://arxiv.org/abs/1606.00180} {arXiv:1606.00180 [astro-ph.CO]}
  \BibitemShut {NoStop}%
\bibitem [{sup()}]{suppl}%
  \BibitemOpen
  \href@noop {} {}\bibinfo {note}
  {\url{http://leandros.physics.uoi.gr/egfs8plots.zip}}\BibitemShut {NoStop}%
\bibitem [{\citenamefont {Song}\ and\ \citenamefont
  {Percival}(2009)}]{Song:2008qt}%
  \BibitemOpen
  \bibfield  {author} {\bibinfo {author} {\bibfnamefont {Yong-Seon}\
  \bibnamefont {Song}}\ and\ \bibinfo {author} {\bibfnamefont {Will~J.}\
  \bibnamefont {Percival}},\ }\bibfield  {title} {\enquote {\bibinfo {title}
  {{Reconstructing the history of structure formation using Redshift
  Distortions}},}\ }\href {\doibase 10.1088/1475-7516/2009/10/004} {\bibfield
  {journal} {\bibinfo  {journal} {JCAP}\ }\textbf {\bibinfo {volume} {0910}},\
  \bibinfo {pages} {004} (\bibinfo {year} {2009})},\ \Eprint
  {http://arxiv.org/abs/0807.0810} {arXiv:0807.0810 [astro-ph]} \BibitemShut
  {NoStop}%
\bibitem [{\citenamefont {Tegmark}\ \emph {et~al.}(2006)\citenamefont {Tegmark}
  \emph {et~al.}}]{Tegmark:2006az}%
  \BibitemOpen
  \bibfield  {author} {\bibinfo {author} {\bibfnamefont {Max}\ \bibnamefont
  {Tegmark}} \emph {et~al.} (\bibinfo {collaboration} {SDSS}),\ }\bibfield
  {title} {\enquote {\bibinfo {title} {{Cosmological Constraints from the SDSS
  Luminous Red Galaxies}},}\ }\href {\doibase 10.1103/PhysRevD.74.123507}
  {\bibfield  {journal} {\bibinfo  {journal} {Phys. Rev.}\ }\textbf {\bibinfo
  {volume} {D74}},\ \bibinfo {pages} {123507} (\bibinfo {year} {2006})},\
  \Eprint {http://arxiv.org/abs/astro-ph/0608632} {arXiv:astro-ph/0608632
  [astro-ph]} \BibitemShut {NoStop}%
\bibitem [{\citenamefont {Davis}\ \emph {et~al.}(2011)\citenamefont {Davis},
  \citenamefont {Nusser}, \citenamefont {Masters}, \citenamefont {Springob},
  \citenamefont {Huchra},\ and\ \citenamefont {Lemson}}]{Davis:2010sw}%
  \BibitemOpen
  \bibfield  {author} {\bibinfo {author} {\bibfnamefont {Marc}\ \bibnamefont
  {Davis}}, \bibinfo {author} {\bibfnamefont {Adi}\ \bibnamefont {Nusser}},
  \bibinfo {author} {\bibfnamefont {Karen}\ \bibnamefont {Masters}}, \bibinfo
  {author} {\bibfnamefont {Christopher}\ \bibnamefont {Springob}}, \bibinfo
  {author} {\bibfnamefont {John~P.}\ \bibnamefont {Huchra}}, \ and\ \bibinfo
  {author} {\bibfnamefont {Gerard}\ \bibnamefont {Lemson}},\ }\bibfield
  {title} {\enquote {\bibinfo {title} {{Local Gravity versus Local Velocity:
  Solutions for $\beta$ and nonlinear bias}},}\ }\href {\doibase
  10.1111/j.1365-2966.2011.18362.x} {\bibfield  {journal} {\bibinfo  {journal}
  {Mon. Not. Roy. Astron. Soc.}\ }\textbf {\bibinfo {volume} {413}},\ \bibinfo
  {pages} {2906} (\bibinfo {year} {2011})},\ \Eprint
  {http://arxiv.org/abs/1011.3114} {arXiv:1011.3114 [astro-ph.CO]} \BibitemShut
  {NoStop}%
\bibitem [{\citenamefont {Hudson}\ and\ \citenamefont
  {Turnbull}(2012)}]{Hudson:2012gt}%
  \BibitemOpen
  \bibfield  {author} {\bibinfo {author} {\bibfnamefont {Michael~J.}\
  \bibnamefont {Hudson}}\ and\ \bibinfo {author} {\bibfnamefont {Stephen~J.}\
  \bibnamefont {Turnbull}},\ }\bibfield  {title} {\enquote {\bibinfo {title}
  {The growth rate of cosmic structure from peculiar velocities at low and high
  redshifts},}\ }\href {http://stacks.iop.org/2041-8205/751/i=2/a=L30}
  {\bibfield  {journal} {\bibinfo  {journal} {The Astrophysical Journal
  Letters}\ }\textbf {\bibinfo {volume} {751}},\ \bibinfo {pages} {L30}
  (\bibinfo {year} {2012})}\BibitemShut {NoStop}%
\bibitem [{\citenamefont {Turnbull}\ \emph {et~al.}(2012)\citenamefont
  {Turnbull}, \citenamefont {Hudson}, \citenamefont {Feldman}, \citenamefont
  {Hicken}, \citenamefont {Kirshner},\ and\ \citenamefont
  {Watkins}}]{Turnbull:2011ty}%
  \BibitemOpen
  \bibfield  {author} {\bibinfo {author} {\bibfnamefont {Stephen~J.}\
  \bibnamefont {Turnbull}}, \bibinfo {author} {\bibfnamefont {Michael~J.}\
  \bibnamefont {Hudson}}, \bibinfo {author} {\bibfnamefont {Hume~A.}\
  \bibnamefont {Feldman}}, \bibinfo {author} {\bibfnamefont {Malcolm}\
  \bibnamefont {Hicken}}, \bibinfo {author} {\bibfnamefont {Robert~P.}\
  \bibnamefont {Kirshner}}, \ and\ \bibinfo {author} {\bibfnamefont {Richard}\
  \bibnamefont {Watkins}},\ }\bibfield  {title} {\enquote {\bibinfo {title}
  {Cosmic flows in the nearby universe from type ia supernovae},}\ }\href@noop
  {} {\bibfield  {journal} {\bibinfo  {journal} {Monthly Notices of the Royal
  Astronomical Society}\ }\textbf {\bibinfo {volume} {420}},\ \bibinfo {pages}
  {447--454} (\bibinfo {year} {2012})}\BibitemShut {NoStop}%
\bibitem [{\citenamefont {Samushia}\ \emph {et~al.}(2012)\citenamefont
  {Samushia}, \citenamefont {Percival},\ and\ \citenamefont
  {Raccanelli}}]{Samushia:2011cs}%
  \BibitemOpen
  \bibfield  {author} {\bibinfo {author} {\bibfnamefont {L.}~\bibnamefont
  {Samushia}}, \bibinfo {author} {\bibfnamefont {W.~J.}\ \bibnamefont
  {Percival}}, \ and\ \bibinfo {author} {\bibfnamefont {A.}~\bibnamefont
  {Raccanelli}},\ }\bibfield  {title} {\enquote {\bibinfo {title} {Interpreting
  large-scale redshift-space distortion measurements},}\ }\href {\doibase
  10.1111/j.1365-2966.2011.20169.x} {\bibfield  {journal} {\bibinfo  {journal}
  {Monthly Notices of the Royal Astronomical Society}\ }\textbf {\bibinfo
  {volume} {420}},\ \bibinfo {pages} {2102--2119} (\bibinfo {year}
  {2012})}\BibitemShut {NoStop}%
\bibitem [{\citenamefont {{Blake}}\ \emph {et~al.}(2012)\citenamefont
  {{Blake}}, \citenamefont {{Brough}}, \citenamefont {{Colless}}, \citenamefont
  {{Contreras}}, \citenamefont {{Couch}}, \citenamefont {{Croom}},
  \citenamefont {{Croton}}, \citenamefont {{Davis}}, \citenamefont
  {{Drinkwater}}, \citenamefont {{Forster}}, \citenamefont {{Gilbank}},
  \citenamefont {{Gladders}}, \citenamefont {{Glazebrook}}, \citenamefont
  {{Jelliffe}}, \citenamefont {{Jurek}}, \citenamefont {{Li}}, \citenamefont
  {{Madore}}, \citenamefont {{Martin}}, \citenamefont {{Pimbblet}},
  \citenamefont {{Poole}}, \citenamefont {{Pracy}}, \citenamefont {{Sharp}},
  \citenamefont {{Wisnioski}}, \citenamefont {{Woods}}, \citenamefont
  {{Wyder}},\ and\ \citenamefont {{Yee}}}]{Blake:2012pj}%
  \BibitemOpen
  \bibfield  {author} {\bibinfo {author} {\bibfnamefont {C.}~\bibnamefont
  {{Blake}}}, \bibinfo {author} {\bibfnamefont {S.}~\bibnamefont {{Brough}}},
  \bibinfo {author} {\bibfnamefont {M.}~\bibnamefont {{Colless}}}, \bibinfo
  {author} {\bibfnamefont {C.}~\bibnamefont {{Contreras}}}, \bibinfo {author}
  {\bibfnamefont {W.}~\bibnamefont {{Couch}}}, \bibinfo {author} {\bibfnamefont
  {S.}~\bibnamefont {{Croom}}}, \bibinfo {author} {\bibfnamefont
  {D.}~\bibnamefont {{Croton}}}, \bibinfo {author} {\bibfnamefont {T.~M.}\
  \bibnamefont {{Davis}}}, \bibinfo {author} {\bibfnamefont {M.~J.}\
  \bibnamefont {{Drinkwater}}}, \bibinfo {author} {\bibfnamefont
  {K.}~\bibnamefont {{Forster}}}, \bibinfo {author} {\bibfnamefont
  {D.}~\bibnamefont {{Gilbank}}}, \bibinfo {author} {\bibfnamefont
  {M.}~\bibnamefont {{Gladders}}}, \bibinfo {author} {\bibfnamefont
  {K.}~\bibnamefont {{Glazebrook}}}, \bibinfo {author} {\bibfnamefont
  {B.}~\bibnamefont {{Jelliffe}}}, \bibinfo {author} {\bibfnamefont {R.~J.}\
  \bibnamefont {{Jurek}}}, \bibinfo {author} {\bibfnamefont {I.-h.}\
  \bibnamefont {{Li}}}, \bibinfo {author} {\bibfnamefont {B.}~\bibnamefont
  {{Madore}}}, \bibinfo {author} {\bibfnamefont {D.~C.}\ \bibnamefont
  {{Martin}}}, \bibinfo {author} {\bibfnamefont {K.}~\bibnamefont
  {{Pimbblet}}}, \bibinfo {author} {\bibfnamefont {G.~B.}\ \bibnamefont
  {{Poole}}}, \bibinfo {author} {\bibfnamefont {M.}~\bibnamefont {{Pracy}}},
  \bibinfo {author} {\bibfnamefont {R.}~\bibnamefont {{Sharp}}}, \bibinfo
  {author} {\bibfnamefont {E.}~\bibnamefont {{Wisnioski}}}, \bibinfo {author}
  {\bibfnamefont {D.}~\bibnamefont {{Woods}}}, \bibinfo {author} {\bibfnamefont
  {T.~K.}\ \bibnamefont {{Wyder}}}, \ and\ \bibinfo {author} {\bibfnamefont
  {H.~K.~C.}\ \bibnamefont {{Yee}}},\ }\bibfield  {title} {\enquote {\bibinfo
  {title} {{The WiggleZ Dark Energy Survey: joint measurements of the expansion
  and growth history at $z 1$}},}\ }\href {\doibase
  10.1111/j.1365-2966.2012.21473.x} {\bibfield  {journal} {\bibinfo  {journal}
  {mnras}\ }\textbf {\bibinfo {volume} {425}},\ \bibinfo {pages} {405--414}
  (\bibinfo {year} {2012})},\ \Eprint {http://arxiv.org/abs/1204.3674}
  {arXiv:1204.3674} \BibitemShut {NoStop}%
\bibitem [{\citenamefont {Tojeiro}\ \emph {et~al.}(2012)\citenamefont
  {Tojeiro}, \citenamefont {Percival}, \citenamefont {Brinkmann}, \citenamefont
  {Brownstein}, \citenamefont {Eisenstein}, \citenamefont {Manera},
  \citenamefont {Maraston}, \citenamefont {McBride}, \citenamefont {Muna},
  \citenamefont {Reid}, \citenamefont {Ross}, \citenamefont {Ross},
  \citenamefont {Samushia}, \citenamefont {Padmanabhan}, \citenamefont
  {Schneider}, \citenamefont {Skibba}, \citenamefont {Sánchez}, \citenamefont
  {Swanson}, \citenamefont {Thomas}, \citenamefont {Tinker}, \citenamefont
  {Verde}, \citenamefont {Wake}, \citenamefont {Weaver},\ and\ \citenamefont
  {Zhao}}]{Tojeiro:2012rp}%
  \BibitemOpen
  \bibfield  {author} {\bibinfo {author} {\bibfnamefont {Rita}\ \bibnamefont
  {Tojeiro}}, \bibinfo {author} {\bibfnamefont {Will~J.}\ \bibnamefont
  {Percival}}, \bibinfo {author} {\bibfnamefont {Jon}\ \bibnamefont
  {Brinkmann}}, \bibinfo {author} {\bibfnamefont {Joel~R.}\ \bibnamefont
  {Brownstein}}, \bibinfo {author} {\bibfnamefont {Daniel~J.}\ \bibnamefont
  {Eisenstein}}, \bibinfo {author} {\bibfnamefont {Marc}\ \bibnamefont
  {Manera}}, \bibinfo {author} {\bibfnamefont {Claudia}\ \bibnamefont
  {Maraston}}, \bibinfo {author} {\bibfnamefont {Cameron~K.}\ \bibnamefont
  {McBride}}, \bibinfo {author} {\bibfnamefont {Demitri}\ \bibnamefont {Muna}},
  \bibinfo {author} {\bibfnamefont {Beth}\ \bibnamefont {Reid}}, \bibinfo
  {author} {\bibfnamefont {Ashley~J.}\ \bibnamefont {Ross}}, \bibinfo {author}
  {\bibfnamefont {Nicholas~P.}\ \bibnamefont {Ross}}, \bibinfo {author}
  {\bibfnamefont {Lado}\ \bibnamefont {Samushia}}, \bibinfo {author}
  {\bibfnamefont {Nikhil}\ \bibnamefont {Padmanabhan}}, \bibinfo {author}
  {\bibfnamefont {Donald~P.}\ \bibnamefont {Schneider}}, \bibinfo {author}
  {\bibfnamefont {Ramin}\ \bibnamefont {Skibba}}, \bibinfo {author}
  {\bibfnamefont {Ariel~G.}\ \bibnamefont {Sánchez}}, \bibinfo {author}
  {\bibfnamefont {Molly E.~C.}\ \bibnamefont {Swanson}}, \bibinfo {author}
  {\bibfnamefont {Daniel}\ \bibnamefont {Thomas}}, \bibinfo {author}
  {\bibfnamefont {Jeremy~L.}\ \bibnamefont {Tinker}}, \bibinfo {author}
  {\bibfnamefont {Licia}\ \bibnamefont {Verde}}, \bibinfo {author}
  {\bibfnamefont {David~A.}\ \bibnamefont {Wake}}, \bibinfo {author}
  {\bibfnamefont {Benjamin~A.}\ \bibnamefont {Weaver}}, \ and\ \bibinfo
  {author} {\bibfnamefont {Gong-Bo}\ \bibnamefont {Zhao}},\ }\bibfield  {title}
  {\enquote {\bibinfo {title} {The clustering of galaxies in the sdss-iii
  baryon oscillation spectroscopic survey: measuring structure growth using
  passive galaxies},}\ }\href {\doibase 10.1111/j.1365-2966.2012.21404.x}
  {\bibfield  {journal} {\bibinfo  {journal} {Monthly Notices of the Royal
  Astronomical Society}\ }\textbf {\bibinfo {volume} {424}},\ \bibinfo {pages}
  {2339--2344} (\bibinfo {year} {2012})}\BibitemShut {NoStop}%
\bibitem [{\citenamefont {de~la Torre}\ \emph {et~al.}(2013)\citenamefont
  {de~la Torre} \emph {et~al.}}]{delaTorre:2013rpa}%
  \BibitemOpen
  \bibfield  {author} {\bibinfo {author} {\bibfnamefont {S.}~\bibnamefont
  {de~la Torre}} \emph {et~al.},\ }\bibfield  {title} {\enquote {\bibinfo
  {title} {{The VIMOS Public Extragalactic Redshift Survey (VIPERS). Galaxy
  clustering and redshift-space distortions at z=0.8 in the first data
  release}},}\ }\href {\doibase 10.1051/0004-6361/201321463} {\bibfield
  {journal} {\bibinfo  {journal} {Astron. Astrophys.}\ }\textbf {\bibinfo
  {volume} {557}},\ \bibinfo {pages} {A54} (\bibinfo {year} {2013})},\ \Eprint
  {http://arxiv.org/abs/1303.2622} {arXiv:1303.2622 [astro-ph.CO]} \BibitemShut
  {NoStop}%
\bibitem [{\citenamefont {Chuang}\ and\ \citenamefont
  {Wang}(2013)}]{Chuang:2012qt}%
  \BibitemOpen
  \bibfield  {author} {\bibinfo {author} {\bibfnamefont {Chia-Hsun}\
  \bibnamefont {Chuang}}\ and\ \bibinfo {author} {\bibfnamefont {Yun}\
  \bibnamefont {Wang}},\ }\bibfield  {title} {\enquote {\bibinfo {title}
  {Modelling the anisotropic two-point galaxy correlation function on small
  scales and single-probe measurements of $h(z)$, $da(z)$ and $f(z)\sigma 8(z)$
  from the sloan digital sky survey dr7 luminous red galaxies},}\ }\href
  {\doibase 10.1093/mnras/stt1290} {\bibfield  {journal} {\bibinfo  {journal}
  {Monthly Notices of the Royal Astronomical Society}\ }\textbf {\bibinfo
  {volume} {435}},\ \bibinfo {pages} {255--262} (\bibinfo {year}
  {2013})}\BibitemShut {NoStop}%
\bibitem [{\citenamefont {Komatsu}\ \emph {et~al.}(2011)\citenamefont {Komatsu}
  \emph {et~al.}}]{Komatsu:2010fb}%
  \BibitemOpen
  \bibfield  {author} {\bibinfo {author} {\bibfnamefont {E.}~\bibnamefont
  {Komatsu}} \emph {et~al.} (\bibinfo {collaboration} {WMAP}),\ }\bibfield
  {title} {\enquote {\bibinfo {title} {{Seven-Year Wilkinson Microwave
  Anisotropy Probe (WMAP) Observations: Cosmological Interpretation}},}\ }\href
  {\doibase 10.1088/0067-0049/192/2/18} {\bibfield  {journal} {\bibinfo
  {journal} {Astrophys. J. Suppl.}\ }\textbf {\bibinfo {volume} {192}},\
  \bibinfo {pages} {18} (\bibinfo {year} {2011})},\ \Eprint
  {http://arxiv.org/abs/1001.4538} {arXiv:1001.4538 [astro-ph.CO]} \BibitemShut
  {NoStop}%
\bibitem [{\citenamefont {Blake}\ \emph {et~al.}(2013)\citenamefont {Blake}
  \emph {et~al.}}]{Blake:2013nif}%
  \BibitemOpen
  \bibfield  {author} {\bibinfo {author} {\bibfnamefont {Chris}\ \bibnamefont
  {Blake}} \emph {et~al.},\ }\bibfield  {title} {\enquote {\bibinfo {title}
  {{Galaxy And Mass Assembly (GAMA): improved cosmic growth measurements using
  multiple tracers of large-scale structure}},}\ }\href {\doibase
  10.1093/mnras/stt1791} {\bibfield  {journal} {\bibinfo  {journal} {Mon. Not.
  Roy. Astron. Soc.}\ }\textbf {\bibinfo {volume} {436}},\ \bibinfo {pages}
  {3089} (\bibinfo {year} {2013})},\ \Eprint {http://arxiv.org/abs/1309.5556}
  {arXiv:1309.5556 [astro-ph.CO]} \BibitemShut {NoStop}%
\bibitem [{\citenamefont {Sanchez}\ \emph {et~al.}(2014)\citenamefont {Sanchez}
  \emph {et~al.}}]{Sanchez:2013tga}%
  \BibitemOpen
  \bibfield  {author} {\bibinfo {author} {\bibfnamefont {Ariel~G.}\
  \bibnamefont {Sanchez}} \emph {et~al.},\ }\bibfield  {title} {\enquote
  {\bibinfo {title} {{The clustering of galaxies in the SDSS-III Baryon
  Oscillation Spectroscopic Survey: cosmological implications of the full shape
  of the clustering wedges in the data release 10 and 11 galaxy samples}},}\
  }\href {\doibase 10.1093/mnras/stu342} {\bibfield  {journal} {\bibinfo
  {journal} {Mon. Not. Roy. Astron. Soc.}\ }\textbf {\bibinfo {volume} {440}},\
  \bibinfo {pages} {2692--2713} (\bibinfo {year} {2014})},\ \Eprint
  {http://arxiv.org/abs/1312.4854} {arXiv:1312.4854 [astro-ph.CO]} \BibitemShut
  {NoStop}%
\bibitem [{\citenamefont {Anderson}\ \emph {et~al.}(2014)\citenamefont
  {Anderson} \emph {et~al.}}]{Anderson:2013zyy}%
  \BibitemOpen
  \bibfield  {author} {\bibinfo {author} {\bibfnamefont {Lauren}\ \bibnamefont
  {Anderson}} \emph {et~al.} (\bibinfo {collaboration} {BOSS}),\ }\bibfield
  {title} {\enquote {\bibinfo {title} {{The clustering of galaxies in the
  SDSS-III Baryon Oscillation Spectroscopic Survey: baryon acoustic
  oscillations in the Data Releases 10 and 11 Galaxy samples}},}\ }\href
  {\doibase 10.1093/mnras/stu523} {\bibfield  {journal} {\bibinfo  {journal}
  {Mon. Not. Roy. Astron. Soc.}\ }\textbf {\bibinfo {volume} {441}},\ \bibinfo
  {pages} {24--62} (\bibinfo {year} {2014})},\ \Eprint
  {http://arxiv.org/abs/1312.4877} {arXiv:1312.4877 [astro-ph.CO]} \BibitemShut
  {NoStop}%
\bibitem [{\citenamefont {Howlett}\ \emph {et~al.}(2015)\citenamefont
  {Howlett}, \citenamefont {Ross}, \citenamefont {Samushia}, \citenamefont
  {Percival},\ and\ \citenamefont {Manera}}]{Howlett:2014opa}%
  \BibitemOpen
  \bibfield  {author} {\bibinfo {author} {\bibfnamefont {Cullan}\ \bibnamefont
  {Howlett}}, \bibinfo {author} {\bibfnamefont {Ashley}\ \bibnamefont {Ross}},
  \bibinfo {author} {\bibfnamefont {Lado}\ \bibnamefont {Samushia}}, \bibinfo
  {author} {\bibfnamefont {Will}\ \bibnamefont {Percival}}, \ and\ \bibinfo
  {author} {\bibfnamefont {Marc}\ \bibnamefont {Manera}},\ }\bibfield  {title}
  {\enquote {\bibinfo {title} {{The clustering of the SDSS main galaxy sample
  – II. Mock galaxy catalogues and a measurement of the growth of structure
  from redshift space distortions at $z = 0.15$}},}\ }\href {\doibase
  10.1093/mnras/stu2693} {\bibfield  {journal} {\bibinfo  {journal} {Mon. Not.
  Roy. Astron. Soc.}\ }\textbf {\bibinfo {volume} {449}},\ \bibinfo {pages}
  {848--866} (\bibinfo {year} {2015})},\ \Eprint
  {http://arxiv.org/abs/1409.3238} {arXiv:1409.3238 [astro-ph.CO]} \BibitemShut
  {NoStop}%
\bibitem [{\citenamefont {Feix}\ \emph {et~al.}(2015)\citenamefont {Feix},
  \citenamefont {Nusser},\ and\ \citenamefont {Branchini}}]{Feix:2015dla}%
  \BibitemOpen
  \bibfield  {author} {\bibinfo {author} {\bibfnamefont {Martin}\ \bibnamefont
  {Feix}}, \bibinfo {author} {\bibfnamefont {Adi}\ \bibnamefont {Nusser}}, \
  and\ \bibinfo {author} {\bibfnamefont {Enzo}\ \bibnamefont {Branchini}},\
  }\bibfield  {title} {\enquote {\bibinfo {title} {{Growth Rate of Cosmological
  Perturbations at $z \approx 0.1$ from a New Observational Test}},}\ }\href
  {\doibase 10.1103/PhysRevLett.115.011301} {\bibfield  {journal} {\bibinfo
  {journal} {Phys. Rev. Lett.}\ }\textbf {\bibinfo {volume} {115}},\ \bibinfo
  {pages} {011301} (\bibinfo {year} {2015})},\ \Eprint
  {http://arxiv.org/abs/1503.05945} {arXiv:1503.05945 [astro-ph.CO]}
  \BibitemShut {NoStop}%
\bibitem [{\citenamefont {Tegmark}\ \emph {et~al.}(2004)\citenamefont {Tegmark}
  \emph {et~al.}}]{Tegmark:2003uf}%
  \BibitemOpen
  \bibfield  {author} {\bibinfo {author} {\bibfnamefont {Max}\ \bibnamefont
  {Tegmark}} \emph {et~al.} (\bibinfo {collaboration} {SDSS}),\ }\bibfield
  {title} {\enquote {\bibinfo {title} {{The 3-D power spectrum of galaxies from
  the SDSS}},}\ }\href {\doibase 10.1086/382125} {\bibfield  {journal}
  {\bibinfo  {journal} {Astrophys. J.}\ }\textbf {\bibinfo {volume} {606}},\
  \bibinfo {pages} {702--740} (\bibinfo {year} {2004})},\ \Eprint
  {http://arxiv.org/abs/astro-ph/0310725} {arXiv:astro-ph/0310725 [astro-ph]}
  \BibitemShut {NoStop}%
\bibitem [{\citenamefont {Okumura}\ \emph {et~al.}(2016)\citenamefont {Okumura}
  \emph {et~al.}}]{Okumura:2015lvp}%
  \BibitemOpen
  \bibfield  {author} {\bibinfo {author} {\bibfnamefont {Teppei}\ \bibnamefont
  {Okumura}} \emph {et~al.},\ }\bibfield  {title} {\enquote {\bibinfo {title}
  {{The Subaru FMOS galaxy redshift survey (FastSound). IV. New constraint on
  gravity theory from redshift space distortions at $z\sim 1.4$}},}\ }\href
  {\doibase 10.1093/pasj/psw029} {\bibfield  {journal} {\bibinfo  {journal}
  {Publ. Astron. Soc. Jap.}\ }\textbf {\bibinfo {volume} {68}},\ \bibinfo
  {pages} {24} (\bibinfo {year} {2016})},\ \Eprint
  {http://arxiv.org/abs/1511.08083} {arXiv:1511.08083 [astro-ph.CO]}
  \BibitemShut {NoStop}%
\bibitem [{\citenamefont {Hinshaw}\ \emph {et~al.}(2013)\citenamefont {Hinshaw}
  \emph {et~al.}}]{Hinshaw:2012aka}%
  \BibitemOpen
  \bibfield  {author} {\bibinfo {author} {\bibfnamefont {G.}~\bibnamefont
  {Hinshaw}} \emph {et~al.} (\bibinfo {collaboration} {WMAP}),\ }\bibfield
  {title} {\enquote {\bibinfo {title} {{Nine-Year Wilkinson Microwave
  Anisotropy Probe (WMAP) Observations: Cosmological Parameter Results}},}\
  }\href {\doibase 10.1088/0067-0049/208/2/19} {\bibfield  {journal} {\bibinfo
  {journal} {Astrophys. J. Suppl.}\ }\textbf {\bibinfo {volume} {208}},\
  \bibinfo {pages} {19} (\bibinfo {year} {2013})},\ \Eprint
  {http://arxiv.org/abs/1212.5226} {arXiv:1212.5226 [astro-ph.CO]} \BibitemShut
  {NoStop}%
\bibitem [{\citenamefont {Chuang}\ \emph {et~al.}(2016)\citenamefont {Chuang}
  \emph {et~al.}}]{Chuang:2013wga}%
  \BibitemOpen
  \bibfield  {author} {\bibinfo {author} {\bibfnamefont {Chia-Hsun}\
  \bibnamefont {Chuang}} \emph {et~al.},\ }\bibfield  {title} {\enquote
  {\bibinfo {title} {{The clustering of galaxies in the SDSS-III Baryon
  Oscillation Spectroscopic Survey: single-probe measurements from CMASS
  anisotropic galaxy clustering}},}\ }\href {\doibase 10.1093/mnras/stw1535}
  {\bibfield  {journal} {\bibinfo  {journal} {Mon. Not. Roy. Astron. Soc.}\
  }\textbf {\bibinfo {volume} {461}},\ \bibinfo {pages} {3781--3793} (\bibinfo
  {year} {2016})},\ \Eprint {http://arxiv.org/abs/1312.4889} {arXiv:1312.4889
  [astro-ph.CO]} \BibitemShut {NoStop}%
\bibitem [{\citenamefont {Alam}\ \emph
  {et~al.}(2017{\natexlab{a}})\citenamefont {Alam} \emph
  {et~al.}}]{Alam:2016hwk}%
  \BibitemOpen
  \bibfield  {author} {\bibinfo {author} {\bibfnamefont {Shadab}\ \bibnamefont
  {Alam}} \emph {et~al.} (\bibinfo {collaboration} {BOSS}),\ }\bibfield
  {title} {\enquote {\bibinfo {title} {{The clustering of galaxies in the
  completed SDSS-III Baryon Oscillation Spectroscopic Survey: cosmological
  analysis of the DR12 galaxy sample}},}\ }\href {\doibase
  10.1093/mnras/stx721} {\bibfield  {journal} {\bibinfo  {journal} {Mon. Not.
  Roy. Astron. Soc.}\ }\textbf {\bibinfo {volume} {470}},\ \bibinfo {pages}
  {2617--2652} (\bibinfo {year} {2017}{\natexlab{a}})},\ \Eprint
  {http://arxiv.org/abs/1607.03155} {arXiv:1607.03155 [astro-ph.CO]}
  \BibitemShut {NoStop}%
\bibitem [{\citenamefont {Beutler}\ \emph {et~al.}(2017)\citenamefont {Beutler}
  \emph {et~al.}}]{Beutler:2016arn}%
  \BibitemOpen
  \bibfield  {author} {\bibinfo {author} {\bibfnamefont {Florian}\ \bibnamefont
  {Beutler}} \emph {et~al.} (\bibinfo {collaboration} {BOSS}),\ }\bibfield
  {title} {\enquote {\bibinfo {title} {{The clustering of galaxies in the
  completed SDSS-III Baryon Oscillation Spectroscopic Survey: Anisotropic
  galaxy clustering in Fourier-space}},}\ }\href {\doibase
  10.1093/mnras/stw3298} {\bibfield  {journal} {\bibinfo  {journal} {Mon. Not.
  Roy. Astron. Soc.}\ }\textbf {\bibinfo {volume} {466}},\ \bibinfo {pages}
  {2242--2260} (\bibinfo {year} {2017})},\ \Eprint
  {http://arxiv.org/abs/1607.03150} {arXiv:1607.03150 [astro-ph.CO]}
  \BibitemShut {NoStop}%
\bibitem [{\citenamefont {Wilson}(2016)}]{Wilson:2016ggz}%
  \BibitemOpen
  \bibfield  {author} {\bibinfo {author} {\bibfnamefont {Michael~J.}\
  \bibnamefont {Wilson}},\ }\emph {\bibinfo {title} {{Geometric and growth rate
  tests of General Relativity with recovered linear cosmological
  perturbations}}},\ \href
  {https://inspirehep.net/record/1494559/files/arXiv:1610.08362.pdf} {Ph.D.
  thesis},\ \bibinfo  {school} {Edinburgh U.} (\bibinfo {year} {2016}),\
  \Eprint {http://arxiv.org/abs/1610.08362} {arXiv:1610.08362 [astro-ph.CO]}
  \BibitemShut {NoStop}%
\bibitem [{\citenamefont {Gil-Marín}\ \emph {et~al.}(2017)\citenamefont
  {Gil-Marín}, \citenamefont {Percival}, \citenamefont {Verde}, \citenamefont
  {Brownstein}, \citenamefont {Chuang}, \citenamefont {Kitaura}, \citenamefont
  {Rodríguez-Torres},\ and\ \citenamefont {Olmstead}}]{Gil-Marin:2016wya}%
  \BibitemOpen
  \bibfield  {author} {\bibinfo {author} {\bibfnamefont {Héctor}\ \bibnamefont
  {Gil-Marín}}, \bibinfo {author} {\bibfnamefont {Will~J.}\ \bibnamefont
  {Percival}}, \bibinfo {author} {\bibfnamefont {Licia}\ \bibnamefont {Verde}},
  \bibinfo {author} {\bibfnamefont {Joel~R.}\ \bibnamefont {Brownstein}},
  \bibinfo {author} {\bibfnamefont {Chia-Hsun}\ \bibnamefont {Chuang}},
  \bibinfo {author} {\bibfnamefont {Francisco-Shu}\ \bibnamefont {Kitaura}},
  \bibinfo {author} {\bibfnamefont {Sergio~A.}\ \bibnamefont
  {Rodríguez-Torres}}, \ and\ \bibinfo {author} {\bibfnamefont {Matthew~D.}\
  \bibnamefont {Olmstead}},\ }\bibfield  {title} {\enquote {\bibinfo {title}
  {{The clustering of galaxies in the SDSS-III Baryon Oscillation Spectroscopic
  Survey: RSD measurement from the power spectrum and bispectrum of the DR12
  BOSS galaxies}},}\ }\href {\doibase 10.1093/mnras/stw2679} {\bibfield
  {journal} {\bibinfo  {journal} {Mon. Not. Roy. Astron. Soc.}\ }\textbf
  {\bibinfo {volume} {465}},\ \bibinfo {pages} {1757--1788} (\bibinfo {year}
  {2017})},\ \Eprint {http://arxiv.org/abs/1606.00439} {arXiv:1606.00439
  [astro-ph.CO]} \BibitemShut {NoStop}%
\bibitem [{\citenamefont {Hawken}\ \emph {et~al.}(2017)\citenamefont {Hawken}
  \emph {et~al.}}]{Hawken:2016qcy}%
  \BibitemOpen
  \bibfield  {author} {\bibinfo {author} {\bibfnamefont {A.~J.}\ \bibnamefont
  {Hawken}} \emph {et~al.},\ }\bibfield  {title} {\enquote {\bibinfo {title}
  {{The VIMOS Public Extragalactic Redshift Survey: Measuring the growth rate
  of structure around cosmic voids}},}\ }\href {\doibase
  10.1051/0004-6361/201629678} {\bibfield  {journal} {\bibinfo  {journal}
  {Astron. Astrophys.}\ }\textbf {\bibinfo {volume} {607}},\ \bibinfo {pages}
  {A54} (\bibinfo {year} {2017})},\ \Eprint {http://arxiv.org/abs/1611.07046}
  {arXiv:1611.07046 [astro-ph.CO]} \BibitemShut {NoStop}%
\bibitem [{\citenamefont {Huterer}\ \emph {et~al.}(2017)\citenamefont
  {Huterer}, \citenamefont {Shafer}, \citenamefont {Scolnic},\ and\
  \citenamefont {Schmidt}}]{Huterer:2016uyq}%
  \BibitemOpen
  \bibfield  {author} {\bibinfo {author} {\bibfnamefont {Dragan}\ \bibnamefont
  {Huterer}}, \bibinfo {author} {\bibfnamefont {Daniel}\ \bibnamefont
  {Shafer}}, \bibinfo {author} {\bibfnamefont {Daniel}\ \bibnamefont
  {Scolnic}}, \ and\ \bibinfo {author} {\bibfnamefont {Fabian}\ \bibnamefont
  {Schmidt}},\ }\bibfield  {title} {\enquote {\bibinfo {title} {{Testing
  $\Lambda$CDM at the lowest redshifts with SN Ia and galaxy velocities}},}\
  }\href {\doibase 10.1088/1475-7516/2017/05/015} {\bibfield  {journal}
  {\bibinfo  {journal} {JCAP}\ }\textbf {\bibinfo {volume} {1705}},\ \bibinfo
  {pages} {015} (\bibinfo {year} {2017})},\ \Eprint
  {http://arxiv.org/abs/1611.09862} {arXiv:1611.09862 [astro-ph.CO]}
  \BibitemShut {NoStop}%
\bibitem [{\citenamefont {Pezzotta}\ \emph {et~al.}(2017)\citenamefont
  {Pezzotta} \emph {et~al.}}]{Pezzotta:2016gbo}%
  \BibitemOpen
  \bibfield  {author} {\bibinfo {author} {\bibfnamefont {A.}~\bibnamefont
  {Pezzotta}} \emph {et~al.},\ }\bibfield  {title} {\enquote {\bibinfo {title}
  {{The VIMOS Public Extragalactic Redshift Survey (VIPERS): The growth of
  structure at $0.5 < z < 1.2$ from redshift-space distortions in the
  clustering of the PDR-2 final sample}},}\ }\href {\doibase
  10.1051/0004-6361/201630295} {\bibfield  {journal} {\bibinfo  {journal}
  {Astron. Astrophys.}\ }\textbf {\bibinfo {volume} {604}},\ \bibinfo {pages}
  {A33} (\bibinfo {year} {2017})},\ \Eprint {http://arxiv.org/abs/1612.05645}
  {arXiv:1612.05645 [astro-ph.CO]} \BibitemShut {NoStop}%
\bibitem [{\citenamefont {Feix}\ \emph {et~al.}(2017)\citenamefont {Feix},
  \citenamefont {Branchini},\ and\ \citenamefont {Nusser}}]{Feix:2016qhh}%
  \BibitemOpen
  \bibfield  {author} {\bibinfo {author} {\bibfnamefont {Martin}\ \bibnamefont
  {Feix}}, \bibinfo {author} {\bibfnamefont {Enzo}\ \bibnamefont {Branchini}},
  \ and\ \bibinfo {author} {\bibfnamefont {Adi}\ \bibnamefont {Nusser}},\
  }\bibfield  {title} {\enquote {\bibinfo {title} {{Speed from light: growth
  rate and bulk flow at $z \approx 0.1$ from improved SDSS DR13 photometry}},}\
  }\href {\doibase 10.1093/mnras/stx566} {\bibfield  {journal} {\bibinfo
  {journal} {Mon. Not. Roy. Astron. Soc.}\ }\textbf {\bibinfo {volume} {468}},\
  \bibinfo {pages} {1420--1425} (\bibinfo {year} {2017})},\ \Eprint
  {http://arxiv.org/abs/1612.07809} {arXiv:1612.07809 [astro-ph.CO]}
  \BibitemShut {NoStop}%
\bibitem [{\citenamefont {Howlett}\ \emph {et~al.}(2017)\citenamefont
  {Howlett}, \citenamefont {Staveley-Smith}, \citenamefont {Elahi},
  \citenamefont {Hong}, \citenamefont {Jarrett}, \citenamefont {Jones},
  \citenamefont {Koribalski}, \citenamefont {Macri}, \citenamefont {Masters},\
  and\ \citenamefont {Springob}}]{Howlett:2017asq}%
  \BibitemOpen
  \bibfield  {author} {\bibinfo {author} {\bibfnamefont {Cullan}\ \bibnamefont
  {Howlett}}, \bibinfo {author} {\bibfnamefont {Lister}\ \bibnamefont
  {Staveley-Smith}}, \bibinfo {author} {\bibfnamefont {Pascal~J.}\ \bibnamefont
  {Elahi}}, \bibinfo {author} {\bibfnamefont {Tao}\ \bibnamefont {Hong}},
  \bibinfo {author} {\bibfnamefont {Tom~H.}\ \bibnamefont {Jarrett}}, \bibinfo
  {author} {\bibfnamefont {D.~Heath}\ \bibnamefont {Jones}}, \bibinfo {author}
  {\bibfnamefont {Bärbel~S.}\ \bibnamefont {Koribalski}}, \bibinfo {author}
  {\bibfnamefont {Lucas~M.}\ \bibnamefont {Macri}}, \bibinfo {author}
  {\bibfnamefont {Karen~L.}\ \bibnamefont {Masters}}, \ and\ \bibinfo {author}
  {\bibfnamefont {Christopher~M.}\ \bibnamefont {Springob}},\ }\bibfield
  {title} {\enquote {\bibinfo {title} {{2MTF VI. Measuring the velocity power
  spectrum}},}\ }\href {\doibase 10.1093/mnras/stx1521} {\bibfield  {journal}
  {\bibinfo  {journal} {Mon. Not. Roy. Astron. Soc.}\ }\textbf {\bibinfo
  {volume} {471}},\ \bibinfo {pages} {3135} (\bibinfo {year} {2017})},\ \Eprint
  {http://arxiv.org/abs/1706.05130} {arXiv:1706.05130 [astro-ph.CO]}
  \BibitemShut {NoStop}%
\bibitem [{\citenamefont {Mohammad}\ \emph {et~al.}(2017)\citenamefont
  {Mohammad} \emph {et~al.}}]{Mohammad:2017lzz}%
  \BibitemOpen
  \bibfield  {author} {\bibinfo {author} {\bibfnamefont {F.~G.}\ \bibnamefont
  {Mohammad}} \emph {et~al.},\ }\bibfield  {title} {\enquote {\bibinfo {title}
  {{The VIMOS Public Extragalactic Redshift Survey (VIPERS): An unbiased
  estimate of the growth rate of structure at $\mathbf{\left<z\right>=0.85}$
  using the clustering of luminous blue galaxies}},}\ }\href@noop {} {\
  (\bibinfo {year} {2017})},\ \Eprint {http://arxiv.org/abs/1708.00026}
  {arXiv:1708.00026 [astro-ph.CO]} \BibitemShut {NoStop}%
\bibitem [{\citenamefont {Wang}\ \emph {et~al.}(2017)\citenamefont {Wang},
  \citenamefont {Zhao}, \citenamefont {Chuang}, \citenamefont
  {Pellejero-Ibanez}, \citenamefont {Zhao}, \citenamefont {Kitaura},\ and\
  \citenamefont {Rodriguez-Torres}}]{Wang:2017wia}%
  \BibitemOpen
  \bibfield  {author} {\bibinfo {author} {\bibfnamefont {Yuting}\ \bibnamefont
  {Wang}}, \bibinfo {author} {\bibfnamefont {Gong-Bo}\ \bibnamefont {Zhao}},
  \bibinfo {author} {\bibfnamefont {Chia-Hsun}\ \bibnamefont {Chuang}},
  \bibinfo {author} {\bibfnamefont {Marcos}\ \bibnamefont {Pellejero-Ibanez}},
  \bibinfo {author} {\bibfnamefont {Cheng}\ \bibnamefont {Zhao}}, \bibinfo
  {author} {\bibfnamefont {Francisco-Shu}\ \bibnamefont {Kitaura}}, \ and\
  \bibinfo {author} {\bibfnamefont {Sergio}\ \bibnamefont {Rodriguez-Torres}},\
  }\bibfield  {title} {\enquote {\bibinfo {title} {{The clustering of galaxies
  in the completed SDSS-III Baryon Oscillation Spectroscopic Survey: a
  tomographic analysis of structure growth and expansion rate from anisotropic
  galaxy clustering}},}\ }\href@noop {} {\  (\bibinfo {year} {2017})},\ \Eprint
  {http://arxiv.org/abs/1709.05173} {arXiv:1709.05173 [astro-ph.CO]}
  \BibitemShut {NoStop}%
\bibitem [{\citenamefont {Shi}\ \emph {et~al.}(2017)\citenamefont {Shi} \emph
  {et~al.}}]{Shi:2017qpr}%
  \BibitemOpen
  \bibfield  {author} {\bibinfo {author} {\bibfnamefont {Feng}\ \bibnamefont
  {Shi}} \emph {et~al.},\ }\bibfield  {title} {\enquote {\bibinfo {title}
  {{Mapping the Real Space Distributions of Galaxies in SDSS DR7: II. Measuring
  the growth rate, linear mass variance and biases of galaxies at redshift
  0.1}},}\ }\href@noop {} {\  (\bibinfo {year} {2017})},\ \Eprint
  {http://arxiv.org/abs/1712.04163} {arXiv:1712.04163 [astro-ph.CO]}
  \BibitemShut {NoStop}%
\bibitem [{\citenamefont {Gil-Marín}\ \emph {et~al.}(2018)\citenamefont
  {Gil-Marín} \emph {et~al.}}]{Gil-Marin:2018cgo}%
  \BibitemOpen
  \bibfield  {author} {\bibinfo {author} {\bibfnamefont {Héctor}\ \bibnamefont
  {Gil-Marín}} \emph {et~al.},\ }\bibfield  {title} {\enquote {\bibinfo
  {title} {{The clustering of the SDSS-IV extended Baryon Oscillation
  Spectroscopic Survey DR14 quasar sample: structure growth rate measurement
  from the anisotropic quasar power spectrum in the redshift range $0.8 < z <
  2.2$}},}\ }\href {\doibase 10.1093/mnras/sty453} {\bibfield  {journal}
  {\bibinfo  {journal} {Mon. Not. Roy. Astron. Soc.}\ }\textbf {\bibinfo
  {volume} {477}},\ \bibinfo {pages} {1604--1638} (\bibinfo {year} {2018})},\
  \Eprint {http://arxiv.org/abs/1801.02689} {arXiv:1801.02689 [astro-ph.CO]}
  \BibitemShut {NoStop}%
\bibitem [{\citenamefont {Hou}\ \emph {et~al.}(2018)\citenamefont {Hou} \emph
  {et~al.}}]{Hou:2018yny}%
  \BibitemOpen
  \bibfield  {author} {\bibinfo {author} {\bibfnamefont {Jiamin}\ \bibnamefont
  {Hou}} \emph {et~al.},\ }\bibfield  {title} {\enquote {\bibinfo {title} {{The
  clustering of the SDSS-IV extended Baryon Oscillation Spectroscopic Survey
  DR14 quasar sample: anisotropic clustering analysis in
  configuration-space}},}\ }\href@noop {} {\  (\bibinfo {year} {2018})},\
  \Eprint {http://arxiv.org/abs/1801.02656} {arXiv:1801.02656 [astro-ph.CO]}
  \BibitemShut {NoStop}%
\bibitem [{\citenamefont {Zhao}\ \emph {et~al.}(2018)\citenamefont {Zhao} \emph
  {et~al.}}]{Zhao:2018jxv}%
  \BibitemOpen
  \bibfield  {author} {\bibinfo {author} {\bibfnamefont {Gong-Bo}\ \bibnamefont
  {Zhao}} \emph {et~al.},\ }\bibfield  {title} {\enquote {\bibinfo {title}
  {{The clustering of the SDSS-IV extended Baryon Oscillation Spectroscopic
  Survey DR14 quasar sample: a tomographic measurement of cosmic structure
  growth and expansion rate based on optimal redshift weights}},}\ }\href@noop
  {} {\  (\bibinfo {year} {2018})},\ \Eprint {http://arxiv.org/abs/1801.03043}
  {arXiv:1801.03043 [astro-ph.CO]} \BibitemShut {NoStop}%
\bibitem [{\citenamefont {Mohammad}\ \emph {et~al.}(2018)\citenamefont
  {Mohammad} \emph {et~al.}}]{Mohammad:2018mdy}%
  \BibitemOpen
  \bibfield  {author} {\bibinfo {author} {\bibfnamefont {F.~G.}\ \bibnamefont
  {Mohammad}} \emph {et~al.},\ }\bibfield  {title} {\enquote {\bibinfo {title}
  {{The VIMOS Public Extragalactic Redshift Survey (VIPERS): Unbiased
  clustering estimate with VIPERS slit assignment}},}\ }\href {\doibase
  10.1051/0004-6361/201833853} {\bibfield  {journal} {\bibinfo  {journal}
  {Astron. Astrophys.}\ }\textbf {\bibinfo {volume} {619}},\ \bibinfo {pages}
  {A17} (\bibinfo {year} {2018})},\ \Eprint {http://arxiv.org/abs/1807.05999}
  {arXiv:1807.05999 [astro-ph.CO]} \BibitemShut {NoStop}%
\bibitem [{\citenamefont {Nadathur}\ \emph {et~al.}(2019)\citenamefont
  {Nadathur}, \citenamefont {Carter}, \citenamefont {Percival}, \citenamefont
  {Winther},\ and\ \citenamefont {Bautista}}]{Nadathur:2019mct}%
  \BibitemOpen
  \bibfield  {author} {\bibinfo {author} {\bibfnamefont {Seshadri}\
  \bibnamefont {Nadathur}}, \bibinfo {author} {\bibfnamefont {Paul~M.}\
  \bibnamefont {Carter}}, \bibinfo {author} {\bibfnamefont {Will~J.}\
  \bibnamefont {Percival}}, \bibinfo {author} {\bibfnamefont {Hans~A.}\
  \bibnamefont {Winther}}, \ and\ \bibinfo {author} {\bibfnamefont {Julian}\
  \bibnamefont {Bautista}},\ }\bibfield  {title} {\enquote {\bibinfo {title}
  {{Beyond BAO: Improving cosmological constraints from BOSS data with
  measurement of the void-galaxy cross-correlation}},}\ }\href {\doibase
  10.1103/PhysRevD.100.023504} {\bibfield  {journal} {\bibinfo  {journal}
  {Phys. Rev.}\ }\textbf {\bibinfo {volume} {D100}},\ \bibinfo {pages} {023504}
  (\bibinfo {year} {2019})},\ \Eprint {http://arxiv.org/abs/1904.01030}
  {arXiv:1904.01030 [astro-ph.CO]} \BibitemShut {NoStop}%
\bibitem [{\citenamefont {Qin}\ \emph {et~al.}(2019)\citenamefont {Qin},
  \citenamefont {Howlett},\ and\ \citenamefont {Staveley-Smith}}]{Qin:2019axr}%
  \BibitemOpen
  \bibfield  {author} {\bibinfo {author} {\bibfnamefont {Fei}\ \bibnamefont
  {Qin}}, \bibinfo {author} {\bibfnamefont {Cullan}\ \bibnamefont {Howlett}}, \
  and\ \bibinfo {author} {\bibfnamefont {Lister}\ \bibnamefont
  {Staveley-Smith}},\ }\bibfield  {title} {\enquote {\bibinfo {title} {{The
  redshift-space momentum power spectrum – II. Measuring the growth rate from
  the combined 2MTF and 6dFGSv surveys}},}\ }\href {\doibase
  10.1093/mnras/stz1576} {\bibfield  {journal} {\bibinfo  {journal} {Mon. Not.
  Roy. Astron. Soc.}\ }\textbf {\bibinfo {volume} {487}},\ \bibinfo {pages}
  {5235--5247} (\bibinfo {year} {2019})},\ \Eprint
  {http://arxiv.org/abs/1906.02874} {arXiv:1906.02874 [astro-ph.CO]}
  \BibitemShut {NoStop}%
\bibitem [{\citenamefont {Icaza-Lizaola}\ \emph {et~al.}(2019)\citenamefont
  {Icaza-Lizaola} \emph {et~al.}}]{Icaza-Lizaola:2019zgk}%
  \BibitemOpen
  \bibfield  {author} {\bibinfo {author} {\bibfnamefont {M.}~\bibnamefont
  {Icaza-Lizaola}} \emph {et~al.},\ }\bibfield  {title} {\enquote {\bibinfo
  {title} {{The clustering of the SDSS-IV extended Baryon Oscillation
  Spectroscopic Survey DR14 LRG sample: structure growth rate measurement from
  the anisotropic LRG correlation function in the redshift range 0.6 < z <
  1.0}},}\ }\href@noop {} {\  (\bibinfo {year} {2019})},\ \Eprint
  {http://arxiv.org/abs/1909.07742} {arXiv:1909.07742 [astro-ph.CO]}
  \BibitemShut {NoStop}%
\bibitem [{\citenamefont {Amon}\ \emph {et~al.}(2018)\citenamefont {Amon} \emph
  {et~al.}}]{Amon:2017lia}%
  \BibitemOpen
  \bibfield  {author} {\bibinfo {author} {\bibfnamefont {A.}~\bibnamefont
  {Amon}} \emph {et~al.},\ }\bibfield  {title} {\enquote {\bibinfo {title}
  {{KiDS+2dFLenS+GAMA: Testing the cosmological model with the $E_{\rm G}$
  statistic}},}\ }\href {\doibase 10.1093/mnras/sty1624} {\bibfield  {journal}
  {\bibinfo  {journal} {Mon. Not. Roy. Astron. Soc.}\ }\textbf {\bibinfo
  {volume} {479}},\ \bibinfo {pages} {3422--3437} (\bibinfo {year} {2018})},\
  \Eprint {http://arxiv.org/abs/1711.10999} {arXiv:1711.10999 [astro-ph.CO]}
  \BibitemShut {NoStop}%
\bibitem [{\citenamefont {Singh}\ \emph {et~al.}(2019)\citenamefont {Singh},
  \citenamefont {Alam}, \citenamefont {Mandelbaum}, \citenamefont {Seljak},
  \citenamefont {Rodriguez-Torres},\ and\ \citenamefont {Ho}}]{Singh:2018flu}%
  \BibitemOpen
  \bibfield  {author} {\bibinfo {author} {\bibfnamefont {Sukhdeep}\
  \bibnamefont {Singh}}, \bibinfo {author} {\bibfnamefont {Shadab}\
  \bibnamefont {Alam}}, \bibinfo {author} {\bibfnamefont {Rachel}\ \bibnamefont
  {Mandelbaum}}, \bibinfo {author} {\bibfnamefont {Uros}\ \bibnamefont
  {Seljak}}, \bibinfo {author} {\bibfnamefont {Sergio}\ \bibnamefont
  {Rodriguez-Torres}}, \ and\ \bibinfo {author} {\bibfnamefont {Shirley}\
  \bibnamefont {Ho}},\ }\bibfield  {title} {\enquote {\bibinfo {title}
  {{Probing gravity with a joint analysis of galaxy and CMB lensing and SDSS
  spectroscopy}},}\ }\href {\doibase 10.1093/mnras/sty2681} {\bibfield
  {journal} {\bibinfo  {journal} {Mon. Not. Roy. Astron. Soc.}\ }\textbf
  {\bibinfo {volume} {482}},\ \bibinfo {pages} {785--806} (\bibinfo {year}
  {2019})},\ \Eprint {http://arxiv.org/abs/1803.08915} {arXiv:1803.08915
  [astro-ph.CO]} \BibitemShut {NoStop}%
\bibitem [{\citenamefont {Blake}\ \emph {et~al.}(2016)\citenamefont {Blake}
  \emph {et~al.}}]{Blake:2015vea}%
  \BibitemOpen
  \bibfield  {author} {\bibinfo {author} {\bibfnamefont {Chris}\ \bibnamefont
  {Blake}} \emph {et~al.},\ }\bibfield  {title} {\enquote {\bibinfo {title}
  {{RCSLenS: Testing gravitational physics through the cross-correlation of
  weak lensing and large-scale structure}},}\ }\href {\doibase
  10.1093/mnras/stv2875} {\bibfield  {journal} {\bibinfo  {journal} {Mon. Not.
  Roy. Astron. Soc.}\ }\textbf {\bibinfo {volume} {456}},\ \bibinfo {pages}
  {2806--2828} (\bibinfo {year} {2016})},\ \Eprint
  {http://arxiv.org/abs/1507.03086} {arXiv:1507.03086 [astro-ph.CO]}
  \BibitemShut {NoStop}%
\bibitem [{\citenamefont {Alam}\ \emph
  {et~al.}(2017{\natexlab{b}})\citenamefont {Alam}, \citenamefont {Miyatake},
  \citenamefont {More}, \citenamefont {Ho},\ and\ \citenamefont
  {Mandelbaum}}]{Alam:2016qcl}%
  \BibitemOpen
  \bibfield  {author} {\bibinfo {author} {\bibfnamefont {Shadab}\ \bibnamefont
  {Alam}}, \bibinfo {author} {\bibfnamefont {Hironao}\ \bibnamefont
  {Miyatake}}, \bibinfo {author} {\bibfnamefont {Surhud}\ \bibnamefont {More}},
  \bibinfo {author} {\bibfnamefont {Shirley}\ \bibnamefont {Ho}}, \ and\
  \bibinfo {author} {\bibfnamefont {Rachel}\ \bibnamefont {Mandelbaum}},\
  }\bibfield  {title} {\enquote {\bibinfo {title} {{Testing gravity on large
  scales by combining weak lensing with galaxy clustering using CFHTLenS and
  BOSS CMASS}},}\ }\href {\doibase 10.1093/mnras/stw3056} {\bibfield  {journal}
  {\bibinfo  {journal} {Mon. Not. Roy. Astron. Soc.}\ }\textbf {\bibinfo
  {volume} {465}},\ \bibinfo {pages} {4853--4865} (\bibinfo {year}
  {2017}{\natexlab{b}})},\ \Eprint {http://arxiv.org/abs/1610.09410}
  {arXiv:1610.09410 [astro-ph.CO]} \BibitemShut {NoStop}%
\bibitem [{\citenamefont {Jullo}\ \emph {et~al.}(2019)\citenamefont {Jullo}
  \emph {et~al.}}]{Jullo:2019lgq}%
  \BibitemOpen
  \bibfield  {author} {\bibinfo {author} {\bibfnamefont {E.}~\bibnamefont
  {Jullo}} \emph {et~al.},\ }\bibfield  {title} {\enquote {\bibinfo {title}
  {{Testing gravity with galaxy-galaxy lensing and redshift-space distortions
  using CFHT-Stripe 82, CFHTLenS and BOSS CMASS datasets}},}\ }\href {\doibase
  10.1051/0004-6361/201834629} {\bibfield  {journal} {\bibinfo  {journal}
  {Astron. Astrophys.}\ }\textbf {\bibinfo {volume} {627}},\ \bibinfo {pages}
  {A137} (\bibinfo {year} {2019})},\ \Eprint {http://arxiv.org/abs/1903.07160}
  {arXiv:1903.07160 [astro-ph.CO]} \BibitemShut {NoStop}%
\end{thebibliography}%

\end{document}